\documentclass[12pt,preprint]{aastex}
\shorttitle{}
\shortauthors{Nesvorn\'y et al.}
\begin{document}
\title{Modeling the Historical Flux of Planetary Impactors}
\author{David Nesvorn\'y$^{1,2}$, Fernando Roig$^2$, William F. Bottke$^1$}
\affil{(1) Department of Space Studies, Southwest Research Institute, 1050 Walnut St., \\Suite 300, 
Boulder, CO 80302, USA} 
\affil{(2) Observat\'orio Nacional, Rua Gal. Jose Cristino 77, Rio de Janeiro, RJ 20921-400, Brazil}
\begin{abstract}
The impact cratering record of the Moon and the terrestrial planets provides important clues about the formation
and evolution of the Solar System. Especially intriguing is the epoch $\simeq$3.8-3.9~Gyr ago (Ga), known as 
the Late Heavy Bombardment (LHB), when the youngest lunar basins such as Imbrium and Orientale formed. The LHB was 
suggested to originate from a slowly declining impactor flux or from a late dynamical instability. Here we develop a 
model for the historical flux of large asteroid impacts and discuss how it depends on various parameters, including 
the time and nature of the planetary migration/instability. We find that the asteroid impact flux dropped by 1 to 2 
orders of magnitude during the first 1 Gyr and remained relatively unchanged over the last 3 Gyr. The early impacts 
were produced by asteroids whose orbits became excited during the planetary migration/instability, and by those 
originating from the inner extension of the main belt (E-belt; semimajor axis $1.6<a<2.1$ au). The inner main belt 
dominated the asteroid impact record after $\sim$3-4~Ga. The profiles obtained for the early and late versions of 
the planetary instability initially differ, but end up being similar after $\sim$3 Ga. Thus, the time of the instability 
can only be determined by considering the cratering and other constraints during the first $\simeq$1.5 Gyr of the 
Solar System history. Our absolute calibration of the impact flux indicates that asteroids were probably {\it not} 
responsible for the LHB, independently of whether the instability happened early or late, because the calibrated flux 
is not large enough to explain Imbrium/Orientale and a significant share of large lunar craters. Comets and leftovers 
of the terrestrial planet formation provided additional, and probably dominant source of impacts during early epochs. 
The cometary impacts occur during a narrow interval after the instability and would not be recorded on surfaces if 
the planetary instability happened early. 
\end{abstract}
\section{Introduction}
The first Solar System solids condensed 4.568 Gyr ago (Kleine et al. 2009). This is considered as time zero in the
Solar System history ($t_0$). Jupiter and Saturn have significant gas envelopes and must have formed within the lifetime 
of the protosolar gas disk, probably in 2-10 Myr after $t_0$ (Haisch et al. 2001, Williams \& Cieza 2011). Geochemical 
constraints and numerical modeling suggest that the terrestrial planet formation ended much later, some 30-100 Myr after 
the gas disk dispersal. After these earliest stages, the basic structure of the Solar System was in place, but the 
Solar System continued to evolve as evidenced by the Late Heavy Bombardment (LHB; $\simeq$3.8-3.9~Ga; see Chapman et al. 
(2007) for a review). 

In principle, the LHB could have been caused by a slowly decaying flux of impactors left among the terrestrial planets after 
their accretion, but modeling indicated that this population may have decayed too fast to produce the youngest lunar basins 
(Bottke et al. 2007). Instead, it has been argued that the impact record can perhaps be best understood as a `sawtooth' profile 
(Bottke et al. 2012, Morbidelli et al. 2012, Marchi et al. 2012), a combination of decaying flux from the terrestrial planet 
accretion leftovers, and a modest increase in the number of impacts (by a factor of 5 or so) produced by a dynamical instability 
in the outer Solar System (the Nice model).

The Nice model is an umbrella term for a broad class of dynamical models in which the giant planets experienced a dynamical instability
(Tsiganis et al. 2005, Morbidelli et al. 2007, Nesvorn\'y \& Morbidelli 2012).
Uranus and Neptune migrated into a trans-planetary disk of icy planetesimals during the instability, 
flinging its members throughout the Solar System. In addition, as Jupiter and Saturn moved toward their present
orbits, asteroid impactors were released from the previously stable reservoirs between Mars and Jupiter. This event 
can be a potential source for the LHB, because previous work indicated that the destabilized asteroids could have been capable 
of producing some of the youngest lunar basins, assuming that the instability happened late (Gomes et al. 2005, Levison et al. 
2011, Bottke et al. 2012).  

The goal of this paper is to model the historical flux of impactors in the inner Solar System, and discuss how various model
fluxes compare with the cratering record of terrestrial worlds (e.g., Neukum et al. 2001, Robbins 2014). Our main focus is 
the asteroid impactors. We consider several different models of planetary instability/migration, including cases where the 
orbital evolution of planets is taken from self-consistent simulations (Nesvorn\'y \& Morbidelli  2012). 
These instability/migration  models were previously constrained from Jupiter Trojans, the moons of the giant planets, the orbital 
structure of asteroids and Kuiper belt objects, etc. We therefore have confidence that the orbital evolution of planets 
in these models is reasonable. 

Section 2 describes our methods. The results are reported in Section 3 and discussed in Section 5. Section 4 points
out several caveats of the numerical scheme adopted in this work. Because the exact time of the instability cannot be determined 
from the numerical modeling alone, we consider cases with the instability occurring at $t_{\rm inst}$ after $t_0$, where $t_{\rm inst}$ 
is a free parameter. We find that the impact flux of asteroids on the terrestrial worlds declines relatively slowly with time. 
With such a slow decay, it would be possible to have large impacts on the Moon happening at $\simeq$3.8-3.9 Ga even if the planetary 
instability happened early ($t_{\rm inst}<100$ Myr). Our absolute calibration of the asteroid impact flux (see Section 3.9), however, 
indicates that the flux of large asteroid impactors was not high enough to account for the formation of the Imbrium and 
Orientale basins $\simeq$3.8-3.9 Ga. Asteroids were thus probably not responsible for the LHB. Different alternatives are 
discussed in Section 5.
\section{Methods}
We performed several numerical integrations. Each of the integrations included planets, which were treated as 
massive bodies that gravitationally interact among themselves and affect the orbits of all other bodies, and 
asteroids, which were assumed to be massless (they do not affect each other and the planets). The integrations were 
performed with the {\it Swift} $N$-body program (Levison \& Duncan 1994), which is an efficient implementation of the 
symplectic Wisdom-Holman map (Wisdom \& Holman 1991). Specifically, we used the code known as {\tt swift\_rmvs3} that 
we adopted for the problem in hand (see below). Non-gravitational forces such as the Yarkovsky effect (Bottke et al. 
2006) were not included. The results are therefore strictly applicable only for asteroids with diameters $D\gtrsim10$ 
km, for which the non-gravitational forces are not important. We also ignored the collisional evolution of the 
asteroid belt, because $D\gtrsim10$-km asteroids have very long collisional lifetimes (Bottke et al. 2005). 

Several different setups were considered for the orbital evolution of planets. In the reference case, the planets 
from Venus to Neptune were placed on their current orbits and integrated for 4.5 Gyr (cf. Minton \& Malhotra 2010). Mercury 
was not included to save the computation time. The time step was 0.03~years or about 11 days, which is roughly 1/20 
of the Venus orbital period. In addition to planets, the integration tracked the orbits of 20,000 asteroids. 
Initially, asteroids were uniformly distributed on orbits with the semimajor axes $1.6 < a < 3.3$ au, eccentricities 
$e<0.4$, perihelion distances $q=a(1-e)>1.6$ au, aphelion distances $Q=a(1+e)<4.5$ au, and inclinations $i<30^\circ$. 
The remaining three orbital angles (mean longitude $\lambda$, perihelion longitude $\varpi$, nodal longitude $\Omega$) 
were chosen on random between 0 and $2\pi$.
\subsection{Impact Flux Computation}    
The orbital elements of planets and asteroids were saved at fixed time intervals ($10^5$ or $10^7$ years) 
and used as an input for an \"Opik-style collisional code (e.g., Greenberg 1982, Bottke \& Greenberg 1993), which allowed us
to compute the impact flux on different target bodies. The effects of gravitational focusing were included.\footnote{A 
double focusing factor was used for the Moon to account for the gravitational focusing of the Earth at the Moon
distance, and for the gravitational focusing of the Moon at its surface.}  
We used two versions of the \"Opik code to estimate the uncertainty due to the approximate nature of the impact probability 
calculation. The standard \"Opik code requires that $\lambda$, $\varpi$ and $\Omega$ are randomly distributed. 
A modified code, developed in Vokrouhlick\'y et al. (2012), takes into account the Kozai cycles (Kozai 1962), which 
can be especially important for highly inclined orbits. The differences between the results obtained with the two codes 
were found to be insignificant. 

The {\it Swift} integrator recorded all impacts of asteroids on the terrestrial planets which occurred during the integration. 
This offers an opportunity to compare the number of recorded impacts with the impact profiles obtained from the \"Opik code. 
Ideally, the results should be identical. In reality, however, there are several complicating factors. First, there 
is only a very small number of recorded impacts for $t>1$ Gyr after the start of each integration, because the initial
number of asteroids used in our simulations was not large enough. It is therefore difficult to compare the impact profiles
obtained from the two methods for $t>1$ Gyr. Second, the \"Opik code provides only an approximate representation
of the impact record, because the orbital longitudes of the impactor and target bodies may not be randomly distributed. 
Some differences between the two approaches are therefore expected. We found that the results for Venus and Earth are consistent
in that the timing of recorded impacts follows, within a factor of $<$2, the impact profiles computed from the \"Opik code.
In contrast, the number of recorded late impacts on Mars can be significantly smaller, by a factor of $\sim$2 in some cases, 
than the expectation from the \"Opik probability. We discuss this issue in Section 3.7. 

The \"Opik codes and {\it Swift}-recorded impacts provide estimates of the (time-dependent) impact flux of asteroids on 
different terrestrial worlds.\footnote{The impacts on Mercury are not considered in this work, because the population
of asteroid orbits below 0.4 au may not be correctly modeled with our 11-day time step.} 
The flux is normalized to one asteroid initially in the source region. An independent calibration 
is needed to obtain the actual flux. To this end we determined the number of asteroids surviving in the asteroid belt at 
$t=4.5$~Gyr after the start of the integration and compared it with the actual number of known asteroids (of a given size). 
This resulted in a factor that was applied to compute an absolutely calibrated historical record of impacts.
Another calibration method is to consider the current impact flux on the Earth inferred from observations of the near-Earth asteroids 
(NEAs; Stuart \& Binzel 2004, Johnson et al. 2016). We discuss the calibrated fluxes in Sections~3.9 and 3.10. 
\subsection{Planetary Instability/Migration}
Our additional integrations used the same initial orbits for asteroids as the reference case (hereafter the REF case)
but different orbital histories of planets featuring an instability. These histories were either idealized using the method 
developed in Bottke et al. (2012) 
or taken directly from self-consistent simulations of the planetary instability/migration (Nesvorn\'y \& Morbidelli 2012, 
Roig et al. 2016). In the former case, the outer planets were placed into an initially resonant configuration taken from 
Nesvorn\'y \& Morbidelli (2012). The terrestrial planets were inserted on orbits with zero eccentricities and zero inclinations. 
The dynamical instability was assumed to have happen at $t_{\rm inst}=400$ Myr (Bottke et al. 2012). We mimicked the 
instability by instantaneously transporting all planets to their present orbits at $t_{\rm inst}$ and continuing the 
integration for additional 4.1 Gyr (hereafter the STEP case). 

In our self-consistent simulations, the orbits of the outer planets were taken from Case~1 described in Nesvorn\'y et al. 
(2013). We used a hybrid integrator (Roig et al. 2016), where the orbits of the outer planets were interpolated
from files recorded previously, while the orbits of the inner planets and asteroids were integrated with 
{\tt swift\_rmvs3}.\footnote{The code accounts for all gravitational interactions between planets except for 
the gravitational effects of the inner planets on the outer ones. We verified that this is a minor 
approximation in what concerns the orbital evolution of asteroids.} Two cases were considered. In the first case, 
the Angular Momentum Deficit (AMD) of the terrestrial planets was initially set to be zero (i.e., zero initial 
eccentricities and zero initial inclinations; CASE1). In the second case, the initial AMD of the terrestrial 
planets was set to be only slightly lower than their present AMD (CASE1B). 

To generate a plausible evolution of the terrestrial planets during the instability, we performed one hundred preliminary 
simulations in each case starting from different initial configurations of the inner planets (asteroids were not included 
at this stage). The initial nodal, perihelion and mean longitudes of the inner planets were chosen at random in different 
trials. We used the hybrid code where the orbits of the outer planets were interpolated from Case~1 (Nesvorn\'y et al. 2013). 
All orbits were propagated 100 Myr past the instability (a longer integration was not needed because the planetary orbits 
were practically unchanging at later times).
We performed the Fourier analysis and checked how well the secular structure of final orbits obtained in each 
trial corresponded to the actual Solar System. Two trials were selected where the correspondence was reasonably 
good (no. 5 in the zero AMD case and no. 41 in the high AMD case). Then, in the main runs, the 
selected integrations were repeated with the hybrid code, extending them to 4.5 Gyr, and this time including 
asteroids. Figure \ref{case1b} shows the orbital history of planets in the selected case with the high initial 
AMD. 
\subsection{Summary of Performed Integrations}
In summary, we performed four full simulations over 4.5 Gyr: the (1) reference integration with no instability (hereafter the 
REF case), (2) integration with an idealized step-like instability at $t=400$ Myr (the STEP case), (3) Case-1 instability 
integration from Nesvorn\'y et al. (2013) with zero initial AMD of the terrestrial planets (CASE1), and (4) Case 1 
instability integration where the initial AMD of the terrestrial planets was set to be slightly lower than their present AMD (CASE1B). 
To improve the statistics in CASE1B we used 50,000 initial asteroid orbits with $i<20^\circ$ (Figure \ref{init}). 
Larger orbital inclinations were ignored, because the primordial asteroid belt most likely did not have a large population
of bodies with $i>20^\circ$ (Morbidelli et al. 2010, Roig \& Nesvorn\'y 2015, Deienno et al. 2016). See Table 1 for basic 
parameters of our four simulations.  

The four cases discussed here represent a very crude sampling of parameter space. The REF and STEP cases 
idealize the orbital evolution of planets, but have the advantage of planets having exactly the correct orbits 
in the end. The long-term depletion of the asteroid belt should therefore be more realistic in these cases. CASE1 
and CASE1B, on the other hand, self-consistently model the evolution of planetary orbits during the instability, 
but the final planetary orbits are not exactly equal to the present ones. These models should thus more closely 
reproduce the effect of the planetary instability on the asteroid orbits, but may show some defects later on. By 
considering all these cases together we are able to identify potential flaws produced by the approximate 
nature of each model (see Section~3), and determine the historical profile of impact flux that should most 
closely represent the reality.
\subsection{Time of the Instability} 

\begin{figure}
\epsscale{0.7}
\plotone{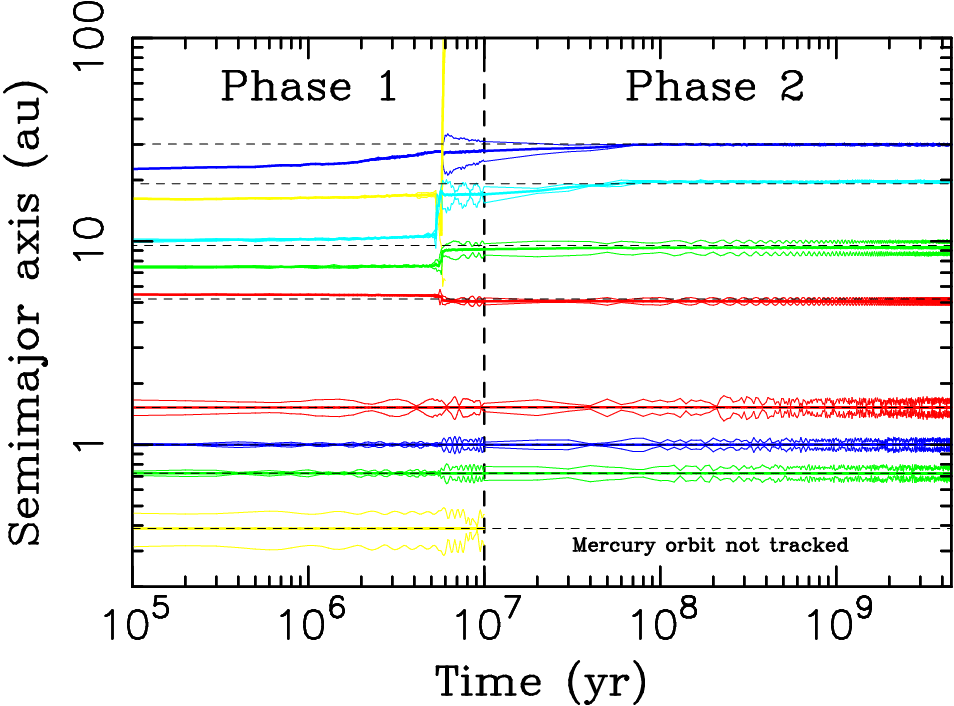}
\caption{The orbital histories of planets in the CASE1B integration. The semimajor axes of planets
are shown by thick solid lines. The thin solid lines are the perihelion and aphelion distances. The 
third ice giant (yellow lines) was initially placed between the orbits of Uranus and Neptune. Phase 1 
is the integration segment where the orbits were evolved with the hybrid code described in Section~2 
and Roig et al. (2016). The instability happened at $t=5.7$ Myr after the start of the integration. The 
third ice giant was ejected from the Solar System soon after that. The remaining outer planets were 
migrated to their present orbits. Mercury's orbit was not tracked during Phase 2 to save the computation 
time.}
\label{case1b}
\end{figure}

\begin{figure} 
\epsscale{0.5}
\plotone{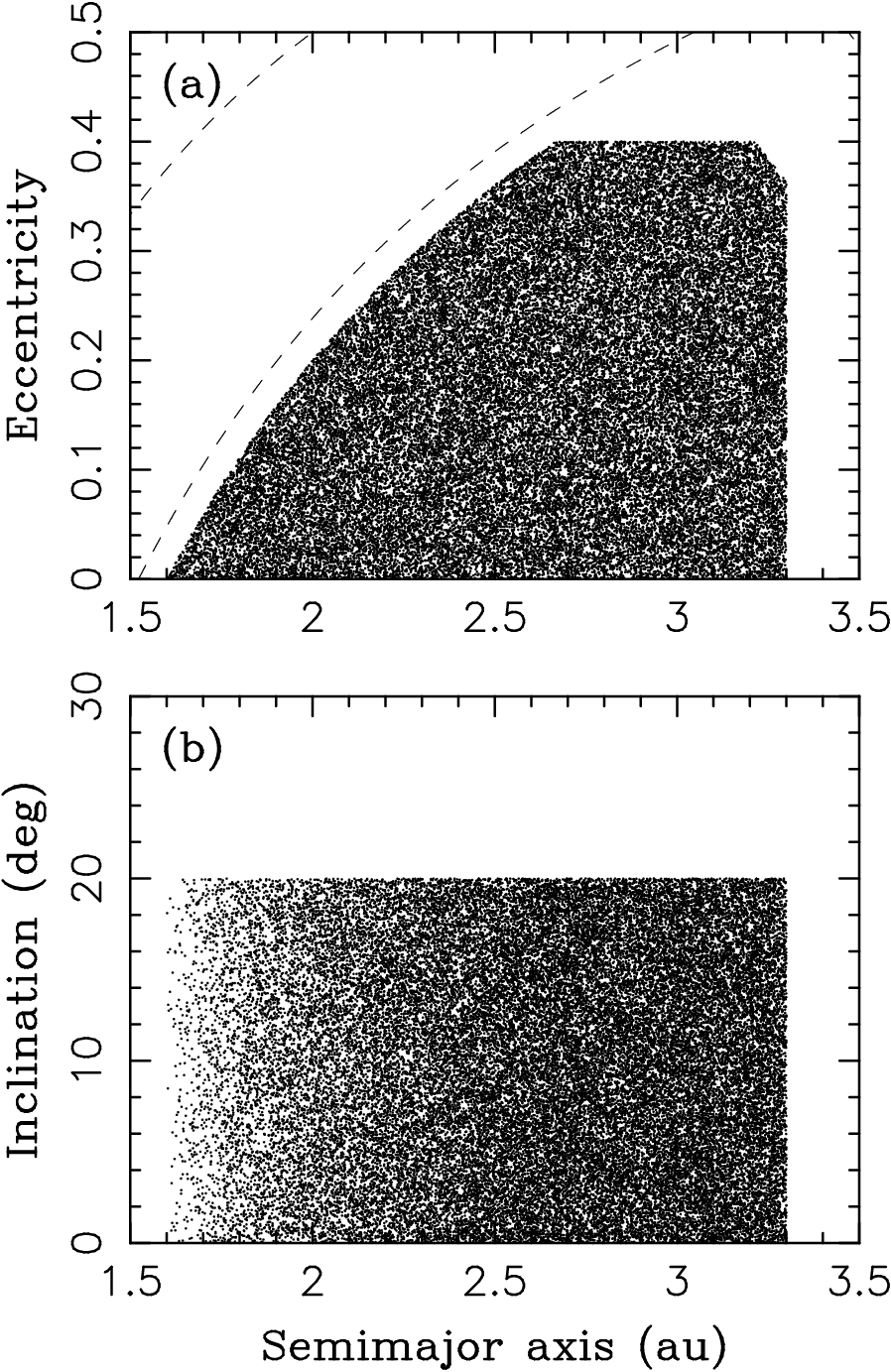}
\caption{The initial distribution of asteroid orbits in CASE1B ($i<20^\circ$ and $N_{\rm ast}=50000$). 
The initial distributions in other cases considered in this work were similar except that we used 
a higher inclination cutoff in REF and CASE1 ($i<30^\circ$) and smaller statistics in REF 
($N_{\rm ast}=20000$) and CASE1 ($N_{\rm ast}=10000$).} 
\label{init}
\end{figure}

Since the time of the planetary instability, $t_{\rm inst}$, cannot be determined from the modeling alone, we considered cases 
with different $t_{\rm inst}$, where $t_{\rm inst}$ was taken as a free parameter. We developed the following method to 
be able to obtain results for any $t_{\rm inst}$. In the original CASE1 and CASE1B integrations, the instability happened 
at $t\simeq 5.7$ Myr after the start of the integration (Figure \ref{case1b}). To be able to use these results with 
any $t_{\rm inst}$, we performed additional simulations designed to represent the situation before the instability 
(Phase 0). The additional integrations started with the same initial orbits of planets and asteroids that we used in CASE1 
and CASE1B, disregarded the instability and migration, and used the standard {\tt swift\_rmvs3} integrator to propagate orbits. 
The integrations covered 1 Gyr allowing us to choose $t_{\rm inst}$ in the $0<t_{\rm inst}<1$ Gyr interval. The results 
were used to compute the impact flux before the instability (i.e., from $t_0$ to $t_{\rm inst}$) following the \"Opik algorithm 
discussed above (or from the number of recorded impacts).  

The Phase0 integrations cannot be directly ``glued'' to the original runs (Phases 1 and 2; Figure \ref{case1b}), because
the planetary orbits at $t=t_{\rm inst}$ in Phase 0 are different from the planetary orbits at $t=0$ in Phase 1. 
Instead, to compute the flux after the instability, we fixed the value of $t_{\rm inst}$, and labeled all asteroids that have 
not been eliminated before $t_{\rm inst}$ in the Phase-0 simulations. We then returned to the original runs 
and computed the impact fluxes with the \"Opik code, but this time {\it only considering the labeled asteroids} 
that we know were not eliminated before $t_{\rm inst}$. This procedure should give approximately correct impact rates after 
$t_{\rm inst}$ assuming that the planetary orbits did not change/migrate much before $t_{\rm inst}$. The beauty of this 
method is that it does not generate any discontinuity in the evolution of planets. 

The same procedure can be applied to the REF simulation to simulate an instability at $t_{\rm inst}$. To do that, we performed 
the Phase-0 simulation for REF, where the initial planetary orbits were taken from STEP. The Phase-0 impact profiles were then 
connected to the REF profiles at $t_{\rm inst}$ using the labeling method described above. To verify that this method works 
properly, we applied it to the REF case with $t_{\rm inst}=400$~Myr (meaning that we used the Phase-0 results for $t<400$ Myr 
and connected them to the Phase-1 \& 2 results for $t>400$ Myr) and compared the results with the STEP case, where we also 
had $t_{\rm inst}=400$ Myr. We found that the results were practically identical thus demonstrating that the labeling 
procedure works well (Section 3.2). 
\subsection{Initial Orbital Distribution of Asteroids}

We developed the following method to test the dependence of the results on the initial orbital distribution of 
asteroids.\footnote{The instability-driven implantation of planetesimals from the trans-Neptunian disk into the main 
belt (Levison et al. 2009, Vokrouhlick\'y et al. 2016) was ignored in this work.} 
First, we chose a mask representing the initial orbits. We tested four different masks, all of which were previously shown to 
lead to the orbital distribution of asteroids that is consistent with the present asteroid belt (Morbidelli et al. 2010, Minton 
\& Malhotra 2011, Roig \& Nesvorn\'y 2015, Deienno et al. 2016). The first one was simply an inclination cutoff $i<20^\circ$
(Morbidelli et al. 2010). The second one was taken from Bottke et al. (2012), where the initial eccentricities and 
initial inclinations were represented by the Gaussian distributions with 0.1 and 10$^\circ$ means, respectively. The third mask,
the Rayleigh distributions in $e$ and $i$ with $\gamma_e=0.1$ and $\gamma_i=10^\circ$, was motivated by the results 
of Roig \& Nesvorn\'y (2015). Finally, we also used a mask from the Grand Tack model (Walsh et al. 2011), where the initial 
eccentricity distribution was approximated by the Gaussian distribution with the mean $\mu_e=0.3$ and standard deviation 
$\sigma_e=0.2$ (Deienno et al. 2016). See Roig \& Nesvorn\'y (2015) for a formal definition of these masks.
The initial distributions of orbits were obtained from the original uniform distributions 
by applying the appropriate weights and propagating this information through the impact flux calculation.   
\section{Results}
We first consider the orbital evolution of the asteroid belt (Section 3.1), and then move to describing the general
appearance of impact profiles (Sections 3.2 and 3.3). The dependence of the impact profiles on the initial distribution of asteroids 
is discussed in Section 3.4. The following sections cover various subjects related to the contribution of different source 
regions (Section 3.5), dependence on $t_{\rm inst}$ (Section 3.6), comparison of the \"Opik-code results with impacts recorded 
by the {\it Swift} integrator (Section 3.7), impact fluxes on different target bodies (Section 3.8), absolute flux 
calibration (Section 3.9), comparison with the current day impact flux (section 3.10), and exponential parametrization 
of the differential impact flux (Section 3.11).
\subsection{Asteroid Belt Evolution}
\begin{figure}
\epsscale{0.23}
\plotone{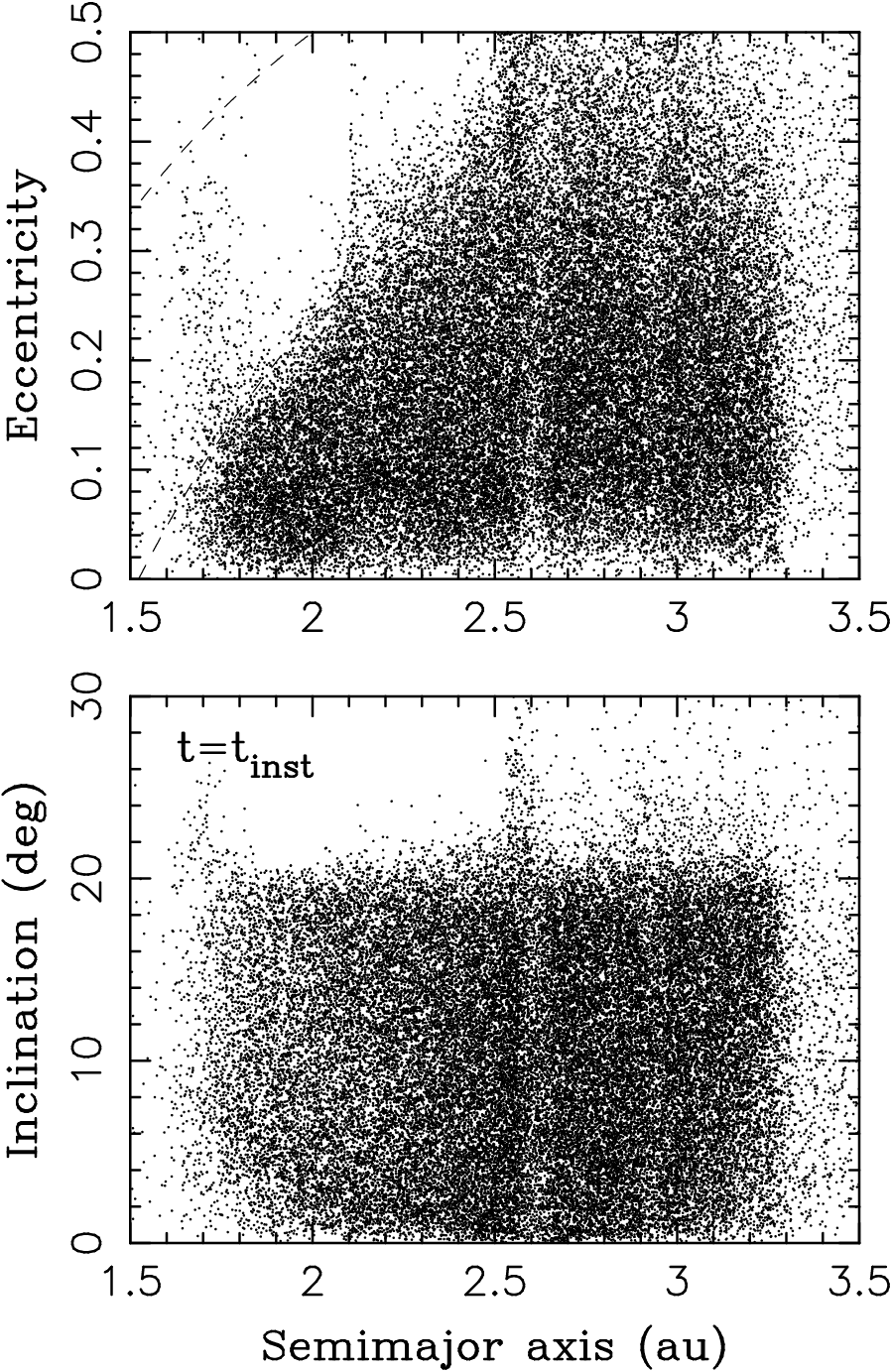}
\plotone{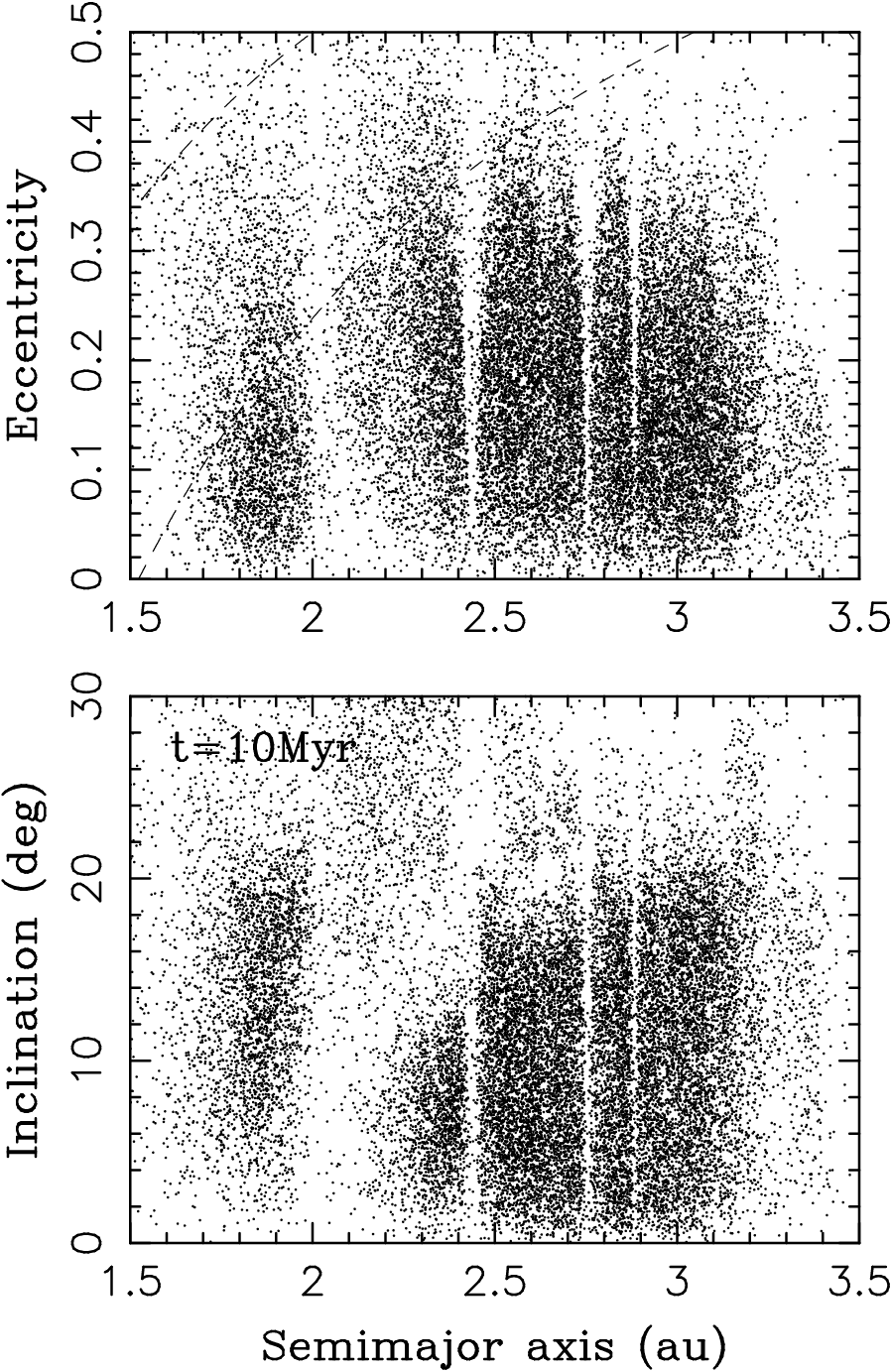}
%\plotone{case1b_0.1Gyr.eps}
\plotone{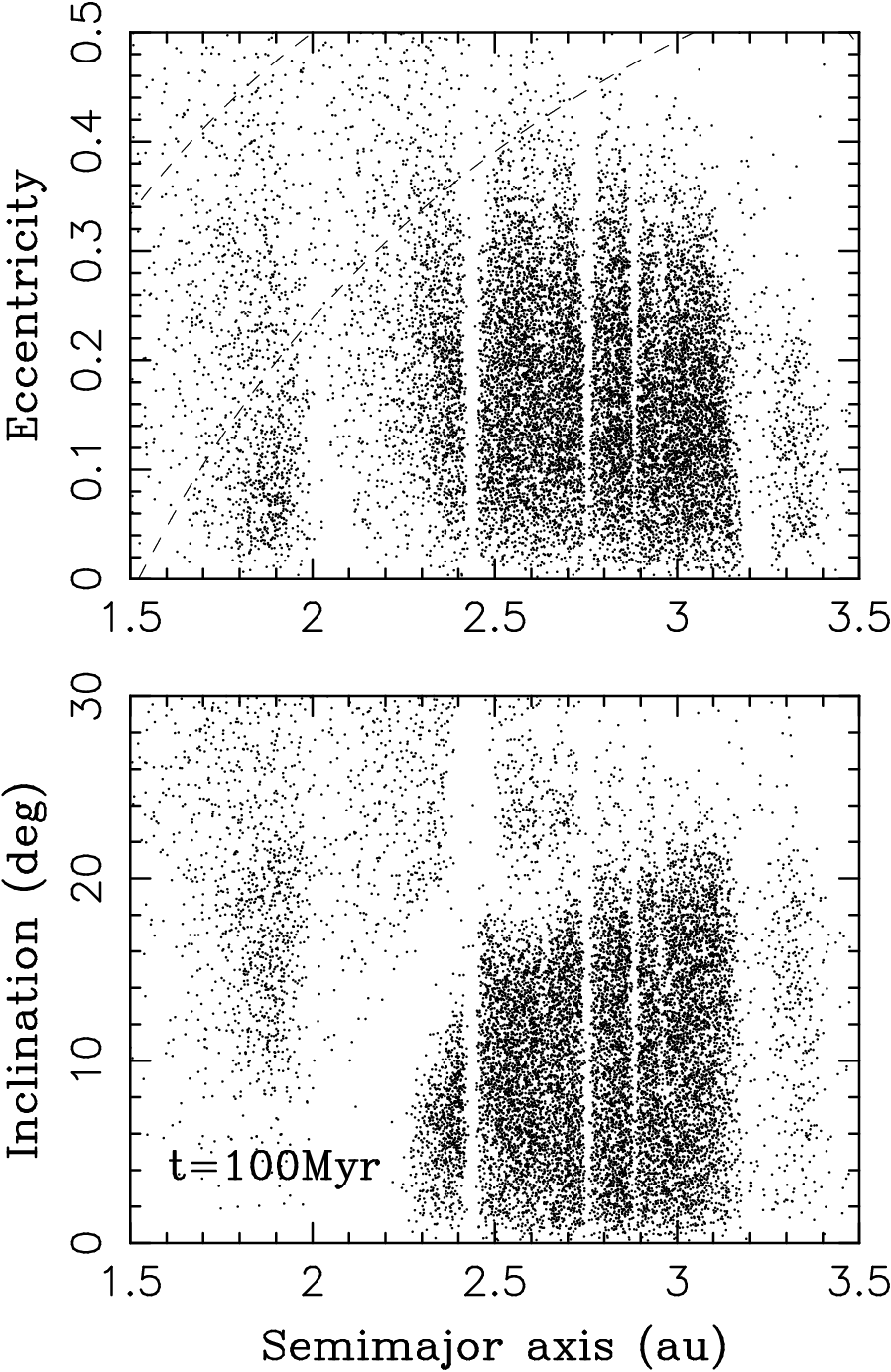}
\plotone{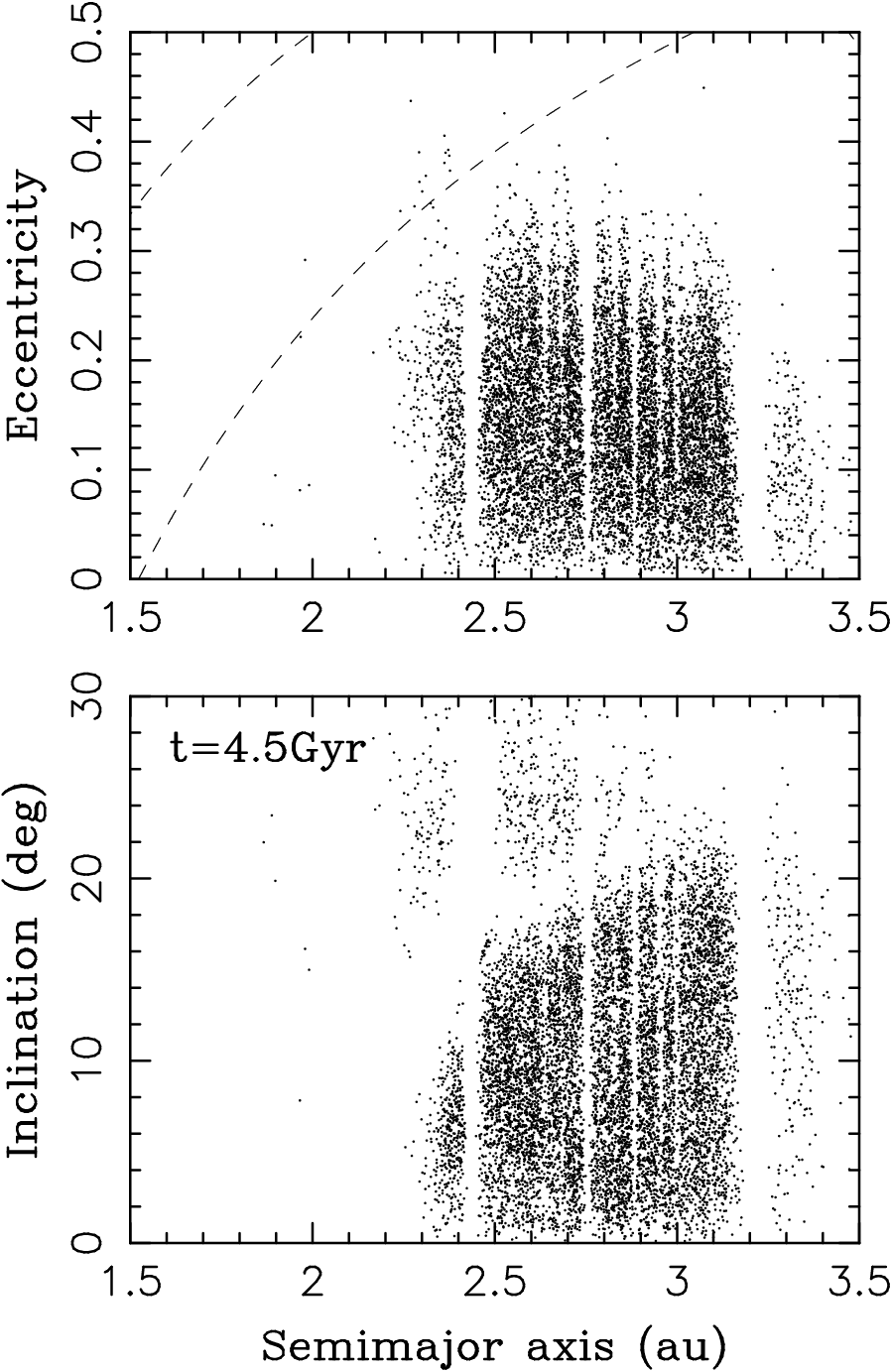}
\caption{The orbital evolution of the asteroid belt in the CASE1B integration. From left to right the 
different panels show snapshots of the orbital distribution at $t=t_{\rm inst}$ (i.e., at the time of 
the instability), $t=t_{\rm inst}+10$ Myr (just after the instability), $t=t_{\rm inst}+100$ Myr and 
$t=t_{\rm inst}+4.5$ Gyr. The rightmost panel is the final orbital distribution obtained in the 
model.}
\label{snaps}
\end{figure}

Figure \ref{snaps} shows several snapshots of the asteroid belt evolution in the CASE1B simulation. Before
the instability, the orbital distribution closely resembles the initial distribution shown in Figure \ref{init}.
The situation dramatically changes during the instability when orbits become excited. The orbital excitation is especially 
noticeable for $a<2$ au, where many orbits reach $q<1.6$ au (see the second panel from the left in Figure \ref{snaps})
and become Mars crossing. Subsequently, with Jupiter and Saturn converging to their current orbits, the mean motion and secular 
resonances move toward their current positions and deplete portions of orbital space. The familiar structure of the asteroid 
belt gradually emerges with practically no objects below 2.2 au, a strongly depleted inner belt, the inclination distribution 
sculpted by the secular resonances $\nu_{6}$ and $\nu_{16}$, and the Kirkwood gaps at the mean motion resonances with 
Jupiter.\footnote{Since Jupiter 
ended $\simeq$0.1 au inside of its current orbit ($a_{\rm Jup}=5.205$ au) in the simulation, the mean motion resonances 
are slightly shifted with respect to their actual positions. We made no effort in Figure \ref{snaps} to correct for 
that.}

\begin{figure} 
\epsscale{0.23}
\plotone{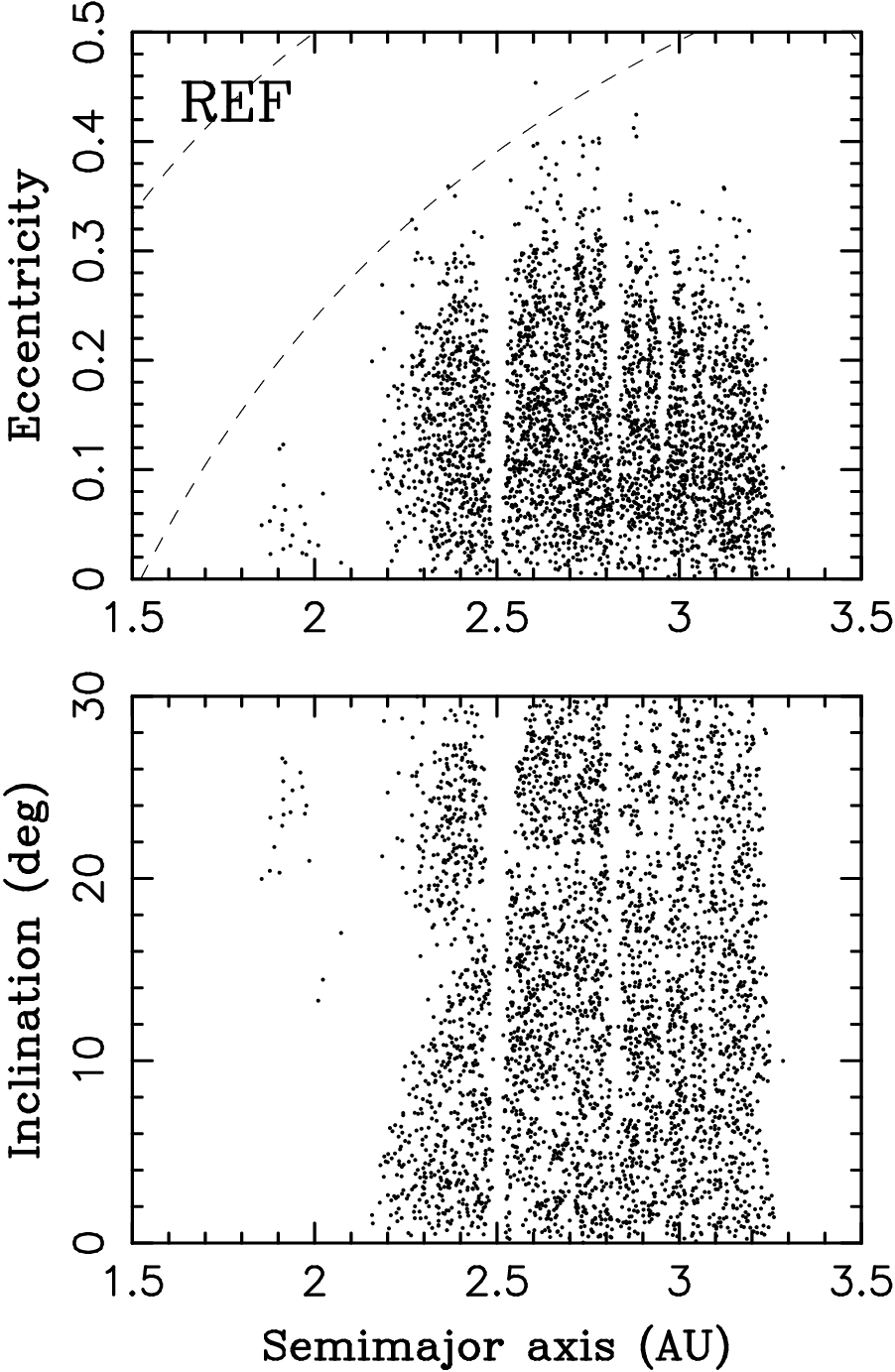}
\plotone{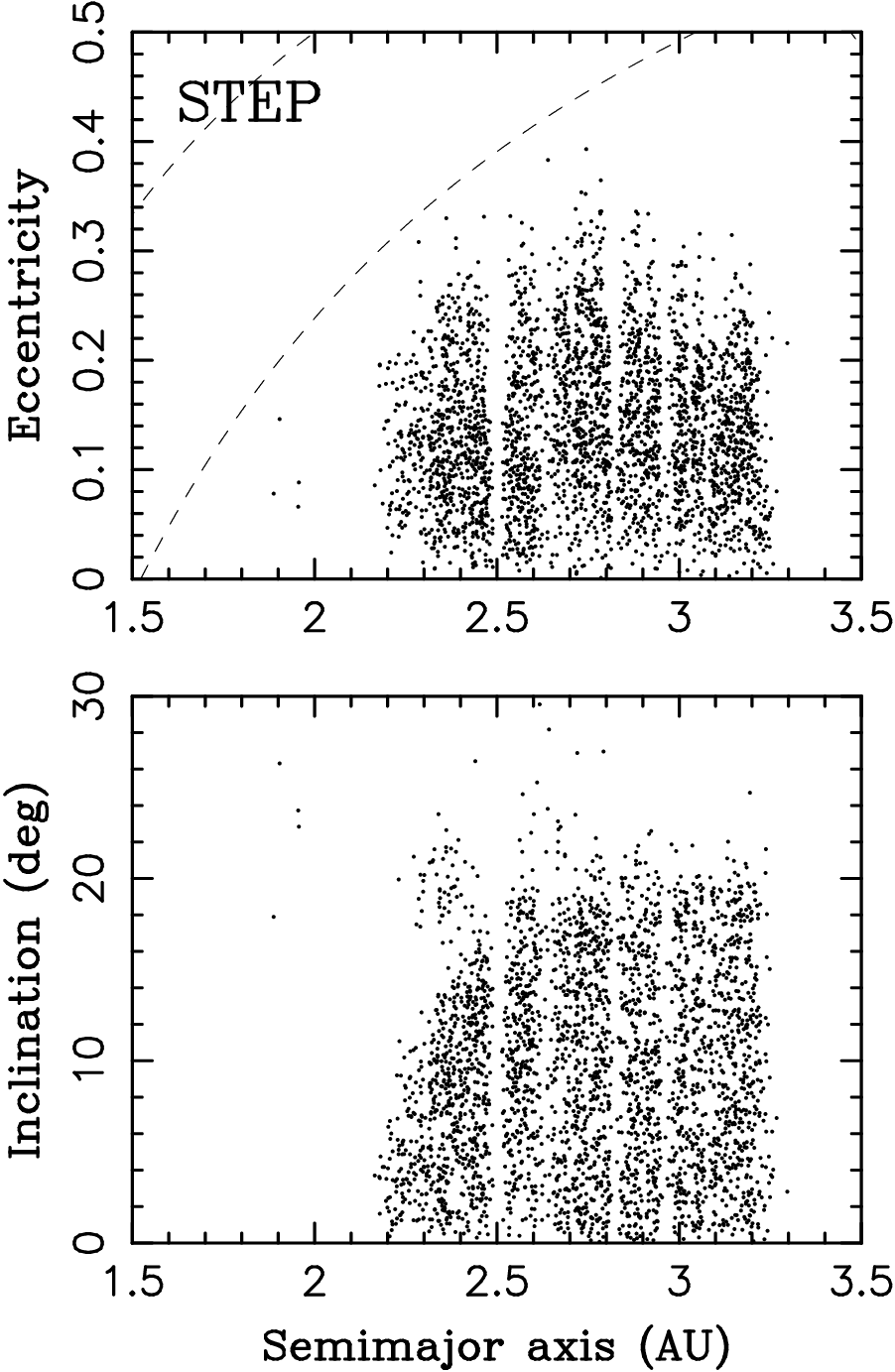}
\plotone{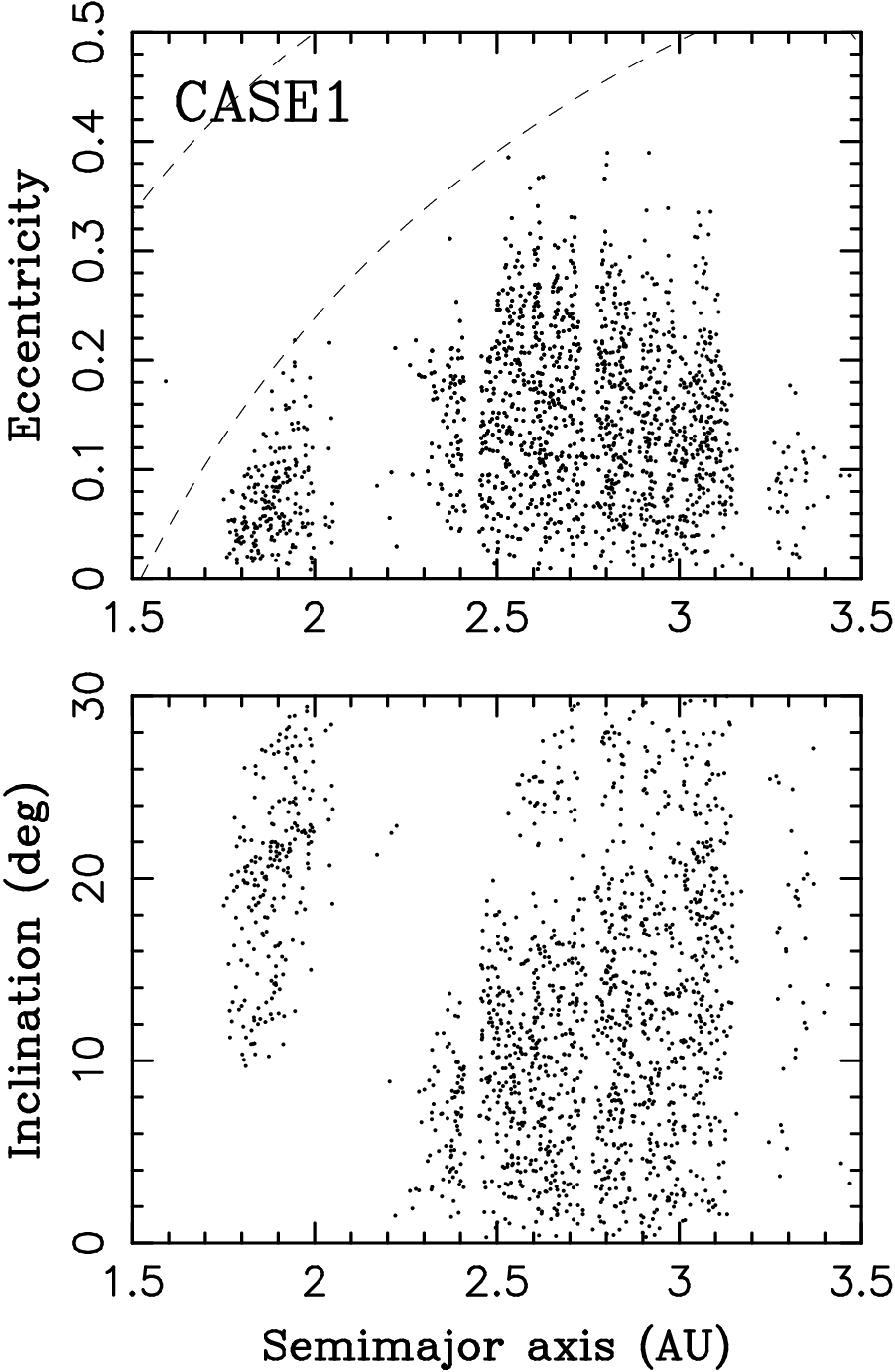}
\plotone{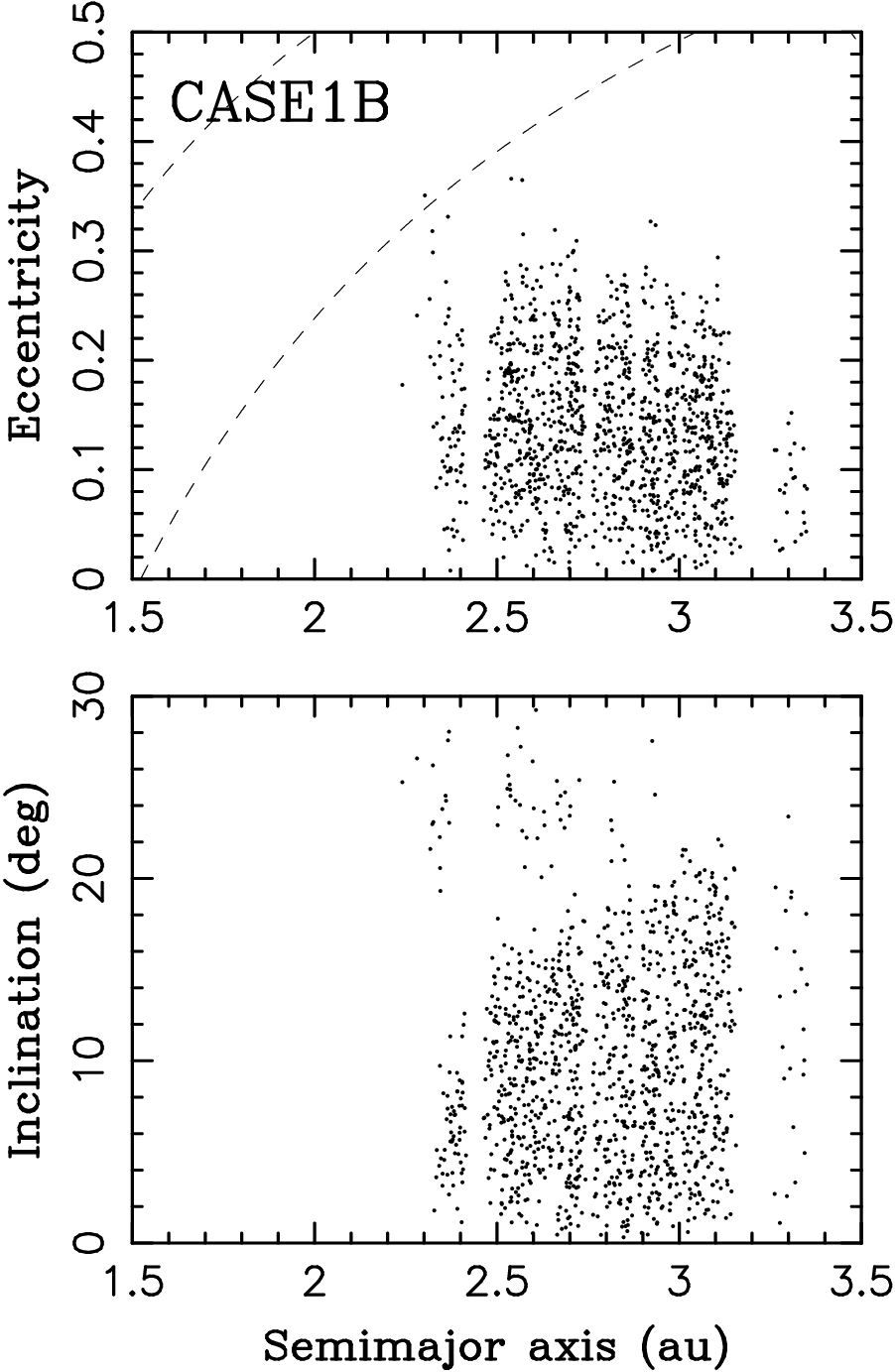}
\caption{The final orbits of asteroids obtained in different models ($t=4.5$~Gyr). From 
left to right the panels show the results of the REF, STEP, CASE1 and CASE1B integrations.
Note that the initial inclination distribution extended to $i=30^\circ$ in REF and CASE1. To ease 
the comparison of different cases shown here, we undersampled the distributions in REF, STEP and CASE1B 
such that they correspond to having 10,000 asteroids originally (i.e., the same statistics as in CASE1).}
\label{final}
\end{figure}

\begin{figure} 
\epsscale{0.5}
\plotone{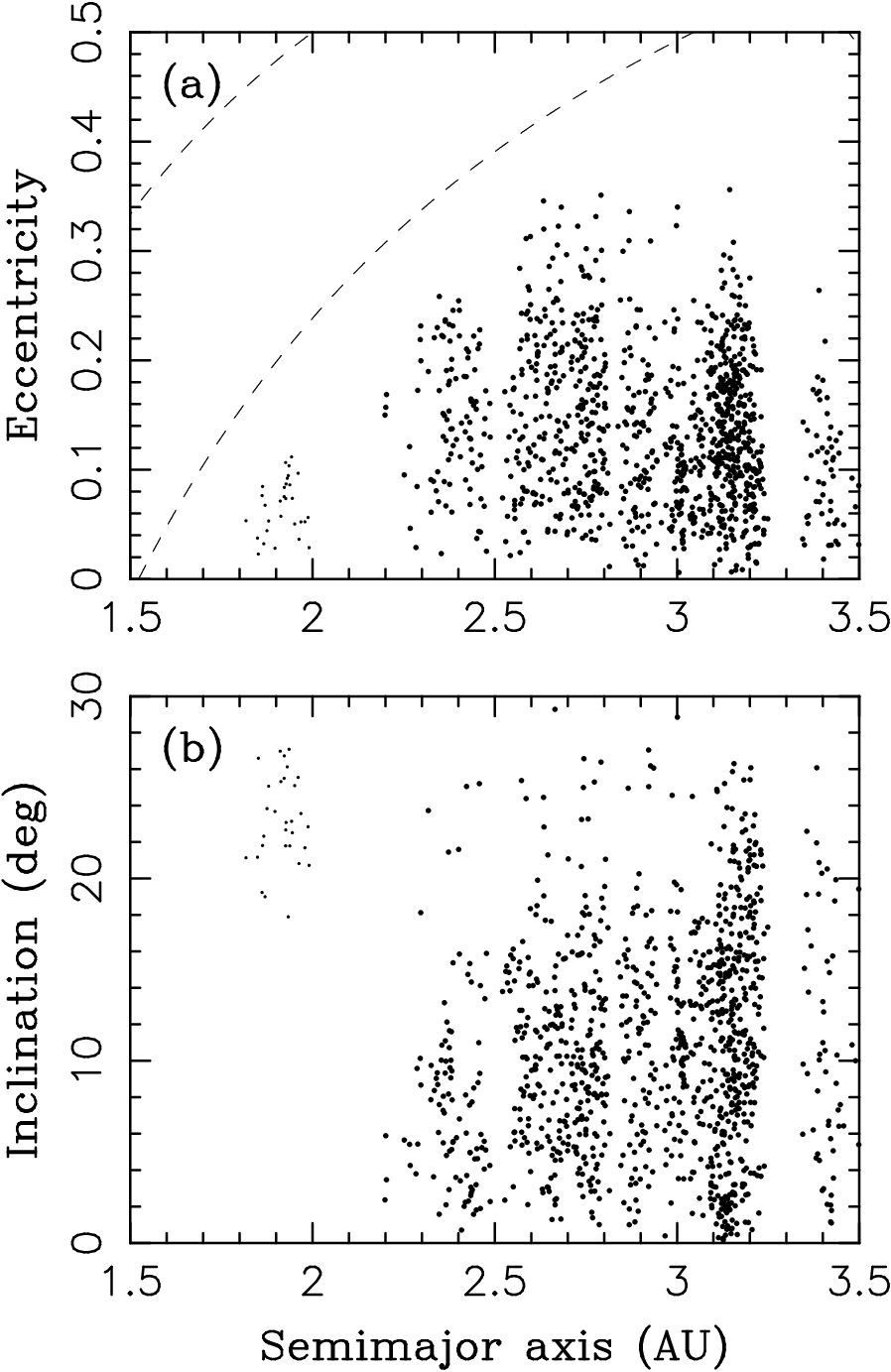}
\caption{The orbital distribution of main belt asteroids with $D>30$ km (see Section 3.9 for details). The 
smaller dots for $a<2.2$ au highlight Hungarias, which are thought to be a remnant of the E-belt (Bottke 
et al. 2012). Since Hungarias are all smaller than 30 km, here we plot known asteroids with $a<2.2$ au 
and $D>5$ km.}
\label{real}
\end{figure}

The final distributions of asteroids obtained in our simulations are shown in Figure \ref{final}. These distributions can be 
compared with the orbital distribution of the main belt asteroids (Figure \ref{real}). Several things can be noted. First, in the 
REF and CASE1 simulations, where the initial distribution of asteroids extended above 20$^\circ$ (Table 1), many asteroids
remain with $i>20^\circ$ at the end of simulations, and this population is disproportionally large when compared to the actual 
distribution of main belt asteroids. Therefore, the original asteroid belt most likely did not have such high orbital 
inclinations (Morbidelli et al. 2010, Roig \& Nesvorn\'y 2015). Second, there is a clear difference between the survival 
of asteroids in the {\it inner} main belt ($2.1<a<2.5$ au) in REF and STEP on one hand (two left columns in Figure 
\ref{final}), and CASE1 and CASE1B on the other hand (two right columns in Figure \ref{final}). Specifically, the survival rate 
is substantially higher in the former cases than in the later cases. This happens because various resonances, which act to
destabilize orbits, have well defined positions in REF and STEP, and affect only a relatively small fraction of 
orbits. In contrast, these resonances move during the instability following the orbital changes of planets in CASE1 and CASE1B;
the orbital depletion caused by them is therefore more pronounced. Overall, we find that roughly 50\% of asteroids
survive 4.5 Gyr of evolution in REF and STEP (cf. Minton \& Malhotra 2010), while only $\sim$20\% survive in CASE1 and CASE1B.   

\begin{figure} 
\epsscale{1.0}
\plotone{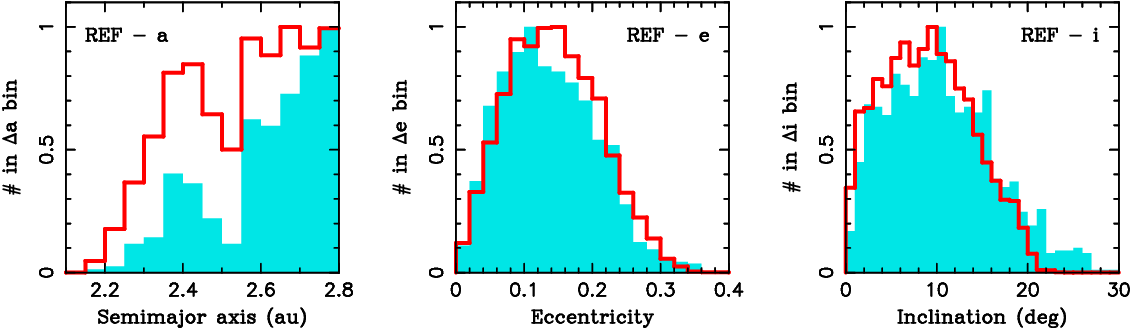}\\[0.5cm]
\plotone{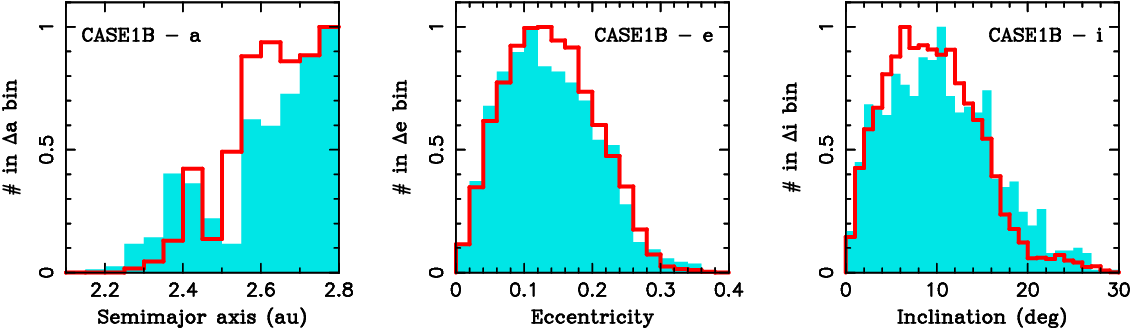}
\caption{A comparison of the orbital distribution of $D>30$-km main belt asteroids (blue histograms) and
the distribution obtained in our model (red lines; REF shown on the top, CASE1B on the bottom). The asteroid 
diameters were collected from all available sources by M. Delbo (see Section 3.9 for more information). We used 
a Gaussian mask in $e$ and $i$ to set up the initial distribution of model orbits. The results for other masks
were similar (see Roig \& Nesvorn\'y 2015). Here we show the final 
model orbits at $t=4.5$~Gyr after the start of the integration. In left panels, the focus is on the 
semimajor axis distribution in the inner part of the main belt. In the REF model, too many bodies ended up with 
$a<2.5$ au leaving the inner main belt overpopulated. The semimajor axis distribution is better in CASE1B.}
\label{histo}
\end{figure}

We determine the number of surviving asteroids for $2.2<a<2.5$ au ($N_{\rm <2.5}$) and $2.5<a<2.8$ 
au ($N_{\rm >2.5}$), and compute their ratio $f=N_{\rm <2.5}/N_{\rm >2.5}$ (a cutoff on the initial inclinations, 
$i<20^\circ$, was applied in all cases for consistency). In the real main belt, $f=0.3$ for $D>10$ km. 
For REF and STEP we obtain $f=0.60$-0.75, which shows that the inner belt 
is not sufficiently strongly depleted (relative to the middle part of the belt; Figure \ref{histo}). For CASE1 and CASE1B 
the ratios are $f\simeq0.24$-0.33, in close correspondence to $f=0.3$ found for the real asteroids. This result could be used to 
favor CASE1 and CASE1B.
On the other hand, the distribution of the inner-belt orbits obtained in CASE1 and CASE1B does not match very well
the observed distribution in that most model orbits are concentrated in the 2.3-2.5 au interval, 
while the observed distribution extends below 2.3 au.
This suggests that too much depletion occurred at $\simeq$2.2-2.3 au in CASE1 and CASE1B (see discussion in Roig
\& Nesvorn\'y 2015). We conclude that the real evolution of planetary orbits was probably somewhat intermediate between 
REF/STEP and CASE1/CASE1B.

As for the orbits in the inner extension of the main belt ($1.6<a<2.1$ au; hereafter E-belt; Bottke et al. 2012), 
the surviving population is sensitive to the initial inclination 
distribution of bodies in this region, and to the orbital histories of planets. In the present E-belt, Hungarias 
represent a minor population of small asteroids ($D\lesssim10$ km) with orbital inclinations between 16$^\circ$ 
and 34$^\circ$ (Figure \ref{real}). The orbits with $i<15^\circ$ and $1.6<a<2.2$ au are unstable in the present 
configuration of planets (Michel \& Froeschl\'e 1997, Milani et al. 2010); this orbital region is thus nearly 
void of asteroids. Since the orbital inclinations of asteroids are {\it not} strongly altered during the instability
in any of our simulations, Hungarias presumably represent remains of the original population with $i\gtrsim15^\circ$. 
Indeed, we find that the vast majority ($>$90\%) of bodies present in the Hungaria region at the end of our simulations 
started on orbits with $i>15^\circ$.

The final orbits of planets are key to the long-term behavior of the E-belt. The population of orbits with $a<2.1$ au 
obtained in CASE1 (third panel in Figure \ref{final}) does not match the observed population. First, too many 
bodies survive with $a<2.1$ au. Second, the CASE1 inclination distribution extends down to $\simeq$10$^\circ$,
while the real one has a lower limit at $\simeq$16$^\circ$ (Figure \ref{real}). We found that this problem is probably 
related to the final orbit of Mars. The final eccentricity and final inclination of Mars in CASE1 are 0.052 and 2.2$^\circ$, 
respectively, with both values being lower than the actual ones ($e_{\rm Mars}=0.069$ and $i_{\rm Mars}=4.4^\circ$). 
In fact, it is difficult to sufficiently excite Mars's orbit during the instability if Mars starts with $e=0$ and $i=0$ (
Brasser et al. 2013, Roig et 
al. 2016, Nesvorn\'y \& Roig 2016). In CASE1B, on the other hand, Mars ends up having the mean eccentricity 0.078 and 
the mean inclination 5.2$^\circ$, both values being only slightly higher than the actual ones. The structure of the 
E-belt looks much better in this case. The E-belt is depleted, just as strongly as in the STEP simulation, and the 
surviving orbits have $i>15^\circ$. 
\subsection{Comparison with Bottke et al. (2012)}
Using the methods described in Section 2 we computed the impact fluxes of asteroids on different target bodies. 
We start by discussing the impacts on the Earth. The impacts on target bodies other than the Earth are considered in 
Section 3.8. We use a cumulative representation of the impact flux, $N_{\rm c}(t)$, with plots showing the expected number of 
impacts in the interval between time $t$ and the present epoch. The actual number of impacts at $t$ is the derivative of the 
cumulative function. The profiles are normalized to one asteroid in the original population. They can therefore be understood 
as the average probability of impact for each asteroid in the original source region. The absolute calibration of impact fluxes 
is the subject of Section 3.9. Here, the normalized flux profiles are compared to each other, using different assumptions, 
to understand the dependence on various parameters.
 
\begin{figure} 
\epsscale{0.6}
\plotone{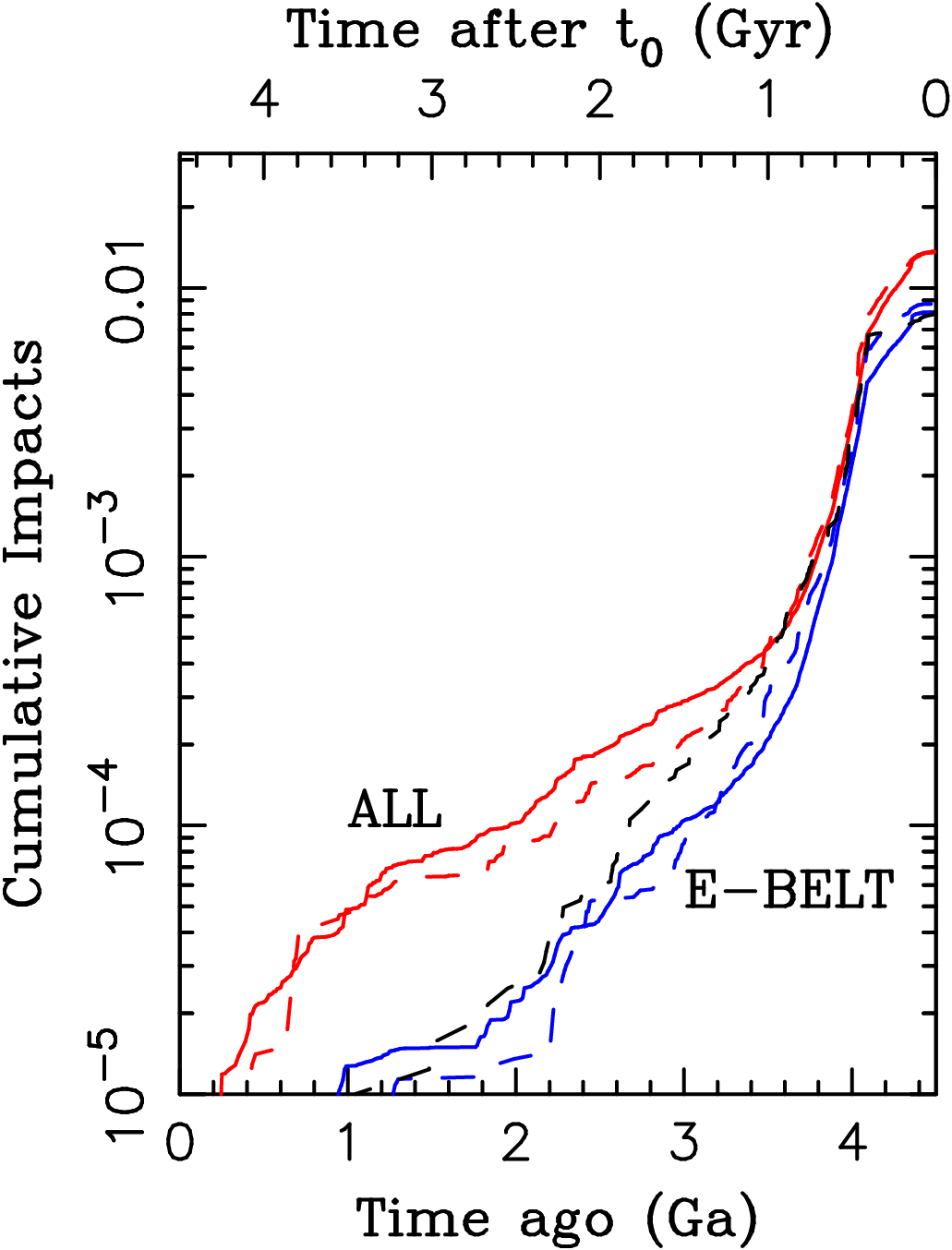}
\caption{The cumulative number of impacts on the Earth. The red lines show STEP (solid), and the REF
case with $t_{\rm inst}=400$ Myr (dashed). The blue lines show the same cases but for the impactors from the E-belt. 
For reference, the black dashed line is the E-belt impact flux from Bottke et al. (2012).}
\label{first}
\end{figure}

First, let us consider the case with $t_{\rm inst}=400$ Myr, because this was the instability time suggested from 
independent constraints by several recent works (e.g., Bottke et al. 2012; Marchi et al. 2012, 2013; Morbidelli et al. 2012). 
Figure \ref{first} shows a comparison of our STEP and REF simulations with $t_{\rm inst}=400$ Myr with the results 
of Bottke et al. (2012). Three time intervals with distinct appearance of the impact profiles can be identified 
in Figure \ref{first}. The first one, between $t=0$ and $t=400$ Myr is characterized by a roughly twofold 
reduction of $N_{\rm c}(t)$ from $\simeq$0.014 to $\simeq$0.007. This means that $\sim$50\% of all impacts 
are expected to happen in the first 400 Myr. In the second interval, just after the instability, the impact profiles 
drop by a factor of $\sim$10 in 500 Myr. This is when a large number of impacts is expected to happen. Finally, during 
the late stage, roughly after 3.5 Ga, the cumulative profiles slowly and steadily decline. 

The results obtained for the STEP and REF simulations with $t_{\rm inst}=400$ Myr are nearly identical (red lines 
in Figure \ref{first}). This demonstrates that the labeling procedure described in Section 2 works well for the late 
instability cases. We will use this method in Section 3.6 to understand the dependence 
of the impact fluxes on $t_{\rm inst}$. To compare our
results with Bottke et al. (2012), we note that  Bottke et al. obtained their results for the E-belt only while 
here we consider the whole main belt. To be able to compare things, 
we select asteroids with initial $1.6<a<2.1$~au in our model and plot $N_{\rm c}(t)$ for them in Figure \ref{first} 
(blue lines). We also apply the same Gaussian mask to this sub-population that was used in Bottke et al. to set up
the initial distribution of E-belt orbits in $e$ and $i$.  In addition, because the normalization of Bottke's 
profiles to one initial asteroid is unavailable, we shift them vertically in the plot such that the overall 
impact probability is equal to that obtained from the E-belt in our STEP simulations. 

With these provisions we obtain the E-belt impact profiles that are similar to those reported in Bottke et al. 
(2012). This verifies, among other things, that the cloning procedure employed in Bottke et al. (2012) worked well 
(we do not use cloning here). A difference is noted in the interval before the instability, where our cumulative 
profiles for E-belt drop steeper than those of Bottke et al. (2012) ($\sim$50\% drop from $t_0$ to $t_{\rm inst}$ compared 
to only 10-30\% decline in Bottke et al.), indicating a larger number of impacts in our case. 
The origin of this difference is understood. It is related to the initial orbits of the giant planets,
which were set up differently in Bottke et al. and here (Section 3.6). 
We verified that various initial distributions of asteroids in the E-belt 
all lead to a similar drop-off. Moreover, the impact profile in the first 400 Myr interval is not sensitive to the 
orbital distribution of asteroids near the Mars-crossing boundary, because similar results are obtained when considering 
impactor populations with various initial perihelion distances (e.g., $q>1.6$, 1.7 or 1.8 au).  

Figure \ref{first} offers the first glimpse into the relative importance of the E-belt and the main belt. As for the 
overall number of impacts over 4.5 Gyr, the E-belt and main belt appear to be nearly equally important each providing 
roughly half of the terrestrial impactors. Their contributions differ in the temporal distribution of impacts. While the 
E-belt provides most impactors in the first 1 Gyr, the main belt takes over at $\simeq$3.5 Ga and provides most impactors 
since then. We will discuss this issue in more detail in Section 3.5. 
\subsection{Impact Flux in Different Instability Cases}
\begin{figure} 
\epsscale{0.6}
\plotone{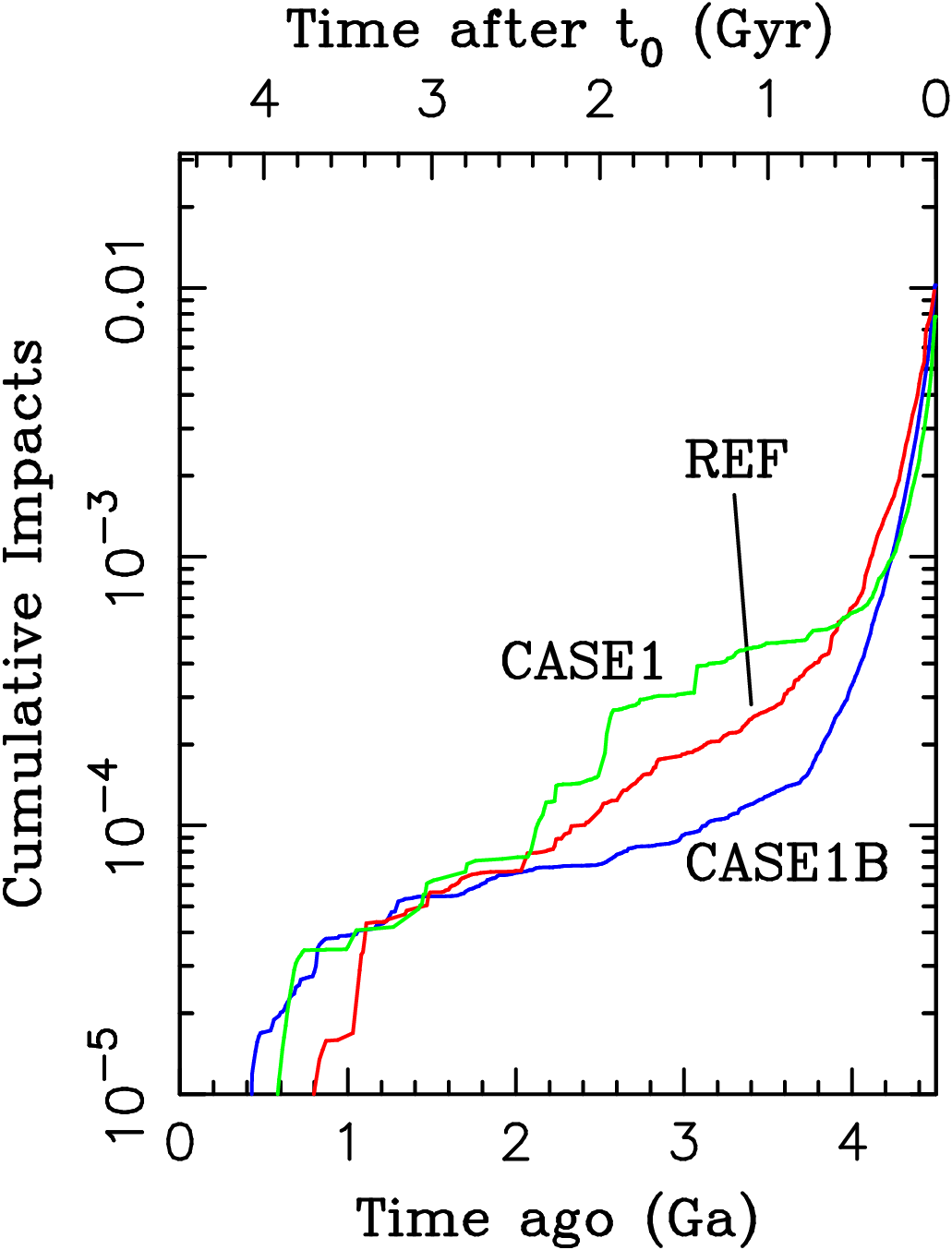}
\caption{The cumulative number of impacts on the Earth in different instability cases. The lines show: the (1) REF 
case (red line), (2) CASE1 with $t_{\rm inst}=0$ (green line), and (3) CASE1B with $t_{\rm inst}=0$ Myr (blue line).
We used a Gaussian mask in $e$ and $i$ to set up the initial distribution of orbits (see Section 2).}
\label{best}
\end{figure}

Figure \ref{best} compares the impact profiles obtained for different cases of planetary evolution during the instability.
Here we use $t_{\rm inst}=0$ (see Section 3.6 for $t_{\rm inst}>0$). The 
results for REF and CASE1 with $t_{\rm inst}=0$ bear resemblance to each other, featuring a fast initial decline and 
turnover to a shallower slope at $\sim$3.5-4 Ga.\footnote{In CASE1, where there were 10,000 asteroids initially, the 
impact profile shows step-like fluctuations after $\sim$2 Gyr after the start of the simulation. 
These fluctuations would disappear with a better statistic.} In a closer look, however, it can be noticed 
that the initial decline in CASE1 is steeper than in REF, and the transition to the shallow slope in CASE1 happens abruptly
at $\simeq$4 Ga. In the REF case, instead, the impact flux declines more monotonically and has a more gradual transition 
to the shallow slope at $\sim$3.5~Ga.  

We performed several tests to understand the origin of these transitions and found that they have different causes 
in REF and CASE1. In REF, most early impacts are produced by the E-belt. The early impact 
flux rapidly declines because the E-belt is becoming depleted on a characteristic timescale of a few hundreds of Myrs. 
Then, at $t\sim1$ Gyr ($\sim$3.5 Ga), a crossover happens. After this time, the impact flux is dominated by bodies 
leaking from the main belt ($a>2.1$ au).\footnote{The crossover from the E-belt to the main belt must have 
happened at some time during the Solar System history, assuming that E-belt was important initially, because the 
main belt is presently the source of the near-Earth asteroids.} The impact flux was more constant over the past 
$\sim$3 Gyr, because the main belt is a large reservoir and acts to supply impactors at a nearly constant rate. 
This projects into a steadily decreasing $N_{\rm c}(t)$ in Figure \ref{best}.  

The crossover from the fast to slow decline in CASE1 has a different explanation. Recall that due to problems
with the Mars orbit in the CASE1 simulation (Section 3.1), the E-belt population decayed more slowly in CASE1 
than it had in reality. The crossover from the E-belt to the main belt in CASE1 therefore happened very late ($t\sim2$ 
Gyr or $\sim$2.5 Ga), and has nothing to do with the transition seen in Figure \ref{best} at 4 Ga. Instead, 
this early transition is related to the excitation of asteroid orbits during the dynamical instability. 
During the instability, a large fraction of asteroids evolve onto planet-crossing orbits and produce a heavy bombardment. 
This population has a relatively short dynamical lifetime and rapidly decays, thus explaining the fast 
decline of the impact flux between 4 and 4.5 Ga (Figure \ref{best} -- green line). We conclude that our CASE1 
model has important limitations. 

Figure \ref{best} also shows the impact profile for CASE1B, which is probably the most realistic case 
considered in this work. There is a significant difference with respect to the profiles discussed previously in 
that the initial decline is stronger and extends longer, to $\simeq$0.8 Gyr after $t_{\rm inst}$. This happens 
because the E-belt population rapidly decays in CASE1B (as it should because planets end up on approximately 
correct orbits), and the orbital excitation during the instability is substantial. Both the E-belt and excited
populations then combine to generate many impacts during the first Gyr, when $N_{\rm c}(t)$ drops by almost two 
orders of magnitude. The crossover to the main belt impactors happens at $\simeq$3.7 Ga. A much slower decline of the impact flux 
follows after the crossover when the impactor population evolves toward a steady state. 
\subsection{Dependence on the Initial Orbital Distribution of Asteroids}
\begin{figure} 
\epsscale{0.6}
\plotone{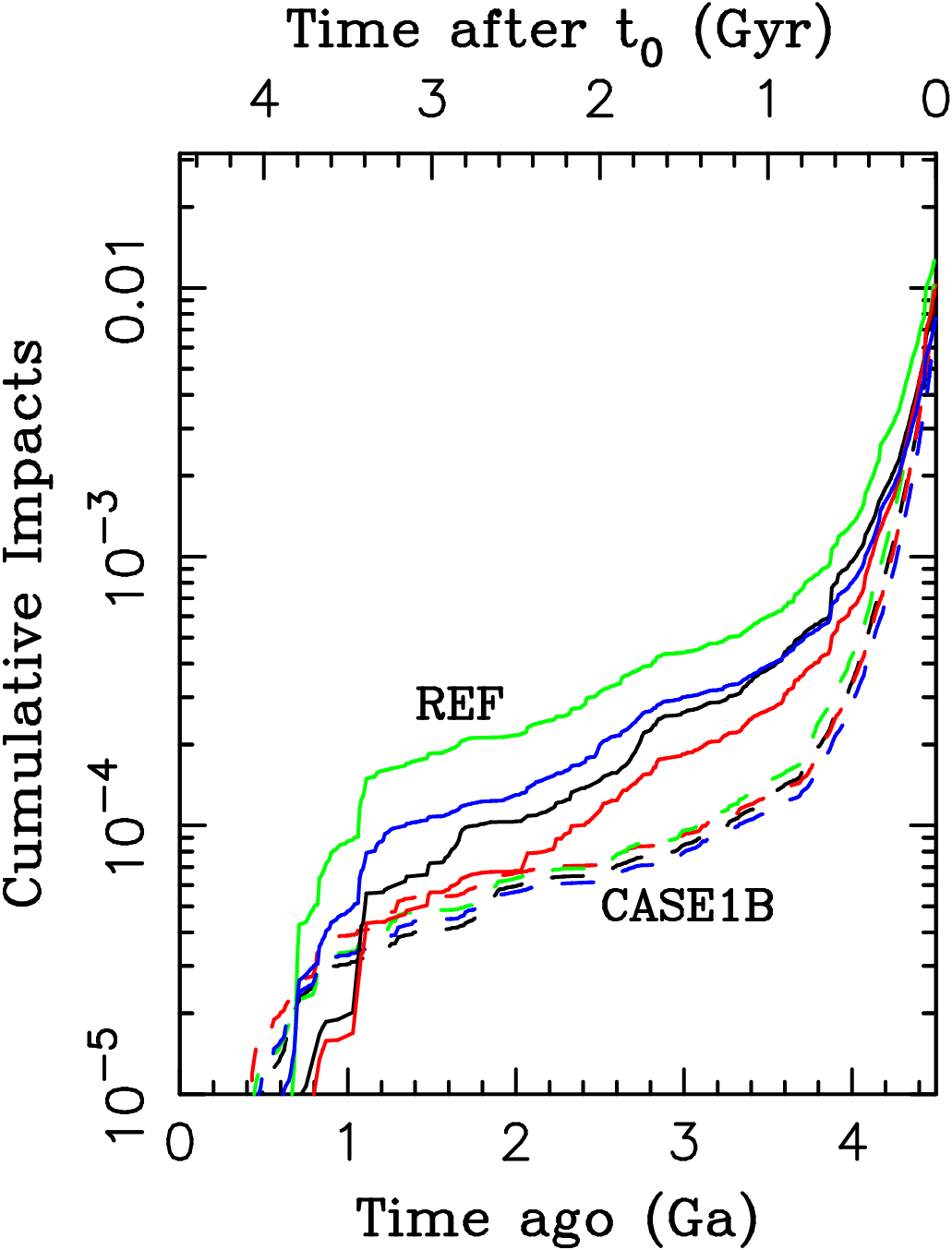}
\caption{The cumulative number of impacts on the Earth in REF (solid lines) and CASE1B (dashed lines) with $t_{\rm inst}=0$. 
Different colors indicate the dependence of the impact flux on the initial distribution of asteroids: the inclination cutoff $i<20^\circ$ 
(black), the Gaussian distributions in $e$ and $i$ from Bottke et al. (2012) (red), the Rayleigh distributions in $e$ and $i$ 
from Roig \& Nesvorn\'y (2015) (green), and the Grand Tack distribution (Walsh et al. 2011, Deienno et al. 2016) (blue).}
\label{mask}
\end{figure}

Figure \ref{mask} illustrates the dependence of the results on the initial distribution of asteroids in $e$ and $i$. 
Four different initial masks were considered (Section 2). In CASE1B, the impact fluxes computed for different masks are 
within $\simeq$20\% of each other at any given time. This shows that the dependence on the initial orbital distribution in 
CASE1B is weak. This is a consequence of the orbital excitation during the instability that wipes out,
to a degree, the memory of the initial distribution. 

The REF case tells a different story. In this case, the
impact fluxes are quite sensitive to the initial orbital distribution of asteroids. The Rayleigh distribution
(green solid line in Figure \ref{mask}) leads to the highest impact flux (overall impact probability of 0.013)   
while the Gaussian mask leads to the lowest impact flux (probability of 0.007). Also, the probability of impact 
in the last 2 Gyr is $4\times10^{-5}$ for the Gaussian mask, while it is $\simeq$4 times higher ($1.6\times10^{-4}$)    
for the Rayleigh mask. This is a significant difference. The inclination cut and Grand Tack mask produce results that 
are intermediate between these two extremes. 

The interpretation of these results is relatively straightforward. The Rayleigh mask, where no inclination cutoff 
was used, allows for higher orbital inclinations than any other initial distribution used here (Section 2.5). With these 
higher inclinations, the region of Hungarias and the high-$i$ main belt are well populated, both being an important 
source of impactors. The former raises the overall number of 
impacts and the later enhances the difference with respect to other masks at late times. Note that this happens because 
the REF simulation used a wide raw distribution of the initial inclinations ($0<i<30^\circ$). The dependence
on the initial inclination distribution is weaker in CASE1B, where we initially had $0<i<20^\circ$ (Table 1).        

We also tested the dependence of the results on the radial distribution of asteroids using different semimajor axis 
masks. For example, a family of impact flux profiles was obtained for the power law distributions $N(a) \propto a^{\gamma}$, 
where $\gamma$ was a free parameter ($\gamma=0$ was used to set up the original distribution). We did not find 
much variation of the impact flux for $-2<\gamma<1$. Profiles with the $\gamma$ values outside this range were not 
considered, because the final semimajor axis distributions obtained with $\gamma<-2$ or $\gamma>1$ did not match the 
present structure of the asteroid belt.        
\subsection{Contribution of Different Source Regions} 
\begin{figure} 
\epsscale{0.4605}
\plotone{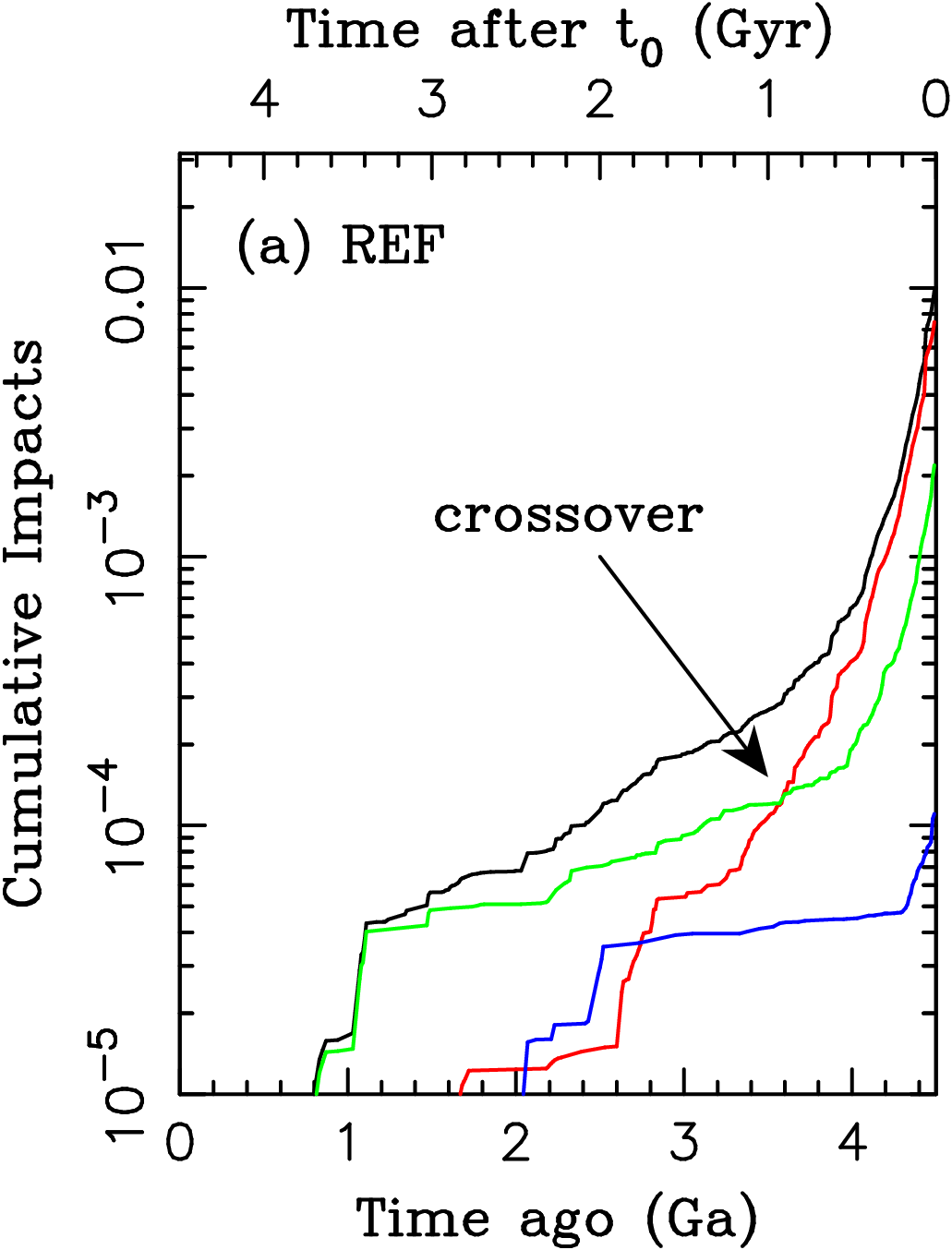}
\epsscale{0.42}
\plotone{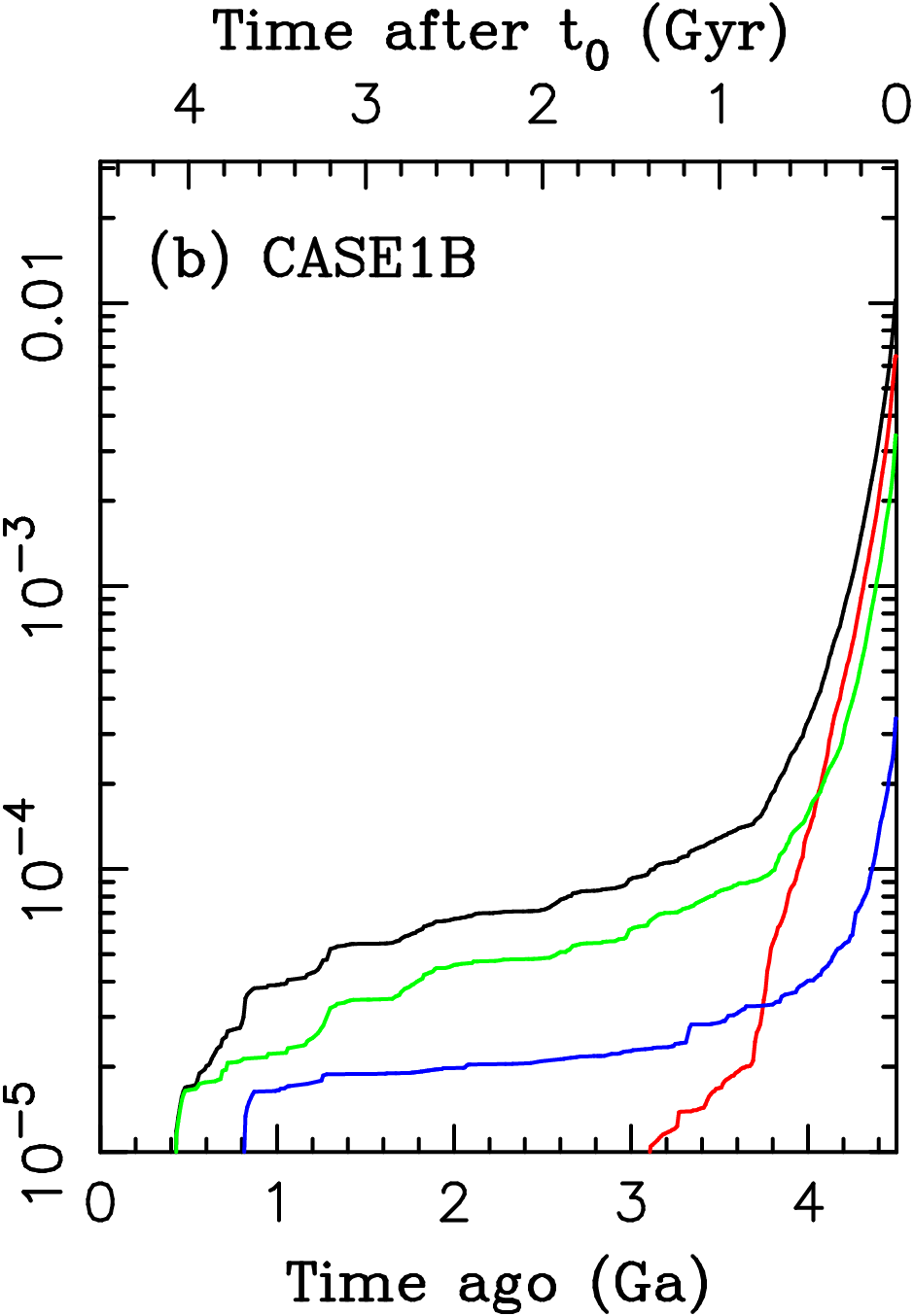}
\caption{The cumulative number of impacts on the Earth in REF (panel a) and CASE1B (panel b). Different lines illustrate the 
dependence of the impact flux on the source location of impactors: the E-belt ($1.6<a<2.1$ au; red lines), inner belt ($2.1<a<2.6$ au; 
green lines), and outer belt ($2.6<a<3.3$ au; blue lines). The black lines show the total fluxes. Here we used a Gaussian 
mask in $e$ and $i$ to set up the initial distribution of orbits. Note that the REF case, where we only used 20,000 initial orbits,
has issues with small statistics, especially in what concerns the late impacts from the outer main belt.}
\label{source}
\end{figure}

Figure \ref{source} shows the dependence of the flux profiles on the provenance of impactors. We split the source
region into three semimajor axis intervals: (1) E-belt ($1.6<a<2.1$ au), inner belt ($2.1<a<2.6$ au), and outer belt 
($2.6<a<3.3$ au; blue lines), and compute the impact fluxes for these regions separately. The fluxes are 
normalized, as before, to one initial asteroid with $1.6<a<3.3$ au. The crossover from the E-belt to main-belt impactors,
already discussed in Section 3.3, is obvious in the REF case. The E-belt dominates until about 3.6 Ga when
the shallower profile of the inner belt takes over. The outer belt has a relatively minor contribution to
the impact flux, but note that the drop of the outer belt profile at 2.5 Ga is caused by the small number statistics.
With better statistics, the outer belt should produce a flatter profile reaching toward more recent epochs.

CASE1B, illustrated in panel (b) of Figure \ref{source}, shows similar trends, but there are also several important
differences. For example, the E-belt is not as dominant during early times as in the REF case. This is because 
the impact flux during early times is contributed by the population exited by the planetary instability.
Since a large part of the excited population started in the inner belt (Section 3.1), the inner belt becomes dominant 
already at 4 Ga, roughly 400~Myr earlier than in REF. The statistics is better in CASE1B (50,000 initial orbits), and we 
are thus able to resolve the contribution of the outer main belt all the way to 1~Ga, where it represents 
roughly one third of the overall number of impacts, with the inner belt being responsible for the other two thirds.

In summary, we find that: (1) the E-belt is important during the first $\simeq$0.5-1.0 Gyr, and becomes a 
minor contribution after that, (2) the inner belt provides most impactors since $\simeq$3.5-4 Ga, and (3) the outer 
belt is insignificant for the early impacts but becomes progressively more important later. The time of the 
crossover from the E-belt to the inner belt, $t_{\rm cross}$, correlates with the degree of excitation 
of the asteroid orbits during the instability, with more excitation implying earlier $t_{\rm cross}$. As a final 
remark, we stress that the CASE1B transition in the impact profiles seen at $\simeq3.7$ Ga (steep profile before,
shallow after) is not related to the E-belt/inner-belt crossover. Instead, as we explained in Section 3.2,
this transition marks the end of dominance of the excited population, and switch to a slower supply of impactors 
from the main belt. 
\subsection{Dependence on $t_{\rm inst}$}
\begin{figure} 
\epsscale{0.4605}
\plotone{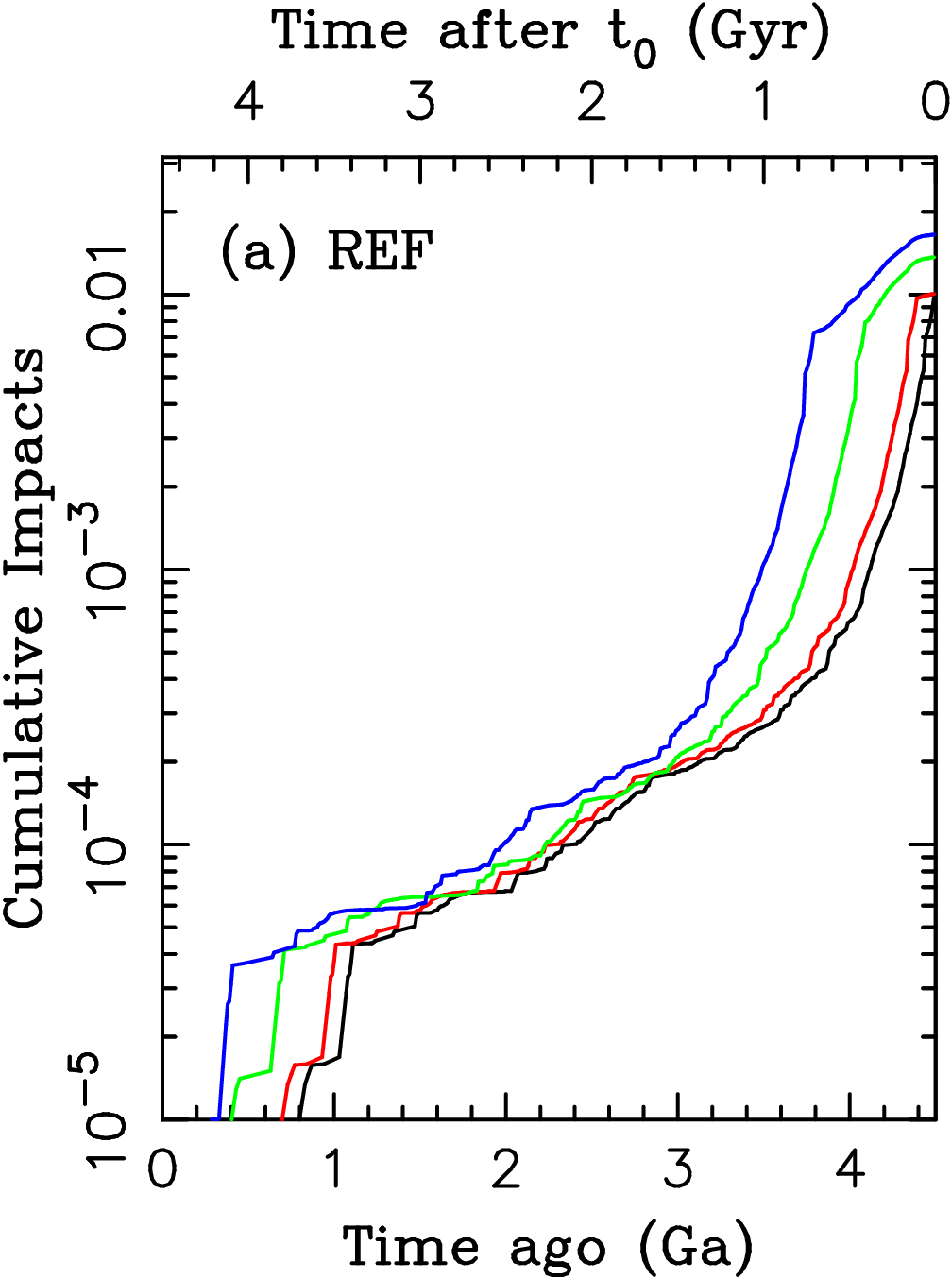}
\epsscale{0.428}
\plotone{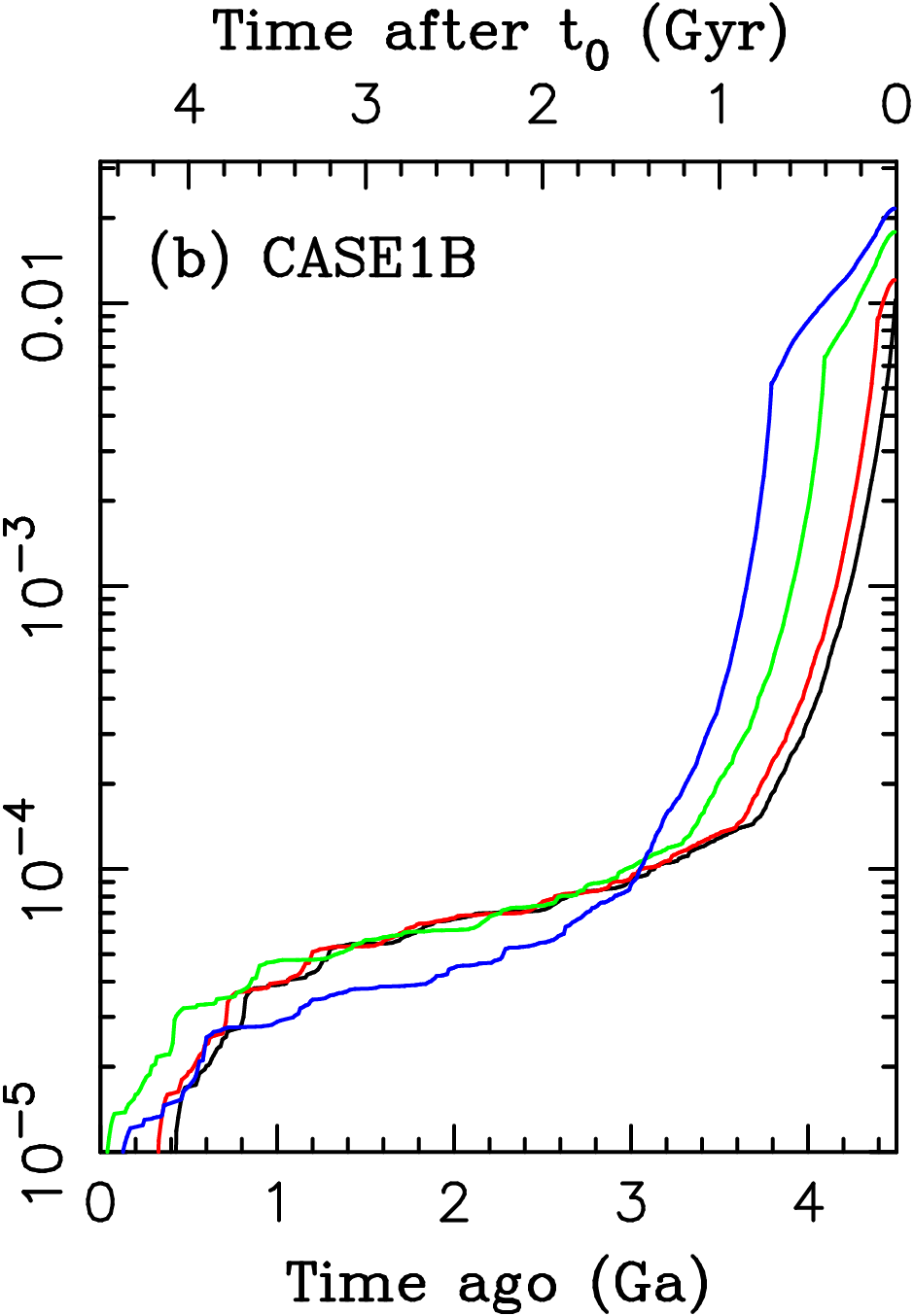}
\caption{The cumulative number of impacts on the Earth in REF (panel a) and CASE1B (panel b). Different lines show the 
dependence of the impact flux on the time of the instability: $t_{\rm inst} = 0$ (black line), $t_{\rm inst} = 100$ Myr (red),
$t_{\rm inst} = 400$ Myr (green), and $t_{\rm inst} = 700$~Myr (blue). The results shown here were obtained with the Gaussian 
mask in $e$ and $i$. }
\label{tinst}
\end{figure}

Figure \ref{tinst} shows the dependence of the results on $t_{\rm inst}$. The total number of impacts over 4.5 Gyr 
correlates with $t_{\rm inst}$, with the late instability cases ($t_{\rm inst}\gtrsim400$ Myr) showing roughly twice 
as many impacts (in total) for each initial asteroid as the early instability cases ($t_{\rm inst}\lesssim100$~Myr). This trend 
is a consequence of different orbital dynamics of asteroids for different $t_{\rm inst}$. If the instability happens 
early, many asteroids are immediately transported by the instability to the short-lived, planetary-crossing orbits. If, 
on the other hand, the instability happens late, the orbits can leak slowly from the E-belt and inner main belt in times
before the instability, and have more chance of hitting planets before they become eliminated (typically by plunging 
into the Sun; Farinella et al. 1994). This initial stage thus raises the overall number of impacts, and the longer
this stage lasts the greater the overall number of impacts is.

The impact flux of asteroids before the instability depends on the orbital structure of the planetary system before
the instability. Here we used the initial configuration of the giant planets from Nesvorn\'y et al. (2013) 
(see Section 2). In this configuration, Jupiter and Saturn are in the 3:2 mean motion resonance, have mean 
eccentricities 0.004 and 0.01, respectively, and significant libration amplitudes in the resonance. We verified that in this 
configuration the asteroid orbits in the mean motion resonances with Jupiter are unstable and the resonant asteroids 
evolve onto planet-crossing orbits, thus contributing to the impact flux. The population of orbits with 
$a>2.5$ au and $e>0.3$ is particularly strongly eroded. This explains why the impact flux {\it before} the
instability is quite high in our simulations. Other planetary configurations, where Jupiter and Saturn are 
locked more deeply in the 3:2 resonance, can lead to a more modest erosion of the asteroid belt before the 
instability, implying lower impact rates.

In both cases shown in Figure \ref{tinst}, the average impact probability on Earth for each asteroid in the original source 
is slightly smaller than 1\% for $t_{\rm inst}=0$, while it is larger than 1\% for the late instability.
These values are intermediate between 0.3\% quoted for the main belt impactors by Minton \& Malhotra (2010) and 
3\% quoted by Bottke et al. (2012) for the E-belt.

The impact profiles shown in Figure \ref{tinst} steeply decline after the instability. This is when a large number 
of impacts is expected. As we already discussed in Section 3.3, during the first 1 Gyr after the instability the 
fluxes drop by a factor of $\sim$100 in CASE1B, and less so, by a factor of $\sim$30, in REF. While the impact 
profiles before 3 Ga are very different for different $t_{\rm inst}$, they become remarkably similar after that.
This is because the excited population and E-belt, which are the main source of impactors after the instability,
are already severely depleted at 3 Ga, and do not represent a major contribution to the impact record 
after that. The structure of the main belt, on the other hand, which is the dominant source of the terrestrial 
impactors after 3 Ga, is not sensitive to $t_{\rm inst}$. It therefore provides roughly equal number of 
impactors at late times, independently of whether the instability happened early or late. This shows that to 
pin down the time of the instability one has to consider the cratering and other constraints during the first 
$\simeq$1.5 Gyr of the Solar System history.

Finally, note that the normalized impact flux in the 1-3 Ga interval in REF is higher than in CASE1B. This is
because the inner main belt is not depleted as strongly during the instability in REF as in CASE1B (compare the 
leftmost and rightmost panel in Figure \ref{final}, and Figure \ref{histo}). Therefore, in the REF case, the inner belt 
is a larger reservoir and supplies more impactors at 1-3 Ga than in CASE1B. It is not obvious which of these cases is more 
realistic. As we discussed in Section 3.1, the REF case ends up with the inner belt that is overpopulated, by a factor of
$\sim$2, relative to the outer main belt, when the model results are compared with the actual distribution of asteroids. 
In the CASE1B model, on the other hand, the inner belt ends up to be too narrow in the semimajor axis, with almost 
no bodies below 2.3 au. We therefore conclude that the real impact flux at 1-3 Ga may have been intermediate between 
the two cases described here.
\subsection{\"Opik Code vs. Impacts Recorded by Swift}
\begin{figure} 
\epsscale{0.4605}
\plotone{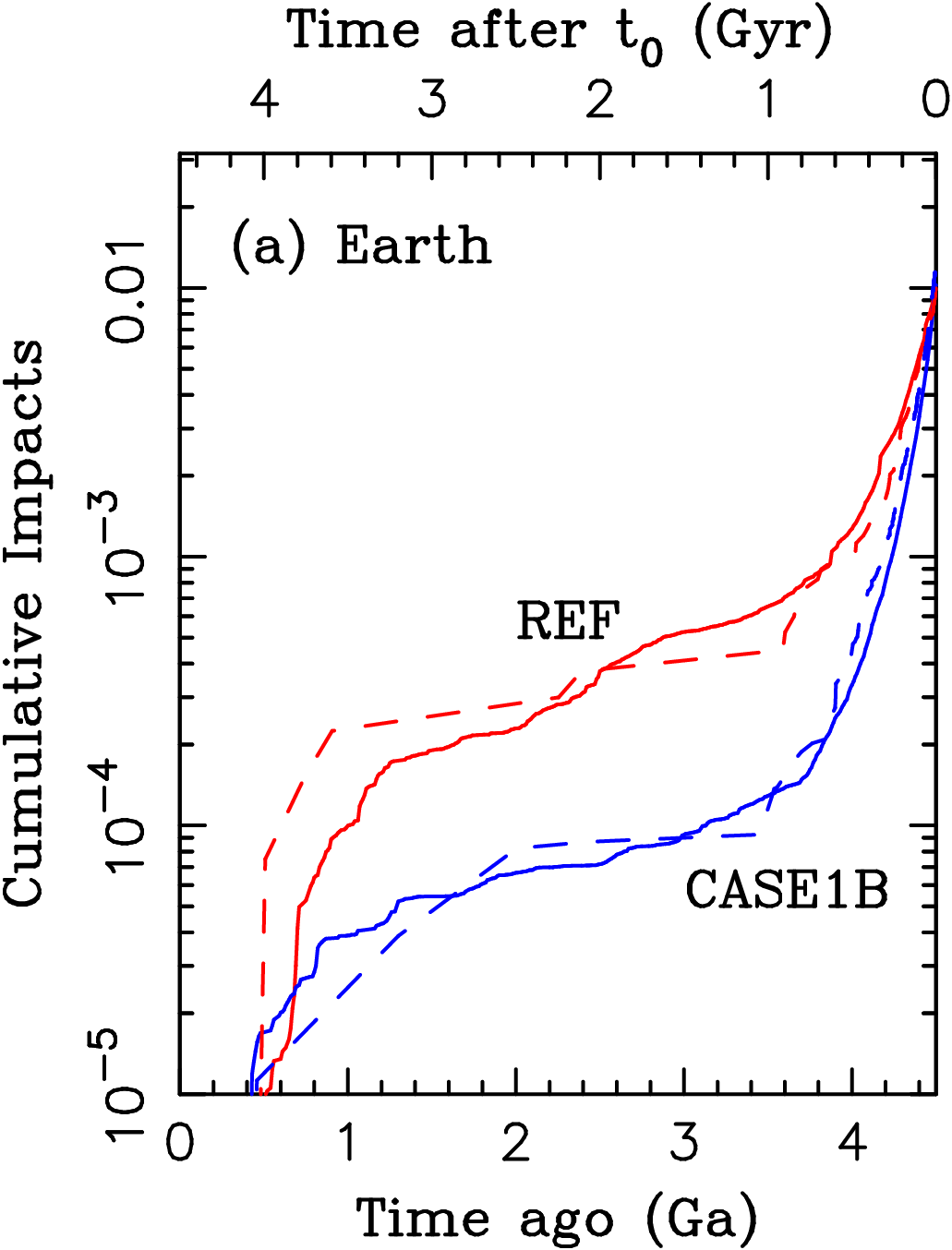}
\epsscale{0.420}
\plotone{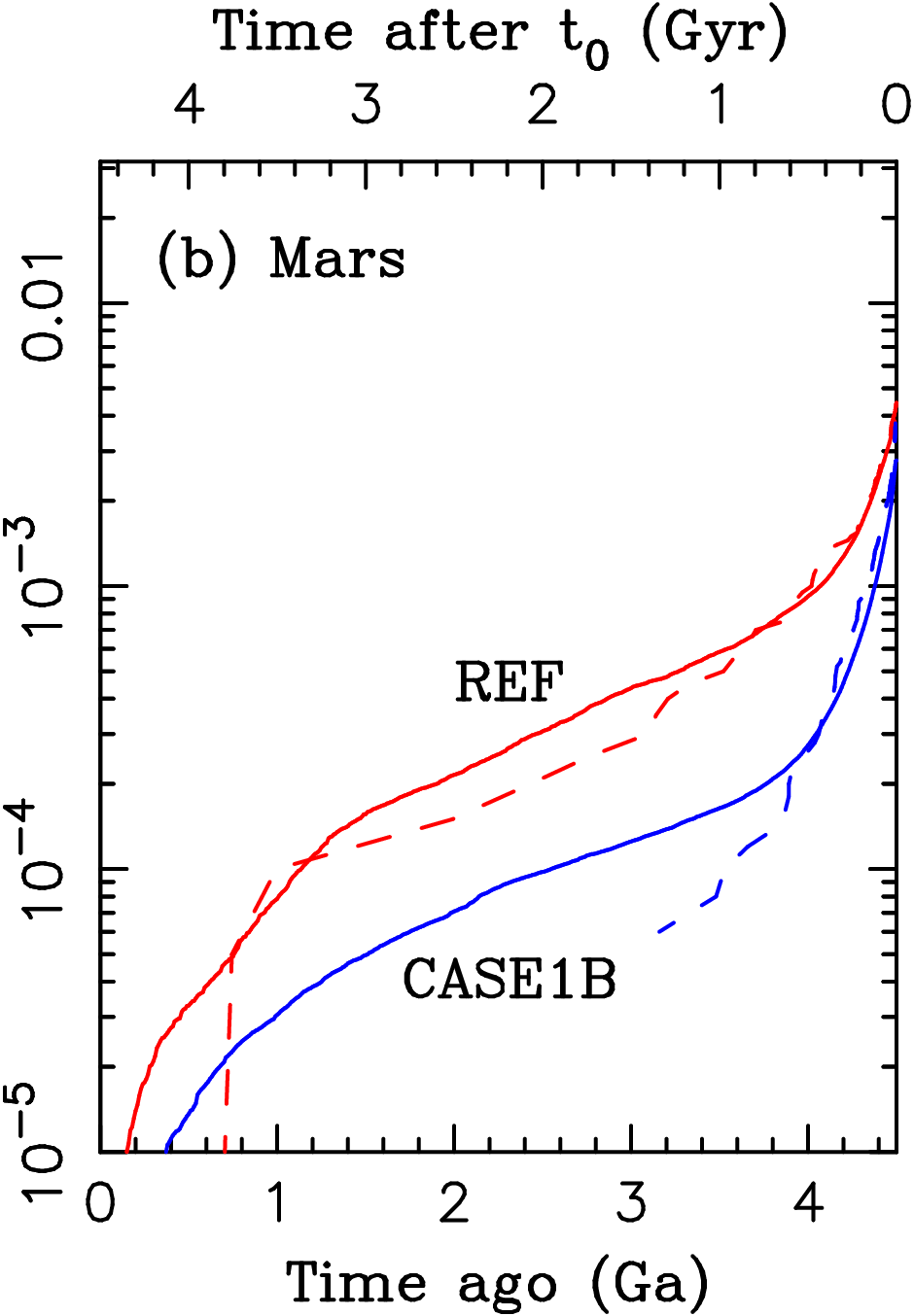}
\caption{A comparison between the results obtained with the \"Opik code (solid lines) and direct impacts recorded by 
the {\it Swift} integrator (dashed lines). The panels show the impact profiles for Earth (panel a) and Mars (panel b).
The red lines in both panels were obtained in the REF case. The blue lines show the results for CASE1B.
To maximize the statistics of direct impacts we used the raw initial orbital distribution of asteroids 
without any inclination cutoff or mask, except for the Earth impacts in CASE1B where we used the usual Gaussian mask. 
We can afford to use a mask in this case because the statistics is better (50,000 asteroids in CASE1B vs. only 20,000 
in REF; Table 1). We point out that there were only a very few recorded impacts in the last 3 Gyr (only one have been 
recorded for Mars in CASE1B). The direct impact profiles after 3 Ga are thus not reliable.}
\label{direct}
\end{figure}

Here we compare the results of the \"Opik code with the number and timing of direct planetary impacts recorded 
by the integrator. This can be done only for the target bodies included in the integration. Figure \ref{direct} 
shows the comparison for the Earth (the results for Venus are similar) and Mars. We find a good agreement for the 
Earth. Both in REF and CASE1B, the total number of recorded impacts, when normalized to one asteroid in the original 
source, closely corresponds to the number of impacts expected from the \"Opik probability profiles. Specifically, 
in CASE1B, there were 572 impacts recorded by the code for 50,000 initial bodies, implying the impact probability 
of 1.1\%, while we obtained 1.0\% from the \"Opik code. In REF, there were 199 impacts for 20,000 bodies, 
suggesting 1.0\% while we obtained 0.98\% from the \"Opik code. These probabilities were obtained for the Gaussian 
mask; the results for other initial masks were nearly identical. The timing of the recorded impacts follows reasonably 
well the \"Opik profiles shown in Figure \ref{direct} (but note that there were only a very few recorded impacts 
in the last 3 Ga).

It is harder to do this comparison for Mars, because Mars is a smaller target and the number of recorded impacts
is small. To maximize our statistics we therefore opted for using the raw initial distributions without applying
any mask or inclination cutoff. The results are shown in Figure \ref{direct}(b). Again, the correspondence
is reasonably good. The total number of recorded impacts was 188 in CASE1B and 89 in REF, which, when normalized,
indicates the impact probabilities of 0.38\% and 0.45\%, respectively. For comparison, the \"Opik probabilities 
were 0.28\% and 0.38\%. A potentially important difference is noted in CASE1B, where the number of recorded impacts
appears to drop more steeply near 3.5 Ga than the \"Opik profiles. Also, there were only 4 recorded impacts on Mars 
in the last 3.5 Gyr in CASE1B, while we would expect $\sim$10 based on the \"Opik probability. At least part of 
this difference is probably related to the small number statistics of recorded impacts. 

We tested this issue in more detail and found that the \"Opik results for Mars are sensitive to small changes of the 
outer boundary below which asteroids are assumed to be Mars crossing (nominally set as $q=Q_{\rm Mars}$, where $q$ and
$Q_{\rm Mars}$ are the pericenter distance of an asteroid and the apocenter distance of Mars, both changing as a function 
of time). This may indicate that the inner belt asteroids diffusing past the Mars-crossing boundary are temporarily 
protected against collisions with Mars, perhaps due to the phase protection mechanism in the Martian mean motion 
resonances (Morbidelli \& Nesvorn\'y 1999). This effect should have been more important after $\sim$3.5 Ga, 
when the slow diffusion in resonances became important (the early epochs were controlled by the excited population 
and E-belt). In any case, we caution that some of our results for Mars may require verification with a better 
statistic and/or different methods. 
\subsection{Impacts on Different Target Bodies}
\begin{figure} 
\epsscale{0.472}
\plotone{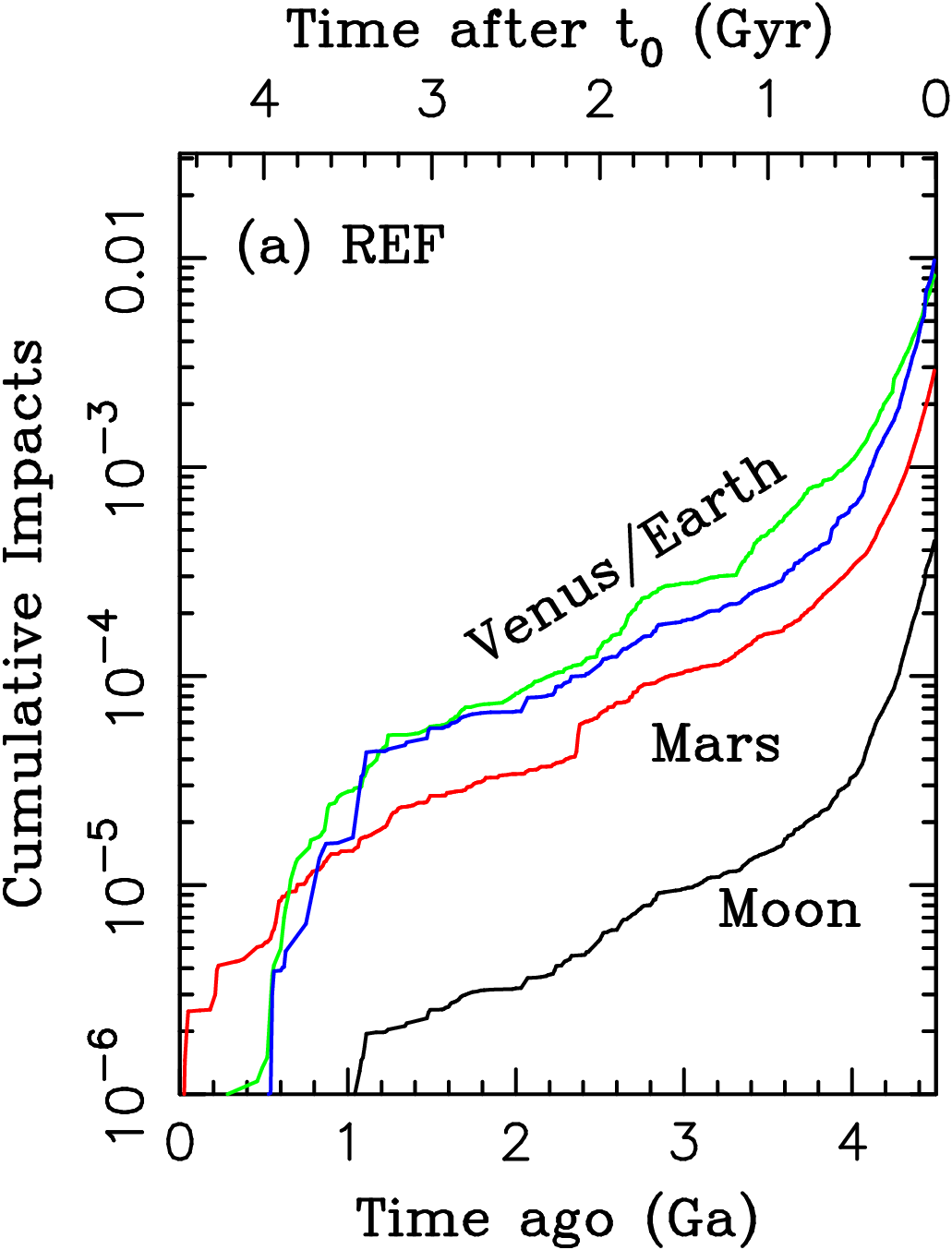}
\epsscale{0.430}
\plotone{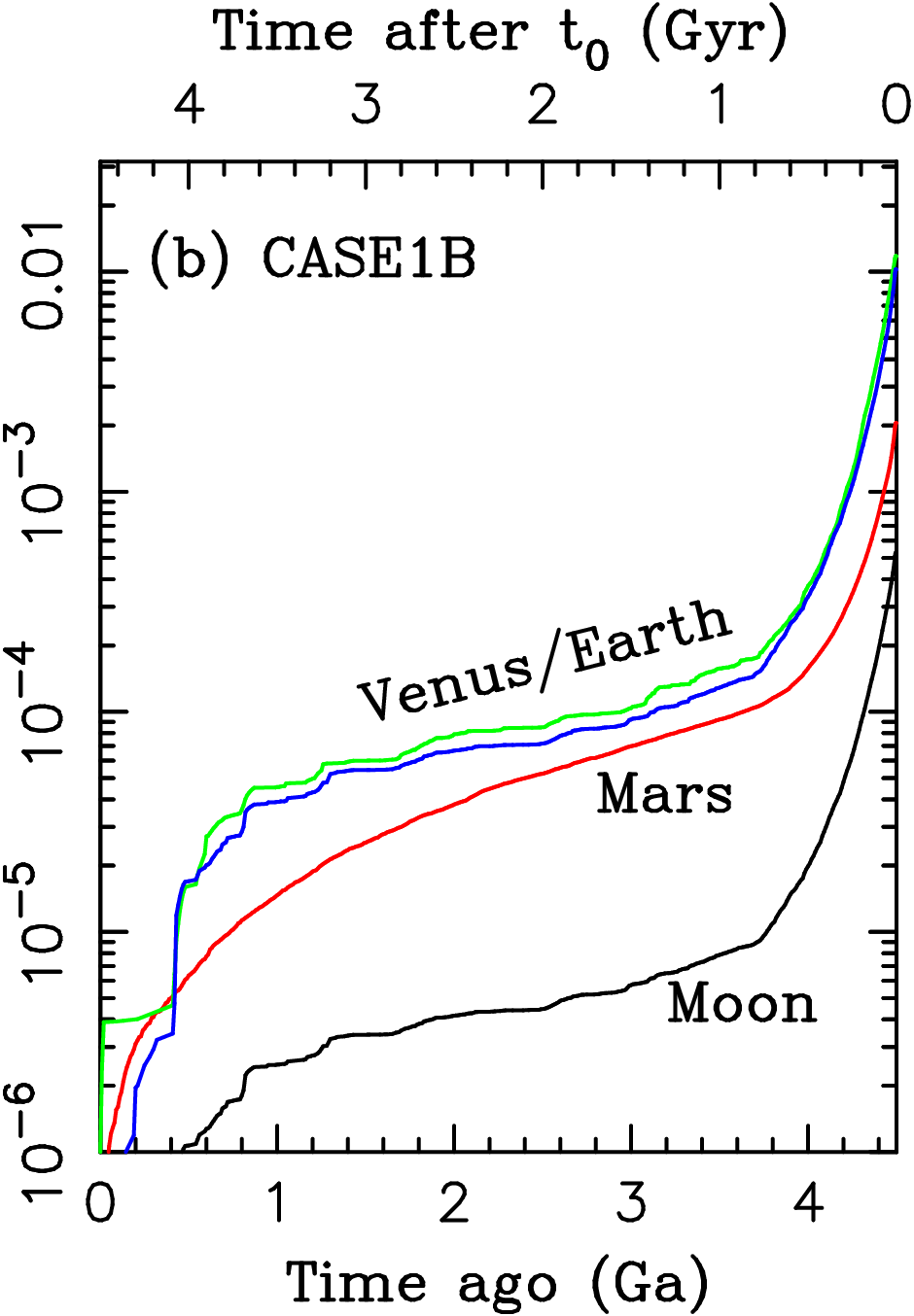}
\caption{The cumulative number of impacts on different target bodies in REF (panel a) and CASE1B (panel b). Different lines 
show the impact profiles for Venus (green), Earth (blue), Moon (black), and Mars (red).  The results shown here were obtained 
with the Gaussian mask in $e$ and $i$.}
\label{target}
\end{figure}

Figure \ref{target} shows the \"Opik-based impact profiles for $t_{\rm inst}=0$ and different target bodies (Venus, Earth, Moon, 
and Mars). The profiles for Venus and Earth are nearly identical in CASE1B. In REF, however, the Venus profile appears to be 
slightly steeper.  Again, things may be influenced by the small statistics in REF. In both REF and CASE1B, Venus receives a 
slightly larger number of impacts, roughly by 10-40\%, than the Earth (the exact value depends on the mask and $t_{\rm inst}$).
The shape of the lunar impact profile is identical to the Earth impact profile, as expected, because these targets share 
the same heliocentric orbit. The Earth-to-Moon ratio in the total number of impacts is 22 in REF and 19 in CASE1B (Table 2). 
These ratios were computed for $t_{\rm inst}=0$ from the \"Opik code. In CASE1B, the Earth-to-Moon ratio increases 
for $t_{\rm inst}>0$ and is $\simeq$22 for $t_{\rm inst}\gtrsim200$ Myr. These ratios are larger than the ratio of the collisional 
cross-sections of these worlds ($\simeq$13.6) due to the larger gravitational focusing factor at the Earth surface. 

The Earth-to-Mars ratio was computed from the number of impacts recorded by the {\it Swift} code, and not from the \"Opik
code which may have issues with Mars (Section 3.7). It depends on $t_{\rm inst}$. For $t_{\rm inst}=0$, for example, 
the Earth-to-Mars ratio is 2.8 in REF and 2.9 in CASE1B (Table 2). For $t_{\rm inst}=400$ Myr, these values drop to 1.5 in REF and
2.3 in CASE1B, while STEP with $t_{\rm inst}=400$ My gives an intermediate value of 1.8.  All three values are lower than 
the cross-section Earth/Mars ratio ($\simeq$3.5) despite stronger gravitational focusing by the Earth. This is related to 
the fact that the Mars-crossing orbits are, in general, more common. The dependence on $t_{\rm inst}$ is caused by a slow 
leakage of asteroids from the main belt before $t_{\rm inst}$, a process that favors Mars impacts over Earth impacts. The 
ratio is higher in CASE1B due to the stronger excitation of orbits during the instability, which favors Earth impacts
over Mars impacts. Our results are broadly consistent with Bottke et al. (2016), where the Earth-to-Mars ratio $\simeq$2 
was reported for the E-belt and $t_{\rm inst}=400$ Myr. 

\begin{figure} 
\epsscale{0.6}
\plotone{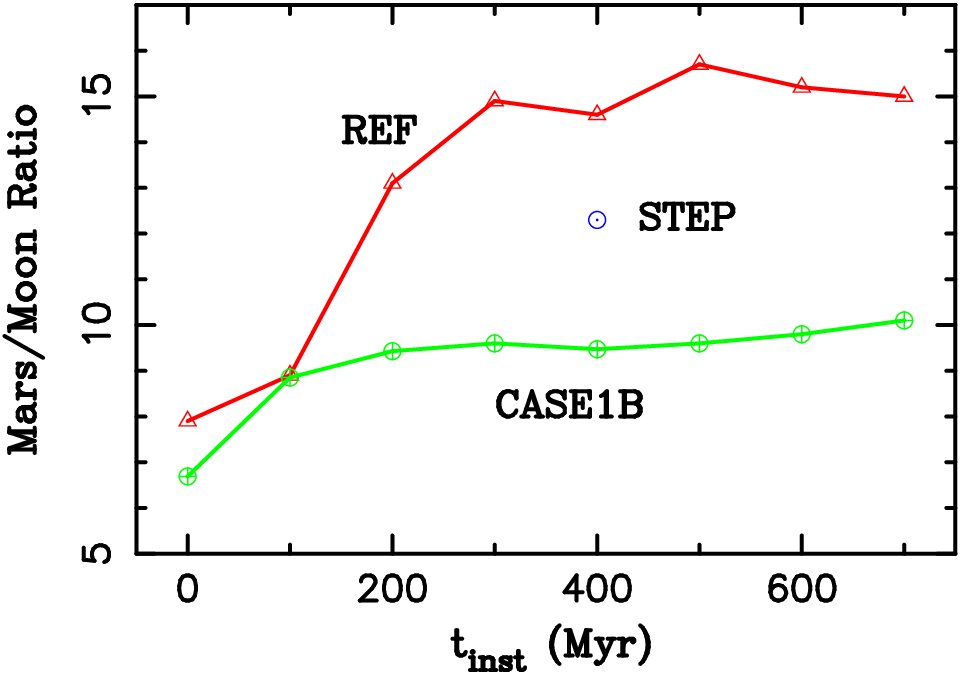}
\caption{The Mars-to-Moon ratio in the total number of impacts since 4.5 Ga as a function of the instability time, 
$t_{\rm inst}$. Different colors corresponds to different instability cases: STEP with $t_{\rm inst}=400$ Myr (blue), 
REF (red) and CASE1B (green). The results shown here were obtained with the Gaussian mask in $e$ and $i$.}
\label{ratio}
\end{figure}

The dependencies discussed above project themselves into the Mars-to-Moon ratio of the total number of impacts. To compute 
the Mars-to-Moon ratio, we combine the Earth-to-Moon ratio from the \"Opik code with the Earth-to-Mars ratio from the 
{\it Swift} record of impacts. Figure \ref{ratio} show how the Earth-to-Moon ratio depends on $t_{\rm inst}$. It is
$\simeq$6.7-7.9 for $t_{\rm inst}=0$ (higher for REF, lower for CASE1B). The REF case shows a stronger dependence on 
$t_{\rm inst}$, because the dynamical effects of the instability are weaker in this case, which favors Mars. The 
Mars-to-Moon ratio is $\simeq$15 in REF and $t_{\rm inst}\gtrsim 300$ Myr, and $\simeq$10 in CASE1B and $t_{\rm inst}\gtrsim 
200$ Myr. The STEP simulation results are intermediate, indicating the ratio 12.3 for $t_{\rm inst}=400$~Myr. All values  
reported here were obtained for the Gaussian mask in $e$ and $i$ (Section 2.5). The results for other masks were similar. 

The impact profiles on all target bodies discussed here show a similar pattern with a faster drop-off during the first $\sim$1 Gyr, 
and a slow later decline. Thus, to zeroth order (e.g., abstracting from the differences in the impact velocities, see Section 3.11, 
and crater scaling relationships for bodies with different gravity), approximate inferences about the impact chronologies 
can be obtained by using a `standard' profile shape that is vertically shifted on a plot like in Figure \ref{target}
by factors discussed above (e.g., Hartmann \& Neukum 2001). Note, however, that the shape of impact profile is specific to 
the instability case (compare the two panels in Figure \ref{target}), and is strongly influenced in the first $\sim$1 Gyr by 
$t_{\rm inst}$ (see Figure~\ref{tinst}). 
\subsection{Absolute Flux Calibration}
We used the method described in Section 2 to obtain an absolute calibration of the impact fluxes. To this end we 
collected all available information about the asteroid diameters from the IRAS, WISE, NEOWISE and AKARI measurements
(Tedesco et al. 2002, Ryan \& Woodward 2010; Masiero et al. 2011, 2012, 2014; Usui et al. 2011; Nugent et al. 2015), 
and cross linked it with the MPC catalog (data kindly provided to us by M. Delbo). The size information is 
available for 133,728 distinct asteroids. M. Delbo's analysis of this data shows that the the catalog is $>$90\% 
complete for main belt asteroids with diameters $D>10$~km. This allows us to calibrate the impact flux to any 
$D>10$ km. 

For example, we find that there are $N(>\!\!10)=7865$ main belt asteroids with $D>10$~km. In our integrations,
using the Gaussian mask, 10,700 asteroids survive in the main belt at $t=4.5$ Gyr in CASE1B, from the original
50,000. This corresponds to a surviving fraction $f_{\rm surv}=10700/50000=0.21$. To translate the fluxes normalized 
to one initial asteroid to the absolutely calibrated flux for $D>10$ km asteroids, we thus apply a multiplication 
calibration factor $f_{\rm calib}(>\!\!10)= N(>\!\!10)/f_{\rm surv}=3.7\times10^4$. For REF, $f_{\rm surv}=0.48$ and 
$f_{\rm calib}(>\!\!10)=1.7\times10^4$. Note that while the normalized fluxes for REF and CASE1B were similar, the 
calibrations are different and favor CASE1B over REF by a factor of $3.7/1.7=2.2$. The calibrated fluxes for CASE1B
are therefore higher. A similar calibration is done for impacts of $D>15$-km, $D>20$-km and $D>50$-km asteroids.

A problem with the straightforward calibration described above arises because the 3-3.3~au main-belt region contains
several very large asteroid families (e.g., Nesvorn\'y et al. 2015), representing a large share of $D>10$ km asteroids
in the main belt. We prefer {\it not} to use this part of the main belt for our calibration, because the existence of many 
$D>10$-km family members at 3-3.3 au would artificially increase the calibration factor. Thus, calibrating on 2-3 au, we obtained 
$N(>\!\!10)=2886$, and $f_{\rm surv}=0.35$ and $f_{\rm calib}(>\!\!10)=8.1\times10^3$ for REF, and $f_{\rm surv}=0.17$ and 
$f_{\rm calib}(>\!\!10)=1.7\times10^4$ for CASE1B. These calibration factors are not very sensitive to the choice
of the semimajor axis interval, assuming that the upper limit is $<$3 au. We also verified that similar calibration 
factors are obtained from the full 2-3.3 au interval when $D>10$~km family members are removed. We prefer to use 
the 2-3 au calibration because it is simpler. 

\begin{figure} 
\epsscale{0.475}
\plotone{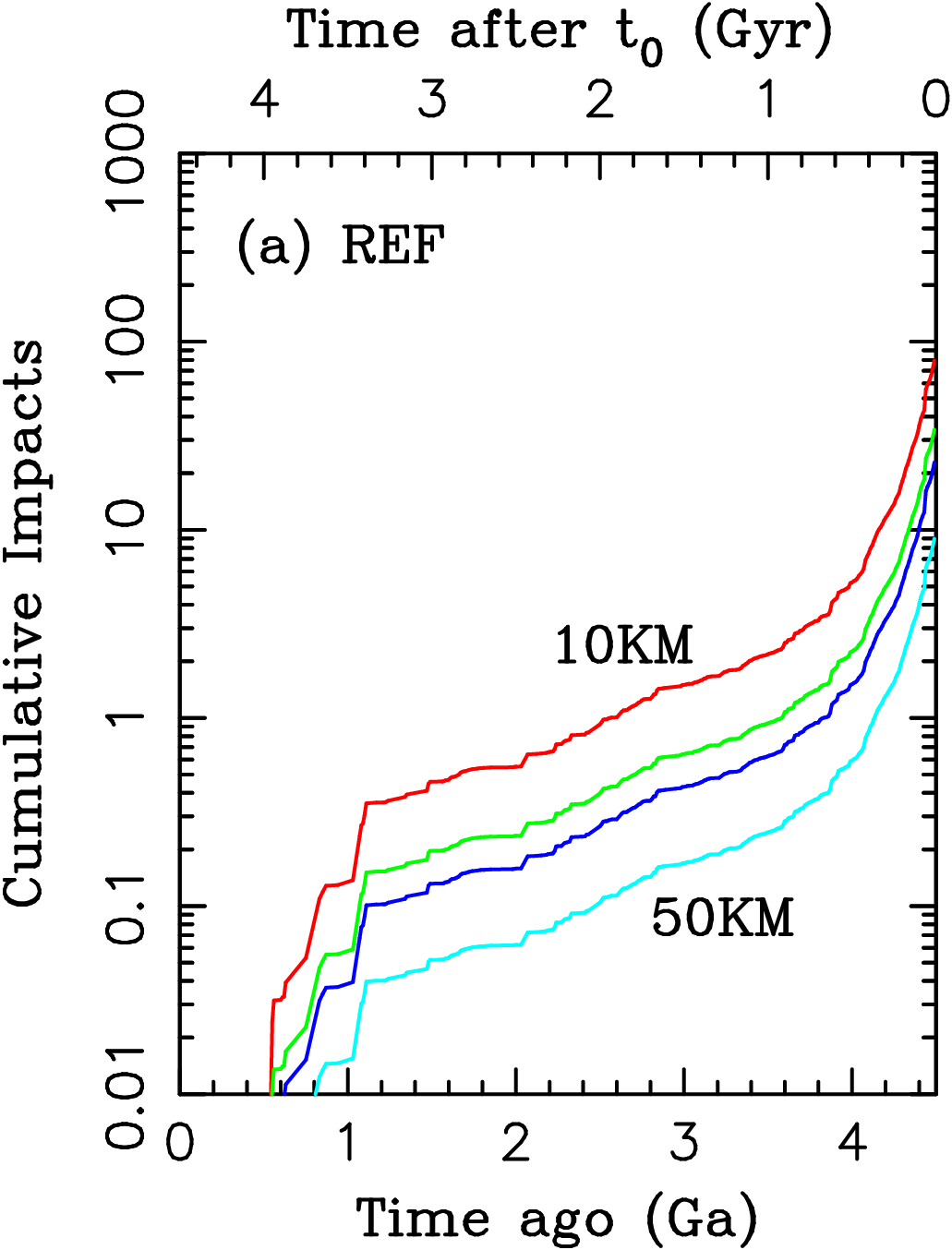}
\epsscale{0.425}
\plotone{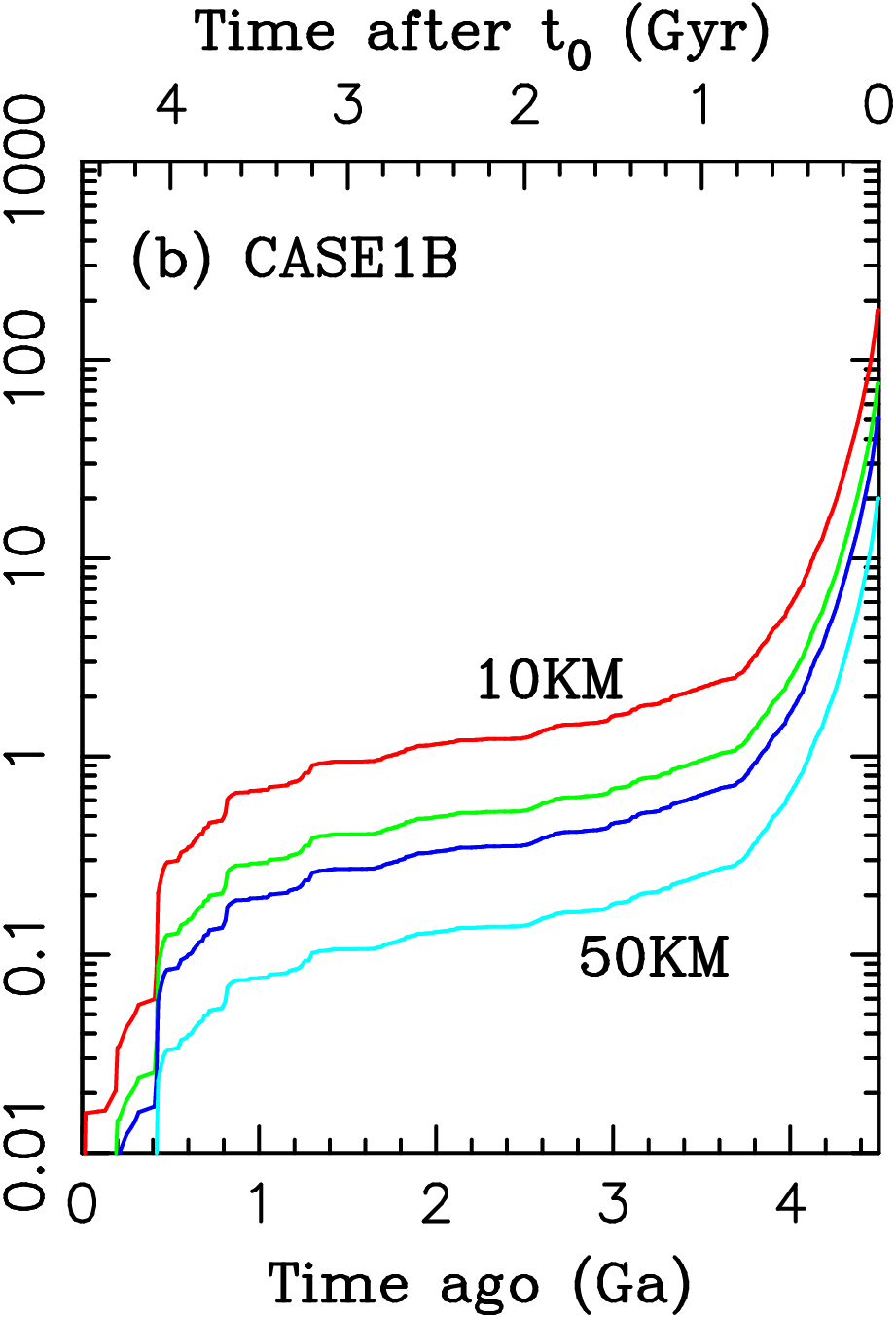}
\caption{The absolutely calibrated impact profiles for the Earth: REF (panel a) and CASE1B (panel b). Different lines 
show the impact profiles for different impactor sizes: $D=10$ km (red), 15 km (green), 20 km (blue), and 50 km 
(turquoise).  The results shown here were obtained with the Gaussian mask in $e$ and $i$.}
\label{earth}
\end{figure}

Figure \ref{earth} shows the result. As for the overall number of impacts on the Earth over 4.5~Gyr, REF shows 79 
impacts of $D>10$ km asteroids, while CASE1B shows 177 impacts, about 2.2 higher than REF, which is mainly contributed
by the larger calibration factor in CASE1B. The number of large impactors closely follows, by design, the size distribution 
of the main belt asteroids with $2<a<3$ au. Specifically, there are 2.3 times fewer impacts for $D>15$ km, 3.5 times 
fewer for $D>20$ km, and 8.9 times fewer for $D>50$ km.  We do not attempt to calibrate the fluxes for $D<10$ km asteroids 
because: (1) the catalog of sizes of main belt asteroids is incomplete for $D<10$ km, and (2) the non-gravitational forces, 
such as the Yarkovsky effect, not modeled here, become important for $D<10$ km. 

An approximate fix to both these problems would be to use the size distribution of NEAs. The size 
distribution of NEAs is well known down to $D<1$ km. Also, since small NEAs escape from the main belt by the Yarkovsky 
effect and resonances, the size distribution of NEAs implicitly includes the Yarkovsky effect. For example,
if there are $\sim$300 times more $D>1$-km NEAs than $D>10$-km NEAs (this factor is uncertain due to the small number 
of large NEAs), then the $D>10$-km asteroid impact flux shown in Figure \ref{earth} should be shifted up by a factor 
of $\sim$300 to yield the $D>1$-km asteroid impact flux. There is, however, at least
one serious problem with this approach. This is because the $\sim$300 factor mixes the effects of 
(1) and (2). When it is properly accounted for (1), independently of (2), the resulting multiplication factor 
could be applied to the normalized flux profile in very much the same as was done above for $D>10$ km. 

The effect of (2), on the other hand, is {\it not}
a simple multiplication factor. This is because in the first 1 Gyr, the Yarkovsky effect is not important. Instead, 
the impact fluxes in the first 1 Gyr are dominated by dynamically unstable populations in the E-belt and on excited 
orbits. The Yarkovsky effect becomes important only after 3 Ga, when the dynamically unstable populations are
strongly depleted and do not produce that many impacts. Thus, 
the multiplication factor of $\sim$300 can only be applied in the last 3 Ga. We can therefore see that the 
{\it shape} of the profiles shown in Figure \ref{earth} should change when extrapolated to $D>1$~km impactors. 
Relative to the $D>10$-km profile, the $D>1$-km profile should be a factor of $\sim$300 higher in the past 3 Ga and 
higher by a smaller factor, estimated to be $\sim$100 here, during the first 1 Gyr. This would very roughly indicate 
that the Earth would receive 8,000 $D>1$-km impacts in total (i.e. since 4.5 Ga) in REF, and 18,000 $D>1$-km impacts 
in CASE1B. Improvements in modeling will be needed to determine these numbers with more confidence. 
\subsection{Comparison with the Current Impact Flux }
An interesting problem is identified when the number of impacts of $D>10$ km asteroids in the last 1 Gyr (Figure 
\ref{earth}) is compared with the current day impactor flux inferred from the observations of NEAs (e.g., 
Stuart \& Binzel 2004, Harris \& D'Abramo 2015). Johnson et al. (2016) reported, assuming that the 
NEA impactor flux was constant over the past 1~Gyr, that $\sim$5 $D>10$-km impacts would be expected on Earth in the 
last 1 Gyr. In contrast, we estimate that $\sim$0.4 such impacts would occur in REF and 
$\sim$0.7 in CASE1B (Figure \ref{earth}). Our model impact rates are therefore $\sim$7-12 times below the expectations. 
Similar discrepancy was noted in Minton \& Malhotra (2010).
This can mean one of several things. First, the current day impactor flux can be, for some reason, larger than the 
average flux over the past Gyr (e.g., Culler et al. 2000, Mazrouet et al. 2016). Alternatively, some model parameters 
need to be tweaked to increase the impact flux in the last 1 Gyr. 

Considering the late instability with $t_{\rm inst}>100$ Myr does not help because, as we discussed in Section 3.6,
the impact profiles in the last 3 Gyr are nearly independent of $t_{\rm inst}$ (assuming that $t_{\rm inst}<1$ Gyr). As for 
the initial distribution of asteroids, the different initial masks give nearly identical results in CASE1B (Section 3.4).
In REF, instead, the Rayleigh and Grand Tack masks give $\sim$1.5 impacts in the last 1 Gyr, roughly 4 times 
more than the Gaussian mask, and closer to the NEA-based estimate. A question arises whether we have 
sufficient statistics in our model to reliably estimate the number of impacts in the last 1 Gyr. We tested this
by using a subset of initial orbits and found that the CASE1B results are reliable. As for the REF case,
where we only had 20,000 initial orbits, we found significant variations of the impact flux in the last 1 Gyr, 
up to a factor of $\sim$2, when we sub-sampled to 10,000 or 5,000 initial orbits. The current flux estimate in REF 
therefore has a significant uncertainty.

Additional issues can be related to some problem with the estimate given in Johnson et al. (2016) or with our model. 
For example, Johnson et al. (2016) assumed that the impact probability of large NEAs is essentially the same as 
the impact probability of small NEAs. This may not be fully accurate because it is not guaranteed
that the orbital distributions of the small and large NEAs are the same (e.g., Valsecchi \& Gronchi 2011). 
On one hand, the large asteroids typically reach the NEA region via slow diffusion in weak resonances 
(Migliorini et al. 1998, Morbidelli \& Nesvorn\'y 1999, Farinella \& Vokrouhlick\'y 1999). The small asteroids, on the other hand, can
drift over a considerable radial distance by the Yarkovsky effect and reach the NEA orbits from the powerful 
$\nu_6$ resonance on the inner edge of the asteroid belt (Bottke et al. 2002). The $\nu_6$ resonance is known to produce 
highly evolved NEA orbits and impact probabilities on Earth in excess of 1\% (Gladman et al. 1997). The small NEAs can therefore have 
larger impact probabilities than the large ones (Valsecchi \& Gronchi 2015). This could mean that the current 
impact flux reported in Fig. 1 in Johnson et al. (2016) should drop more steeply for $D\gtrsim5$ km, therefore
implying fewer than 5 $D>10$-km asteroid impacts in the last 1 Gyr.      

Finally, it is also possible that our model is producing too few large NEAs in the last 1~Gyr. This could be related 
to some problem with the orbital structure of the source regions in the model, or to some dynamical process neglected
in our simulations. Although the main belt structure at $t=4.5$ Gyr is clearly not exactly correct (Section 3.1), 
we find it unlikely that this is causing a difference. We reason that the REF and CASE1B models end up with very 
different orbital distributions in the inner main belt (overpopulated in REF and anemic below 2.3 au in CASE1B), 
but both give the impact fluxes that are significantly below the estimate of Johnson et al. (2016). It is thus 
more likely that the difference stems from some feature that is common to the two models. 

As for the dynamical processes, neglecting the Yarkovsky effect is probably the most important approximation in our model. 
According to Bottke et al. (2006), the maximum drift rate of a $D=10$ km inner-belt asteroid is $\simeq2\times10^{-5}$ 
au Myr$^{-1}$. Thus, the expected maximum drift over 3 Gyr is roughly 0.06 au. This seems small but can be significant, 
because weak diffusive resonances in the inner main belt, which provide the main escape routes for large asteroids, are 
very dense, and the nearest one may be only a tiny fraction of au away. The Yarkovsky effect may thus be 
significant, to some degree, even for $D=10$ km asteroids. We tested this and found that the Yarkovsky effect 
increases to the present day impact flux of $D>10$-km asteroid by 20-50\% relative to a case where the non-gravitational 
forces are neglected. This is not large enough to resolve the discrepancy. The results of these tests will be reported 
elsewhere.  
\subsection{Exponential Parametrization of the Differential Impact Flux, \\and Impact Speeds}
\begin{figure} 
\epsscale{0.475}
\plotone{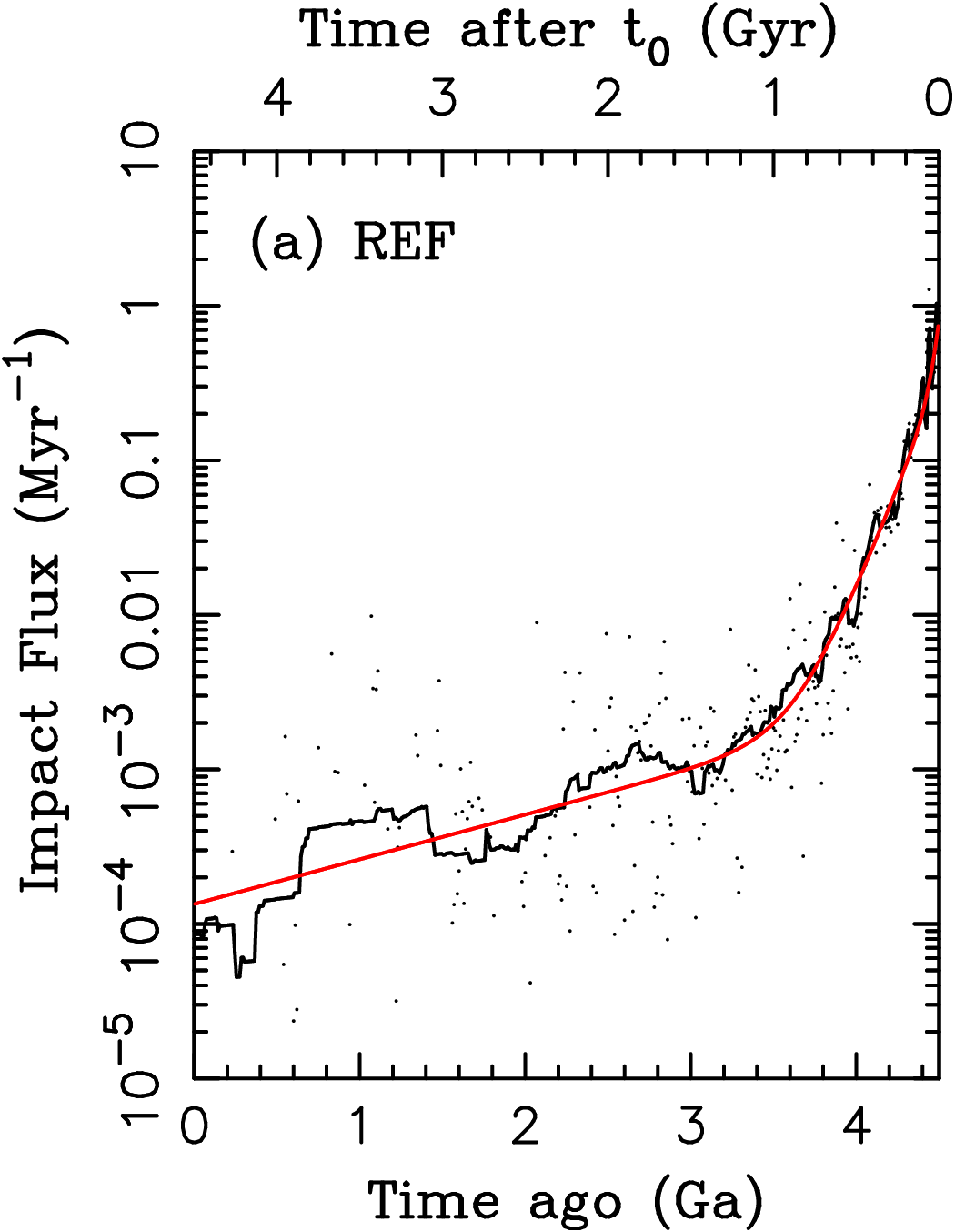}
\epsscale{0.425}
\plotone{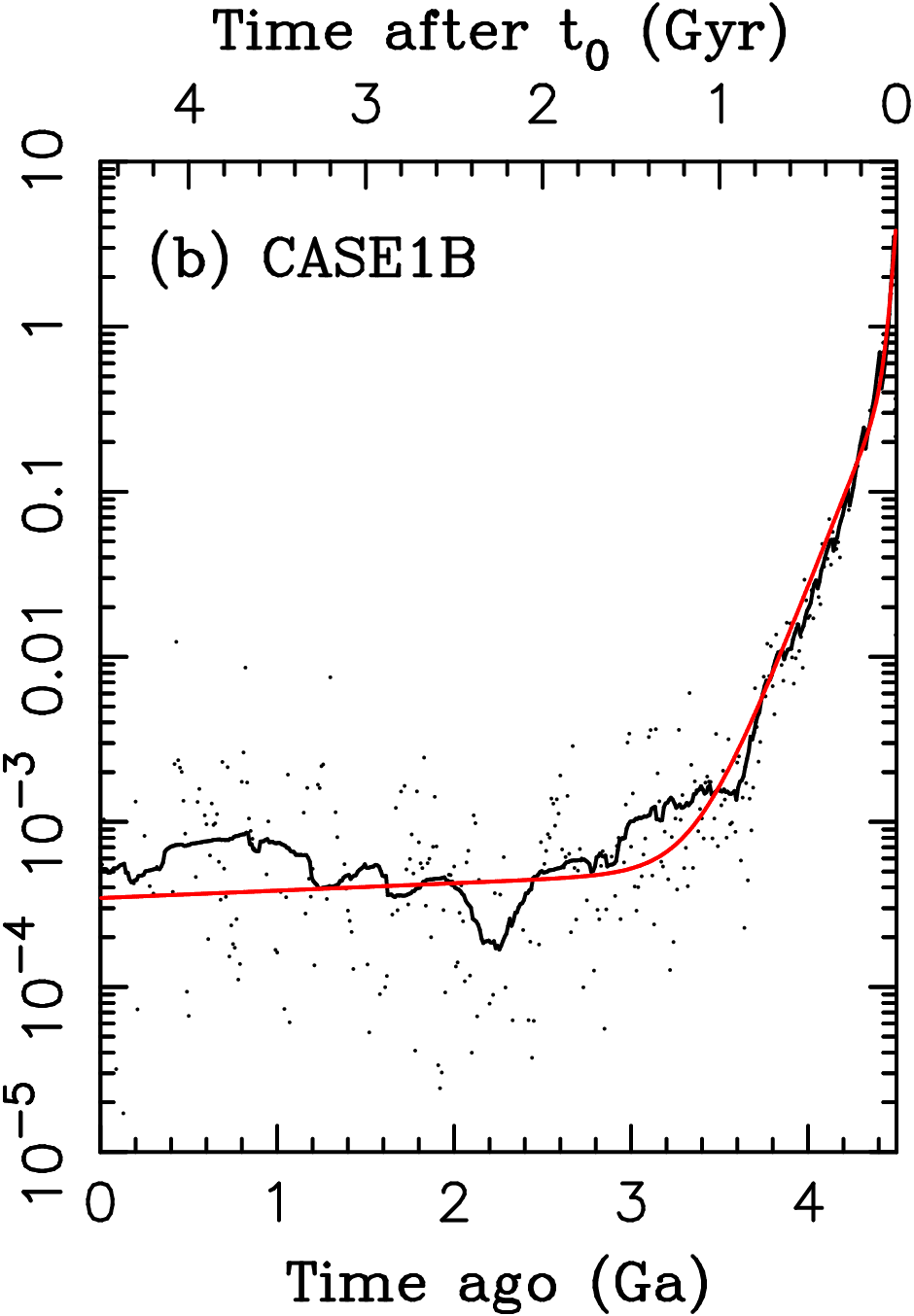}
\caption{A calibrated differential flux of $D>10$ km impactors on the Earth: REF (panel~a) and CASE1B (panel b). The 
black dots are the actual values obtained from the \"Opik code. The solid black 
lines are the moving-window averages. In REF, the statistics in the last $\sim$2 Gyr was poor. The red lines are the best 
exponential fits discussed in the main text.}
\label{diff}
\end{figure}

Figure \ref{diff} shows the calibrated impact flux in a differential plot (the number of $D>10$-km impacts per Myr).
This figure can be compared to Figure \ref{earth}, where the number of $D>10$-km impacts on the Earth was reported 
in a cumulative plot. Perhaps the most interesting thing to be noted in Figure \ref{diff} is that the number of 
impacts per Myr in the last 3 Gyr in CASE1B is relatively constant, while that in the REF case shows a significant 
decline from 3 Ga to the present. To quantify that, we fitted the following functional form to the the differential 
impact profiles 
\begin{equation}
F(D,t)=C_{\rm s}(D) \sum_i F_i \exp \left(- {t \over \tau_i} \right)\ , 
\label{exp1}
\end{equation}
where $C_{\rm s}$ is a size scaling factor, $F_i$ are constants, $\tau_i$ are the characteristic timescales of the 
impact flux decline, and $t$ runs forward from $t=0$ at $t_0$ to $t=4.5$ Gyr at the present epoch. By definition, 
$C_{\rm s}(D)=1$ for $D=10$ km. We found that $C_{\rm s}(D)=(D/10\ {\rm km})^{-\gamma}$ with $\gamma \simeq 2.1$ 
provides an adequate scaling from $D=10$ km to $10<D\lesssim15$ km, and $\gamma \simeq 1.4$-1.8 from $D=10$ km to 
$20 \lesssim D \lesssim 50$ km. The shallower index at larger sizes reflects the shallower size distribution of 
the main belt asteroids with $20 \lesssim D \lesssim 50$ km.  

We experimented with the number of exponential terms in Eq. (\ref{exp1}) and found that at least three terms are
needed to provide an accurate representation of the impact flux. The first term with very short $\tau_1$ is needed to
mimic the drop of the impact flux in the first 100~Myr, when parts of the E-belt and excited population rapidly decay.
The second term  with $\tau_2\simeq0.16$ Gyr is required to reproduce the flux decline in the first $\sim$ 1 Gyr. Finally, 
the third term with longer $\tau_3$ expresses the slow decline of the impact flux in the last $\sim$3 Gyr. 

\begin{figure} 
\epsscale{0.485}
\plotone{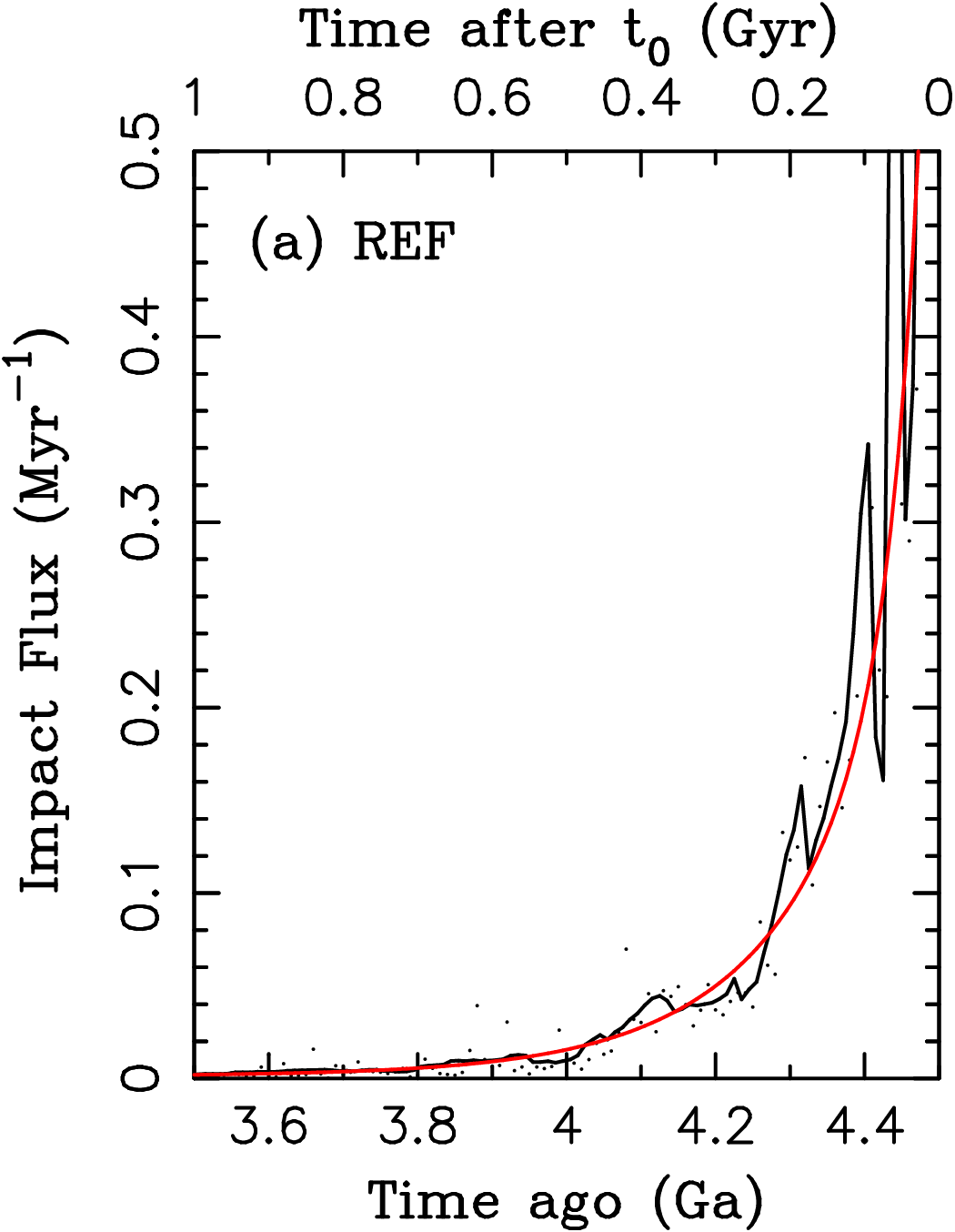}
\epsscale{0.425}
\plotone{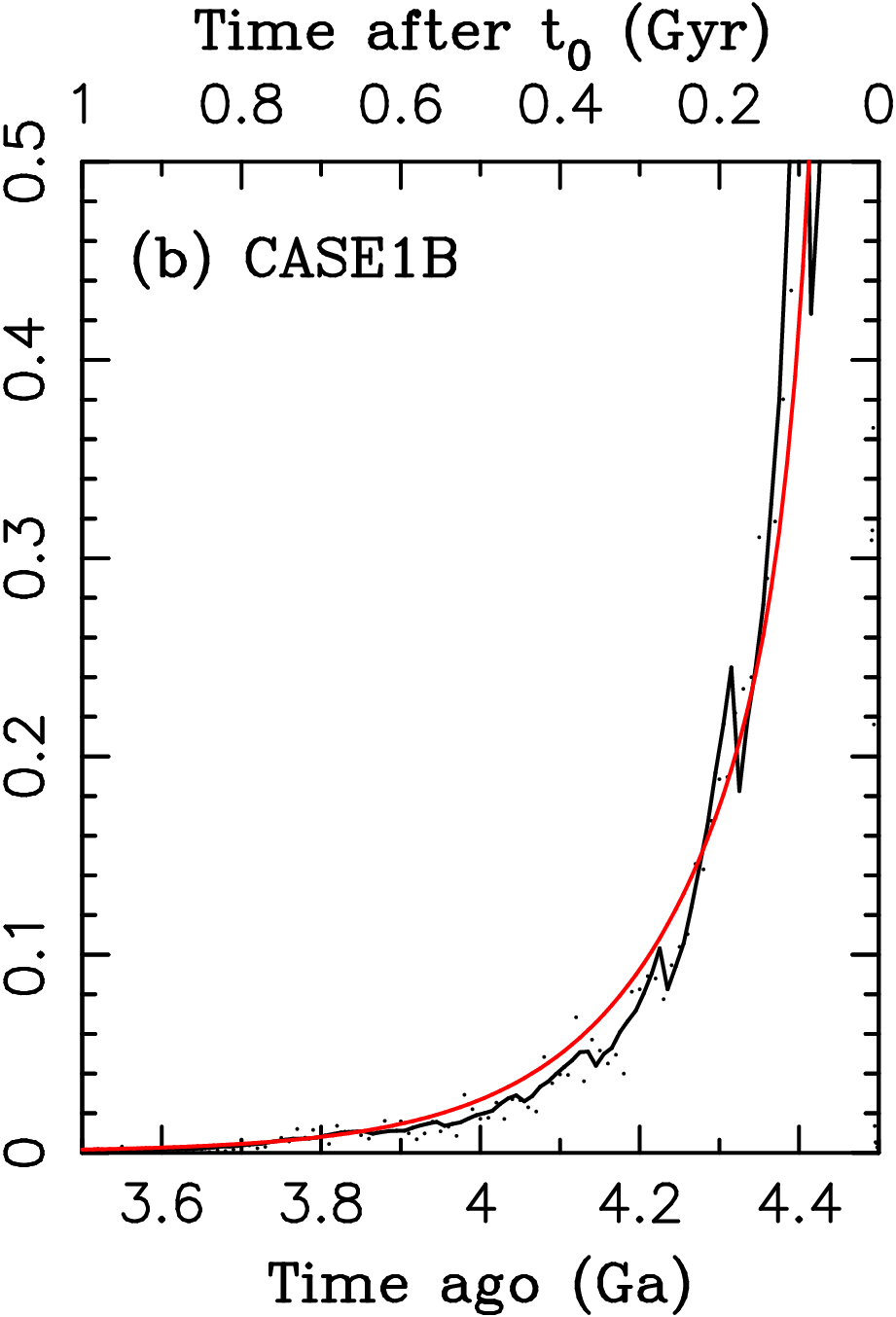}
\caption{The same as Figure \ref{diff} but without the logarithmic scale on the Y axis. Here the focus is on the impact
flux in the first 1.5 Gyr.}
\label{nolog}
\end{figure}

The best-fit parameters are reported in Table 3. Notably, $\tau_3\sim1.5$ Gyr in REF, while $\tau_3 \gg 10$ Gyr in 
CASE1B, corresponding to a much slower decline of the impact flux in CASE1B.
Equation (\ref{exp1}), together with the parameters given in Table 3, can be used to compute the 
flux of $D>10$ km impactors on the Earth at any given time of the Solar System history. The impact flux for a different 
target body can be obtained from the values listed in Table~2. As we discussed in Section 3.8, for $t_{\rm inst}=0$,
the Earth impact flux needs to be divided by a factor of $\simeq$20 to obtain the lunar impact flux, and by a factor of 
$\simeq2.8$-2.9 to obtain the impact flux on Mars. The Earth-to-Mars ratio decreases with $t_{\rm inst}$ and becomes
$\simeq1.5$-2.3 for $t_{\rm inst}\gtrsim200$ Myr. Figure \ref{nolog} shows the differential impact profiles in the first 
1.5~Gyr of the Solar System history. The impact flux steeply declines during this period. Both in REF and CASE1B, 
the exponential form provides a good representation of the impact flux. 

Neukum et al. (2001) reported the impact flux in terms of $N(1)$, the number of $D>1$~km craters per square kilometer
on the Moon, with $N(1)=\alpha[\exp(\beta T)-1]+\gamma T$, where time $T$ is in Gyr before the present time. A detail 
comparison of our profiles with $N(1)$ is difficult because it would require an uncertain extrapolation to the sub-km 
impactors that live in the Yarkovsky world. Their dynamics is not well captured in our model that ignores the Yarkovsky 
effect. Instead, we only point out two similarities. First, Neukum's $N(1)$ is unchanging in the past 3 Ga, which is 
consistent with $\tau_3 \gg 10$ Myr obtained in CASE1B, also implying a nearly unchanging impact flux (at least 
within a factor of $2$ or so). The results for REF, instead, suggest a declining impact flux in the last 3 Ga. 
This is related to a gradual depletion of the inner belt, which was left overpopulated in REF (Section 3.1). 

Second, $\beta=6.93$ Gyr$^{-1}$ in Neukum et al. (2001), suggesting a characteristic timescale of the LHB-flux decline of 
$\tau=1/\beta=0.14$ Gyr. This is similar to $\tau_2=0.16$ Gyr obtained in this work (Table 3). Note that this 
is an amazing correspondence given that Neukum et al. results were obtained from the crater analysis, while our
results build on dynamical modeling. It has to be noted, however, that Robbins (2014), using their new crater counts 
from the {\it Lunar Reconnaissance Orbiter} images, produced a new chronology with $\tau=1/\beta=0.06$ Gyr, a value
that is intermediate between our $\tau_1$ and $\tau_2$ values. This would imply a significantly steeper decline of the 
impact flux at 3.5-4 Ga, and could be taken as a sign of $t_{\rm inst}\simeq4$~Ga. Note that Neukum's and Robbins' 
chronologies cannot be reliably extrapolated to before 4 Ga, where there are no solid constraints. Our dynamical 
modeling suggests that, moving back in time toward $t_{\rm inst}$, the impact fluxes should rise more strongly relative to 
a simple exponential. 
    
\begin{figure} 
\epsscale{0.485}
\plotone{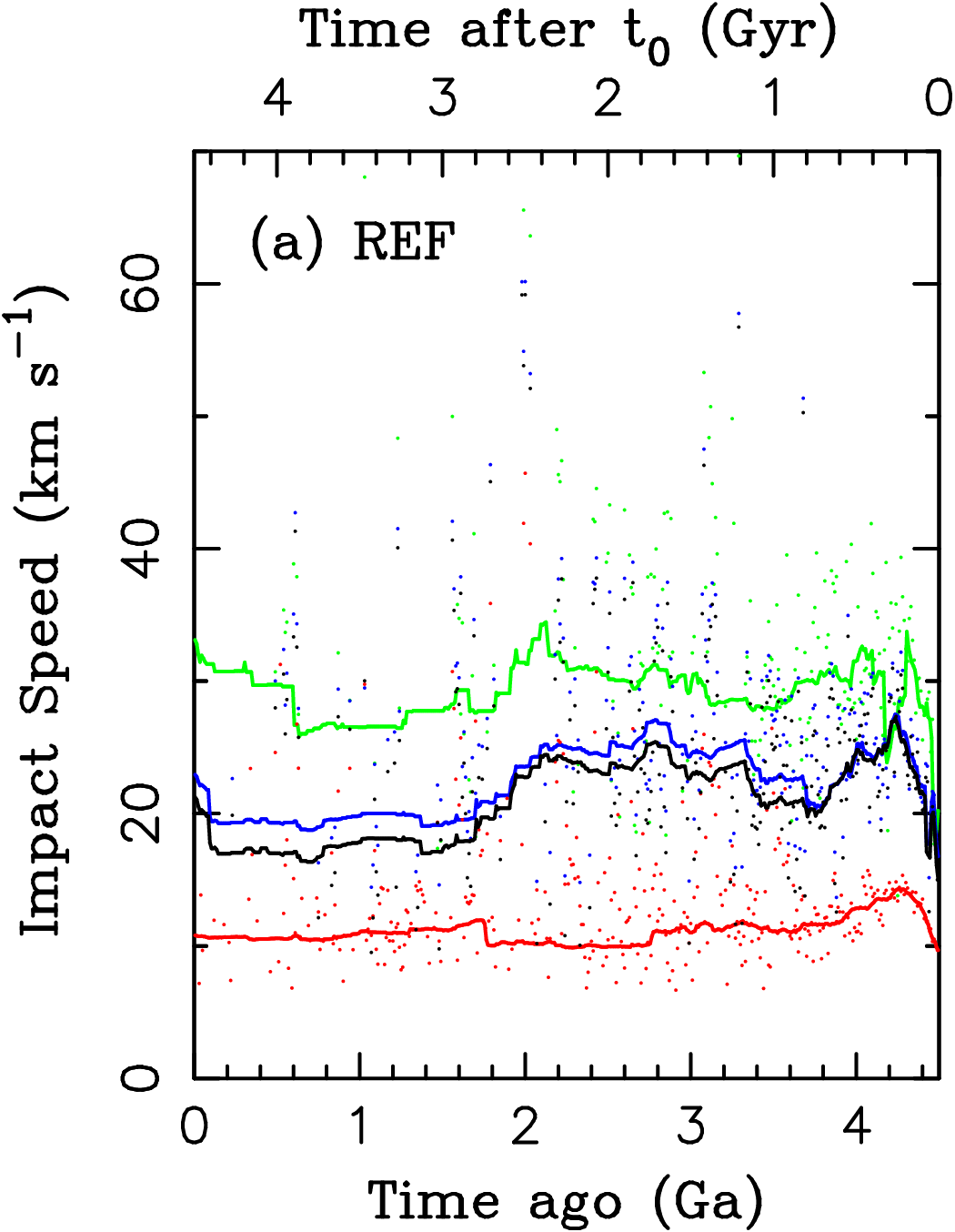}
\epsscale{0.425}
\plotone{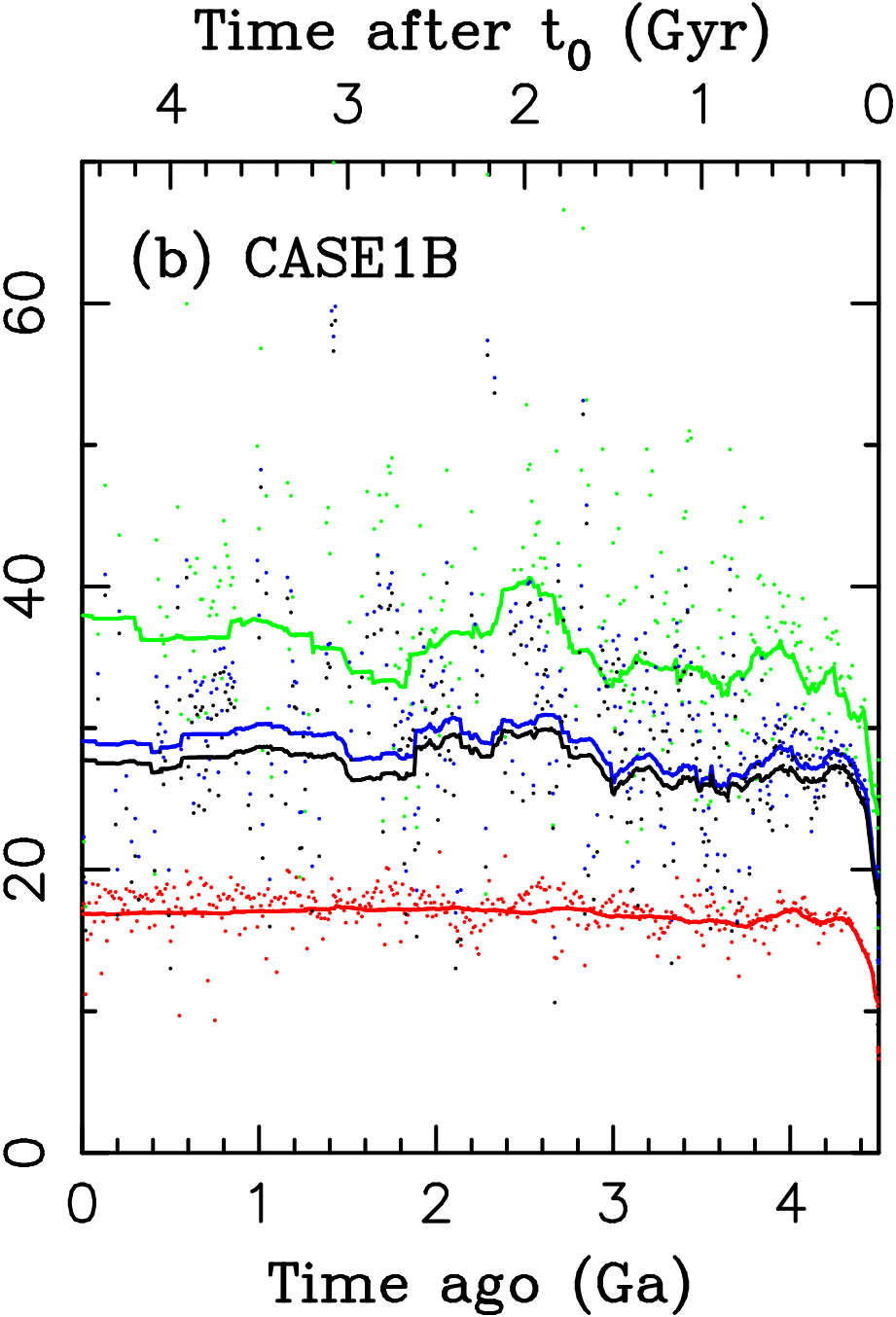}
\caption{The impact speeds of large impactors on different target bodies in REF (panel a) and CASE1B (panel b). 
The dots show the values obtained from the \"Opik code for Venus (green), Earth (blue), Moon (black), and Mars (red).
The lines are the moving-window averages.}
\label{vel}
\end{figure}

The impact speeds on different target bodies are shown in Figure \ref{vel} and the mean values are reported in Table 4. 
For the Earth, the mean impact speed is 21.0 km s$^{-1}$ in REF and 23.5 km s$^{-1}$ in CASE1B, which is somewhat higher than 
22 km s$^{-1}$ found for the E-belt impactors in Bottke et al. (2012). The impact speeds on the Moon are slightly lower 
than on the Earth, while those on Mars are significantly lower (12.0 km s$^{-1}$ in REF and 13.7 km s$^{-1}$ in CASE1B). 
This reflects both the lower orbital velocity of Mars and its smaller gravitational focusing factor. Figure \ref{vel} shows 
that the impact speeds increased during the first 100 Myr, and then remained relatively constant. In the late instability 
cases, the impact speeds before and after the instability are significantly different. For example, in CASE1B, 
the characteristic impact speeds on Earth are $\sim$18 km s$^{-1}$ for $t<t_{\rm inst}$ and $\sim$25 km s$^{-1}$ 
for $t>t_{\rm inst}$. This change could be important for the interpretation of the impact cratering record in the inner 
Solar System (Marchi et al. 2012).     
\section{Caveats for $t_{\rm inst}<100$ Myr}
Here we discuss some of the caveats inherent to the modeling scheme adopted here. In our simulations we assumed 
that the terrestrial planets were fully formed at $t_{\rm inst}$. This may not be correct if $t_{\rm inst}<100$ Myr, because 
the terrestrial planet formation may have taken 30-100 Myr to reach completion. This is important in several different 
ways. First, if $t_{\rm inst}<100$~Myr, the instability might have happened while the terrestrial planet accretion was still 
in progress, perhaps before the Moon-forming impact that is thought to have occurred 30-100 Myr after $t_0$ (e.g., 
Taylor et al. 2009, All\`egre et al. 2008, Jacobson et al. 2014). If that's the case, the instability integrations 
would need to be combined with a model of the terrestrial planet accretion (Walsh \& Morbidelli 2011). For that, however, 
the terrestrial planet formation would need to be better understood (e.g., Walsh \& Levison 2016). Here we just point 
out that Mars may have formed early (e.g., Dauphas \& Pourmand 2011) and probably did not 
participate in the late accretion processes that ended up, after a prolonged stage of giant impacts, producing Earth 
and Venus. This would mean that Mars was already present even if the instability happened early, and the depletion of 
the E-belt and inner part of the main belt produced by Mars may have been reminiscent to what we obtained in our model.

Second, the terrestrial planets represent an important constraint on the orbital evolution of the giant planets during
the instability. This constrain is especially important if the instability happened late, because in this case
it is strictly required that the orbits of Jupiter and Saturn experienced a discontinuity during their encounters with 
a third, planetary-size object (the so-called jumping-Jupiter model; Morbidelli et al. 2009, 2010; Brasser et al. 2009). 
Moreover, it was shown that 
only a small measure of the jumping-Jupiter models is fully satisfactory (e.g., Kaib \& Chambers 2016, Roig et al. 2016), 
while most lead to an excessive excitation of the terrestrial planet orbits. The constraint on the jumping-Jupiter model 
would be relaxed if the instability happened early. 

On one hand, the jumping-Jupiter model is still required from the dynamical structure of the asteroid belt (e.g., 
Morbidelli et al. 2010, Roig \& Nesvorn\'y 2015, Toliou et al. 2016). Planetary encounters are also needed to explain the secular structure 
of the giant planet orbits (Morbidelli et al. 2009, Nesvorn\'y \& Morbidelli 2012), and the Kuiper belt kernel (Nesvorn\'y 
2015b). Moreover, at least some of the terrestrial planets, such as Mars, and perhaps also Mercury, may have been 
present during the instability, even if $t_{\rm inst}<100$ Myr. 

On the other hand, the dynamical structure of the terrestrial planet system was probably different at $t_{\rm inst}$ if 
$t_{\rm inst}<100$ Myr. Some of the secular resonances (mainly $g_1=g_5$ and $g_2=g_5$, where $g_1$, $g_2$ and $g_5$ are the 
apsidal precession frequencies of Mercury, Venus and Jupiter), which were shown to be capable of driving 
strong excitation of the terrestrial planet orbits (Brasser et al. 2009, Agnor \& Lin 2012), 
may have been avoided even if Jupiter and Saturn did not strictly follow 
the orbital evolution in the standard jumping-Jupiter model. Moreover, even if some resonances occurred and the orbits 
were strongly excited, this is not a problem, because the traces of this event would have been erased by the subsequent 
stage of the terrestrial planet formation (Walsh \& Morbidelli 2011). 

Another concern is the crater retention age of different surfaces. For example, following the Moon-forming impact,
both the Earth and proto-Moon, which accreted from the impact-generated disk outside the Roche radius (Canup et al. 2013), 
were in a molten state and may have remained molten for tens to hundreds of Myr. For example, the lunar magma ocean is thought  
to have crystallized only after $\sim$100-200 Myr after the Moon-forming impact (Elkins-Tanton et al. 2011) due to
insulation provided by a thin surface crust. If so, the lunar surface would be incapable of recording large impacts during
the first $\sim$100-200 Myr. Instead, the impactors would penetrate through the crust, exposing the sub-surface magma, 
which would then flow to erase any signs of the impact. A different issue applies to Mars, where the whole pre-existing 
crater record may have been erased by the Borealis-basin impact (Marinova et al. 2008, 2011; Andrews-Hanna et al. 2008;
Nimmo et al. 2008), whose formation time is debated.

\begin{figure} 
\epsscale{0.485}
\plotone{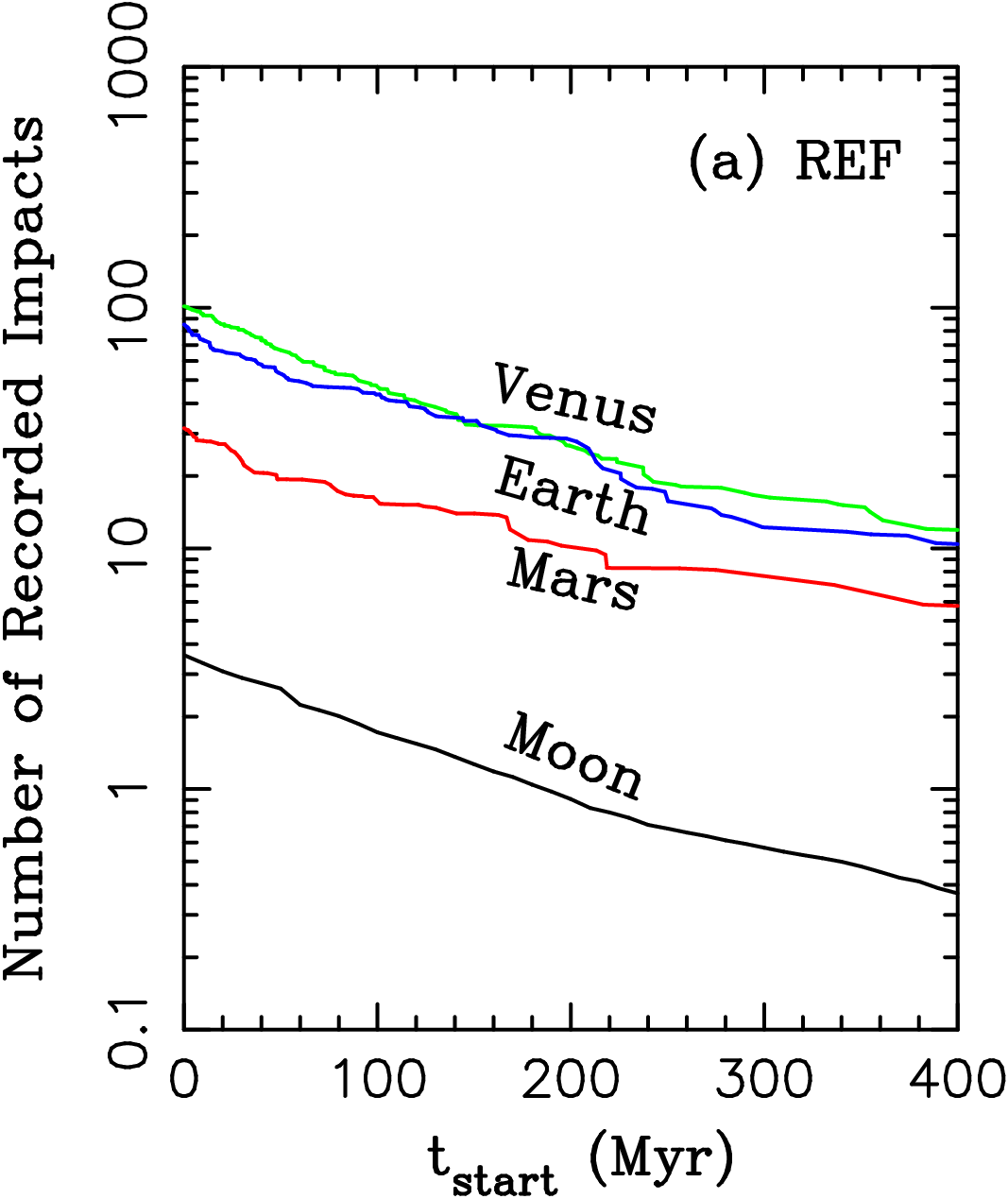}
\epsscale{0.435}
\plotone{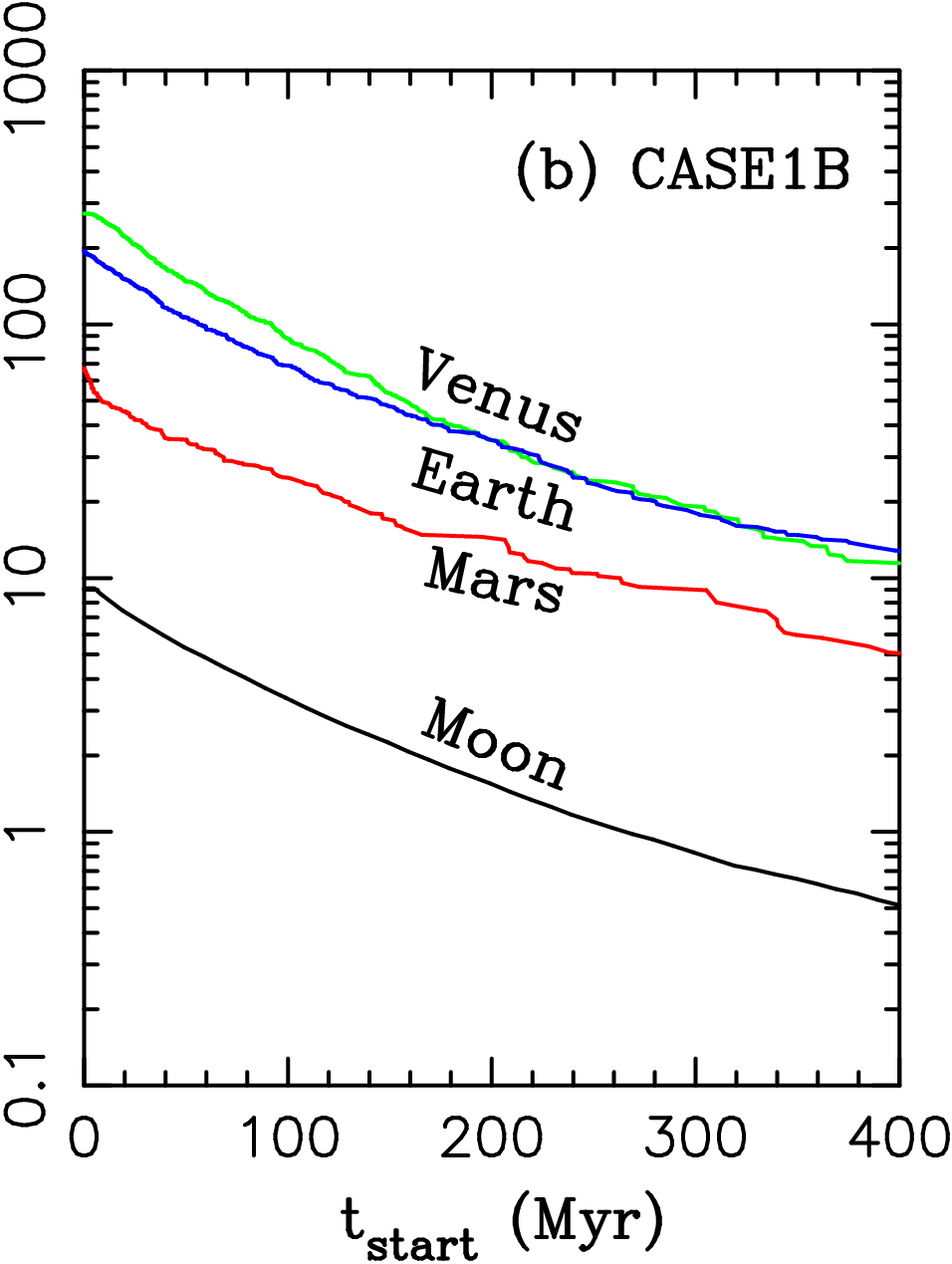}
\caption{The total number of $D>10$-km asteroid impacts recorded on a surface for $t>t_{\rm start}$: Venus (green), 
Earth (blue), Moon (black), and Mars (red). Here we assume that $t_{\rm inst}=0$. The results shown here for Venus,
Earth and Mars were obtained from the planetary impacts recorded by the {\it Swift} integrator; the results
from the \"Opik code are similar. The results shown here for the Moon were obtained from the \"Opik code.}
\label{start}
\end{figure}

To account for the crater retention age in our model, we introduce a new parameter $t_{\rm start}$, and record all impacts
on different terrestrial worlds after $t_{\rm start}$. Figure \ref{start} shows that total number of impacts recorded 
on different surfaces as a function of $t_{\rm start}$. The case with $t_{\rm start}=0$, corresponding to an idealized case
with all impacts since $t_0$ being retained, was already discussed. The surface of the Moon, for example, would record 
$\simeq$10 $D>10$-km asteroid impacts in CASE1B if $t_{\rm start}=0$. With $t_{\rm start}=200$ Myr, on the other hand, the Moon
would record only $\simeq$1-2 $D>10$-km asteroid impacts. Similar inferences can be made for Mars. If the Borealis 
basin formed at 4.4 Ga, for example, then $t_{\rm start}\simeq170$ My, and we would expect $\simeq$10-20 $D>10$-km 
asteroid impacts to be recorded after the Borealis basin formation (while $\simeq$30-60 would be recorded if 
$t_{\rm start}=0$).    
\section{Discussion}
Given the results reported in Section 3 we find it unlikely that asteroids could be responsible for the LHB, independently 
of whether the dynamical instability in the outer Solar System happened early or late. This is because the constraint from 
the Imbrium and Orientale basins implies that very large bodies impacted on the Moon at $\simeq$3.8-3.9 Ga (Wilhelms et 
al. 1987, St\"offler \& Ryder 2001). For example, the Imbrium basin, with the likely formation age $\simeq$3.9 Ga, 
have been excavated by a $D\sim130$-km impactor (Miljkovi\'{c} et al. 2016) or by a 
$D \sim 250$-km impactor (Schultz \& Crawford 2016), with the later estimate giving larger size mainly because
of the assumption of an oblique impact. The younger and smaller Orientale basin was probably made by a $D\simeq50$-100 
km impactor (Zhu et al. 2015, Miljkovi\'{c} et al. 2016) with the lower values in this range probably being more 
reasonable.

\begin{figure} 
\epsscale{0.485}
\plotone{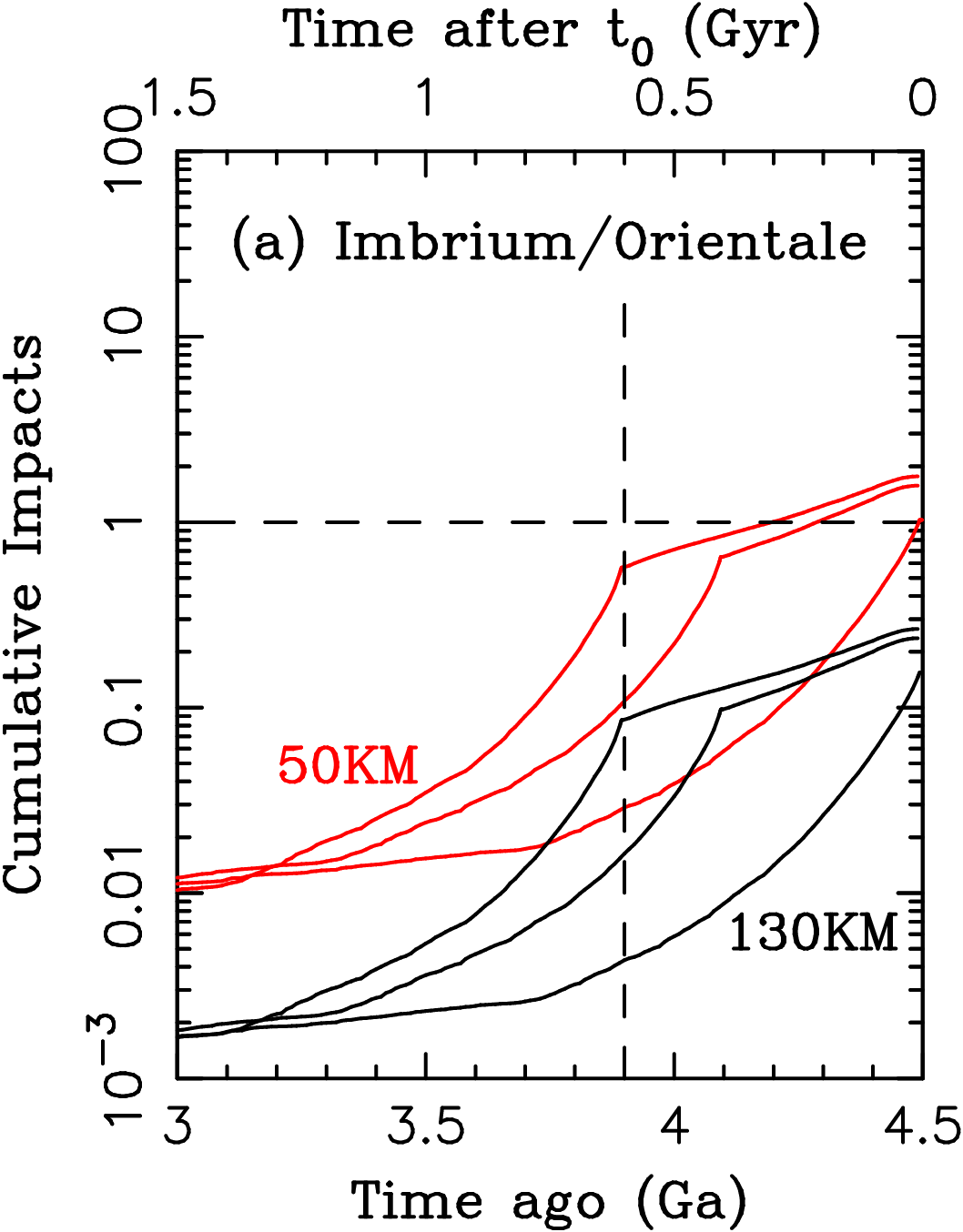}
\epsscale{0.435}
\plotone{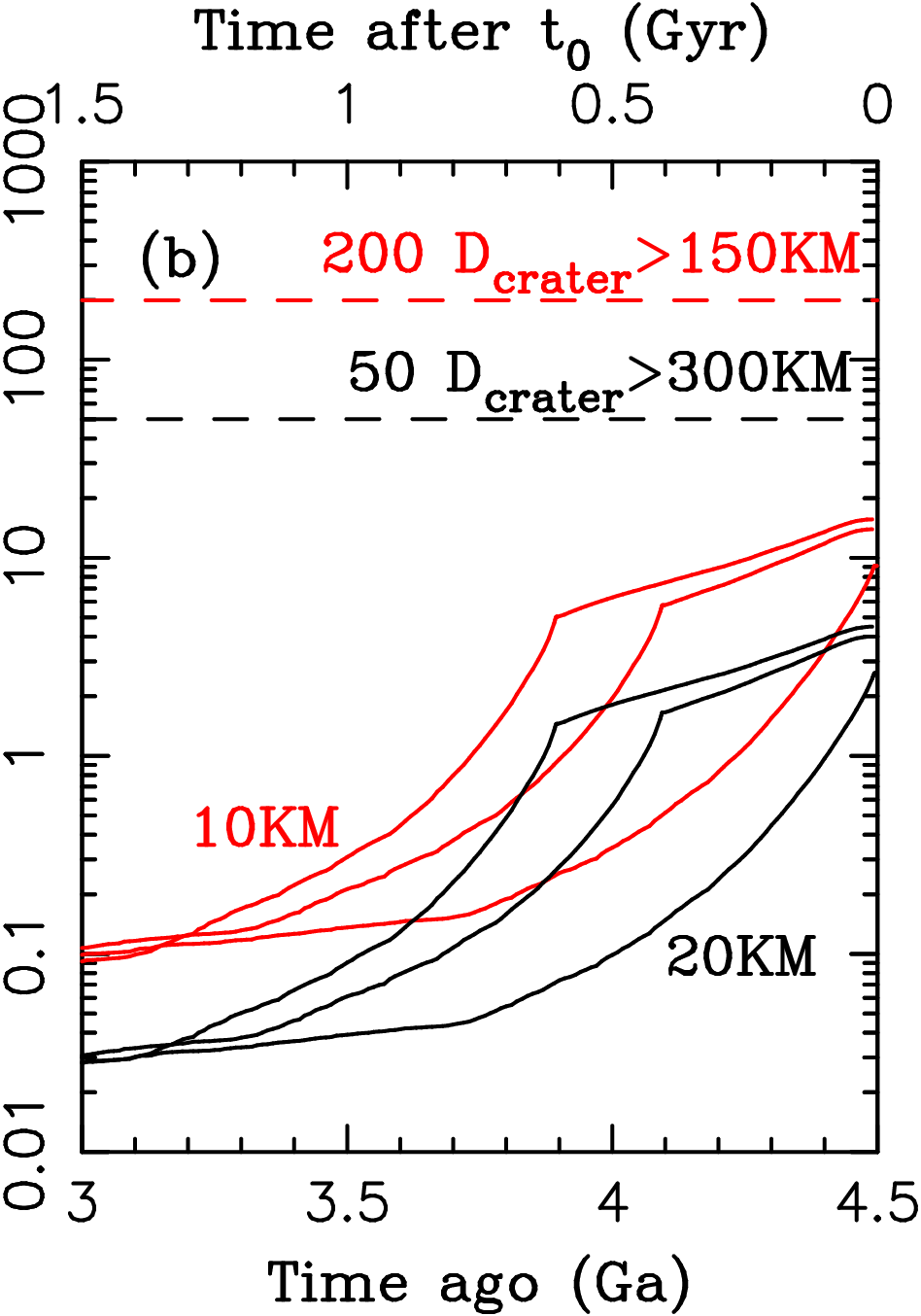}
\caption{A comparison of constraints from the large lunar craters and basins with the impact fluxes of asteroid
impactors in CASE1B. Panel (a) highlights the constraint from the Orientale and Imbrium basins. The family of solid black
curves shows the calibrated impact profile of $D=130$-km impactors for three different values of $t_{\rm inst}$ 
($=3.9$, 4.1 and 4.5 Ga). The red curves show the same for the $D=50$-km impactors. Panel (b) reports the results 
relevant for the $D>150$-km lunar craters and $D>300$-km lunar basins, here assumed to require $D>10$ km
and $D>20$ km asteroid impacts at 23 km s$^{-1}$. Note that panels (a) and (b) have different ranges on the Y axis.}
\label{moon}
\end{figure}

For a comparison, our absolute calibration of the impact flux implies that only $\sim$1-2 $D>50$ km asteroids 
hit the Moon over the whole Solar System history (Figure \ref{moon}). If the instability happened early ($t_{\rm inst}<100$ 
Myr), these large impacts would have happened early as well, leaving Orientale and Imbrium unexplained. If the 
instability happened late ($t_{\rm inst}>100$ Myr), the large impacts would be expected to occur any time between 
$t_0$ and $t_{\rm inst}$, when our cumulative impact profiles decline by a factor of $\sim$2 (Figure \ref{tinst}), 
and not be clustered near $t_{\rm inst}$. In addition, if $t_{\rm inst}\simeq3.9$ Ga, for example, which is the most 
favorable case for obtaining large impacts at 3.9 Ga, the probability of having one $D \geq 130$ km asteroid impact on 
the Moon at or after 3.9 Ga is only $\sim$9\%. 

At least $\sim$40 and up to $\sim$90 lunar basins (crater $D>300$ km) have been recognized or proposed (Wilhelms et al. 1987, 
Spudis 1993, Frey 2011, Fassett et al. 2012). From the crater scaling law appropriate for the lunar gravity
(e.g., Johnson et al. 2016), and using the mean impact speed $V_{\rm impact}=22.9$ km s$^{-1}$ from Table 4, we find 
that a $D\simeq20$ km impactor is needed to excavate a $D>300$ km crater. Figure \ref{moon}(b) shows the impact flux 
of $D > 20$-km asteroids on the Moon in our CASE1B model. In total, there are only $\sim$2-5 $D\simeq20$-km impacts, at 
least a factor of $\sim$8-20 below the lunar basin constraint discussed above. Perhaps the problem is with the existing 
scaling laws, which may be inappropriate for the basin-scale impacts, but Miljkovi\'c et al. (2016), modeling
the GRAIL data with the iSALE-2D hydrocode, found that the lunar record requires $\simeq$52 impacts of $D>20$-km bodies 
(we assumed the impact speed 23 km s$^{-1}$ to convert the $C$ parameter reported in their Table 2 to impactor's $D$). 
This is consistent with the estimate based on the existing crater scaling laws. 

Bottke et al. (2016) found, scaling the crater densities from the heavily cratered, ancient terrains on the far side of 
the Moon to the whole lunar surface, that $\sim$200 $D>150$-km craters should have formed in total. 
Assuming that a $D=150$-km lunar requires a $D\simeq 10$ km asteroid impactor, we find 
$\sim$200 $D>10$-km impactors would be needed to explain the lunar record. In contrast, the modeling work presented 
here suggests only $\sim$10-20 $D>10$-km impactors, at least an order of magnitude below the needed level. 
Assuming instead that smaller, $D=7$ km impactors can produce $D=150$-km lunar craters, and $t_{\rm inst}=4.1$-4.2 Ga, 
which was the preferred timing from Bottke et al. (2012), we find from CASE1B that $\simeq$34 $D>150$-km lunar craters 
would be made in total, of which only $\sim$5 would happen in the last 4 Ga.  

We conclude that the main belt asteroids were probably not a significant source of large lunar impacts before 3.5 Ga and 
some other, more powerful source needs to be invoked to explain the intense lunar bombardment history before 3.5 Ga. 
Note that CASE1B, which was used as a base of different estimates above, gives a larger impact flux than the REF case, 
roughly by a factor of 2 (Figure \ref{earth}). The results for the REF case are even more pessimistic.    

Additional constraints are provided by the impact record on Mars. Bottke et al. (2016) estimated that the
Mars-to-Moon ratio in the number of $D>150$ km craters is $\sim$1-3. Using the surface gravity dependence
of the crater scaling law (e.g., Housen et al. 1983), we estimate that a $D\simeq13$-km impactor is needed to make 
a $D=150$ km Martian crater. Since the number ratio of $D>10$ km to $D>13$ km main belt asteroids is $\simeq2$, we 
find that the Mars-to-Moon ratio in the {\it number of impactors} with $D>10$ km is $\sim$3-6. In contrast, 
we estimated in Section 3.8 that the Mars-to-Moon ratio expected for asteroid impactors is $\simeq7$ if 
$t_{\rm inst}=0$ or $\simeq10$-15 if $t_{\rm inst}\gtrsim300$ Myr (Figure \ref{ratio}). Thus, while the early 
instability case is at the upper limit of the Mars/Moon constraint, the late instability appears to be ruled 
out by this argument.

To make things work, perhaps one could assume that Mars's early craters have not been preserved on the 
surface, for example, because of the formation of the Borealis basin. According to Figure \ref{start}, and 
assuming that $t_{\rm start} \simeq 0$ for the Moon, roughly 50\% of the asteroid impacts on Mars would not be recorded if 
$t_{\rm start}=100$-200 Myr for Mars. This would make the asteroid impact flux more compatible with 
the Mars/Moon constraint. On the other hand, if the lunar magma ocean have crystallized only after $\sim$100-200 
Myr after the Moon-forming impact (Elkins-Tanton et al. 2011), $t_{\rm start}=100$-200 Myr for the Moon. This 
would make things worse. To compensate, an uncomfortably large $t_{\rm start}$ for Mars would have to be invoked.     

Comets (Gomes et al. 2005) and leftovers of the terrestrial planet accretion (Morbidelli et al. 2001, 2012) provided 
additional, and perhaps dominant source of impacts on the terrestrial worlds during early epochs. The leftovers are the 
subject of a separate work (Morbidelli et al. 2016). Here we used the numerical scheme described in Section 2 to 
calculate the impact flux of comets in CASE1B. The only change to this scheme was to distribute the initial 
orbits in the trans-Neptunian disk between 20 and 30 AU (Nesvorn\'y et al. 2013). We found that the overall impact 
probabilities of comets on Venus, Earth, Mars and Moon are $3.7\times10^{-7}$, $5.0\times10^{-7}$, $9.1\times10^{-8}$ 
and $2.6\times10^{-8}$, respectively. 

Nesvorn\'y \& Vokrouhlick\'y (2016) estimated that the trans-Neptunian disk contained $\sim$$3\times10^9$ comets 
with $D>10$ km (and $\sim$$1.5 \times 10^8$ comets with $D>50$ km). These numbers are uncertain by at least a 
factor $\sim$2. Using them we can roughly estimate that comets would be capable of producing $\sim$1,500 $D>10$-km impacts 
on the Earth (and $\sim$75 $D>50$-km impacts). This is $\sim$2.5-5 times more than the expected number of 
asteroid impacts in the same size range (Figure \ref{earth}). We thus see that comets could be very important for 
the cratering record in the inner solar system. According to our calculation, however, most cometary impacts 
occurred during a very narrow time interval ($\sim$20 Myr) after the instability, and may not be recorded on surfaces 
of the terrestrial worlds, if the instability happened early (i.e., before the surface crust grew thick enough to 
record large impacts). 

The analyses of Highly Siderophile Elements (HSEs; Kring \& Cohen 2002) and oxygen isotopes (Joy et al. 2012)
do not provide any firm evidence for cometary impactors. This would seem
surprising if the instability happened late, because, according to the above discussion, the cometary impact flux 
should dwarf the asteroid impact flux. A possible solution to this problem would be to postulate that comets 
disrupt before their orbits reach down to 1 AU (i.e., before they can impact on the terrestrial planets).
While there is an abundant evidence for bursts and disruptions of cometary nuclei, it is uncertain whether this
process can be efficient enough to eliminate {\it most} potential cometary impactors. A more straightforward solution
to this problem, in our opinion, is to assume that the dynamical instability happened early, and the cometary 
shower was over before the terrestrial worlds were able to record impacts. 
\section{Conclusions}
We find that asteroids were probably {\it not} responsible for the LHB, independently of whether the dynamical 
instability in the outer Solar System happened early or late, because the calibrated flux is not large enough 
to explain the ancient lunar record. Comets and leftovers of the terrestrial planet accretion probably provided 
a dominant source of impacts during the early epochs. In this case, it may not be helpful to invoke the {\it late} 
instability, because the existing geochemical evidence argues against comets being the main source of the LHB.
Thus, in the spirit of ``when you have eliminated the impossible, whatever remains, however improbable, must 
be the truth'' (Doyle 1890), we identify the leftovers of the terrestrial planet accretion to be the chief suspect. 
The recent work of Morbidelli et al. (2016) models the leftover contribution in detail and shows that it can 
match many constraints, including the Mars-to-Moon ratio discussed in Section 5. 

The early version of instability offers several advantages. First, the instabilities in dynamical systems 
generally happen early, not late. To trigger a late instability in the outer Solar System, and obtain 
planetary evolution histories that are compatible with the observed structure of the Kuiper belt (e.g., 
Nesvorn\'y 2015a), the parameters of the outer planetesimal disk need to be fine tuned (Gomes et al. 2005, 
Levison et al. 2011, Deienno et al. 2016), which is not very satisfactory. Second, the early version 
of the dynamical instability relaxes the terrestrial planet constraint (e.g., Agnor \& Lin 2012, Kaib \& 
Chambers 2016), because the secular resonances would sweep through the inner Solar System {\it before}
the terrestrial system was in place. Third, there is no problem with the cometary impactors if 
the instability happened early, because early cometary impacts would not be recorded in the geochemical 
markers.
 
As a final note, we point out that even if the dynamical instability in the outer Solar System was not the actual 
cause of the LHB, as we argued here, there is plenty of other evidence that the instability must have happened. 
For example, the instability is needed to extract giant planets from their original, presumably resonant configuration 
(e.g., Masset \& Snellgrove 2001, Morbidelli et al. 2007) and establish the secular structure of their present 
orbits (Tsiganis et al. 2005, Morbidelli et al. 2010). The populations of small bodies in the Solar System 
provide additional strong constraints on the planetary instability (and migration). The instability helps to explain 
the orbital distribution of the asteroid belt (Minton \& Malhotra 2009, Morbidelli et al. 2010, Roig \& Nesvorn\'y 
2015, Toliou et al. 2016), capture of Jupiter Trojans (Morbidelli et al. 2005, Nesvorn\'y et al. 2013) and irregular satellites 
(Nesvorn\'y et al. 2007, 2014), and various properties of the Kuiper belt (Levison et al. 2008, Nesvorn\'y 2015a,b, 
Nesvorn\'y \& Vokrouhlick\'y 2016). 
  
\acknowledgments
This work was supported by NASA's SSERVI and CNPq's Ciencias sem Fronteiras programs. The CPU-expensive simulations 
were performed on NASA's Pleiades Supercomputer. One 50,000 asteroid simulation over 4.5 Gyr required
about 720 hours on 125 Ivy-Bridge nodes (20 cores each). We greatly appreciate the support of the NASA Advanced
Supercomputing Division. We thank M. Delbo for kindly providing the catalog of sizes of the main belt asteroids, and 
S. Marchi, A. Morbidelli and D. Vokrouhlick\'y for numerous useful and stimulating discussions. We also thank 
an anonymous reviewer for helpful comments on the manuscript.

\clearpage
\begin{table}
\centering
{
\begin{tabular}{lrrrr}
\hline \hline
Case                  & $N_{\rm ast}$                & $a$     & $e$ & $i$        \\  
                      &                           & (au)    &     & (deg)    \\  
\hline
REF                   & 20,000 & 1.6-3.3    & 0-0.4   & 0-30   \\     
STEP                  & 50,000 & 1.6-3.3    & 0-0.4   & 0-20   \\    
CASE1                 & 10,000 & 1.6-3.3    & 0-0.4   & 0-30   \\             
CASE1B                & 50,000 & 1.6-3.3    & 0-0.4   & 0-20   \\                   
\hline \hline
\end{tabular}
}
\caption{A summary of numerical integrations performed in this work. In addition to the initial 
ranges of asteroid orbits given here, we also imposed that $q=a(1-e)>1.6$ au and $Q=a(1+e)<4.5$ 
au such that the initial orbits were {\it not} planet crossing. In the second column, $N_{\rm ast}$ denotes
the initial number of asteroids used in each case.}
\end{table}

\clearpage
\begin{table}
\centering
{
\begin{tabular}{lrr}
\hline \hline
Case                  & REF                & CASE1B           \\  
                      & ($10^{-2})$          & ($10^{-2})$          \\    
\hline
Venus (direct)        & 1.2    & 1.5   \\
Venus (\"Opik)        & 1.3    & 1.2   \\    
Earth (\"Opik)        & 1.0    & 1.0   \\      
Earth (direct)        & 1.0    & 1.1   \\
Mars (direct)         & 0.36   & 0.39   \\   
Moon (\"Opik)         & 0.044  & 0.053  \\                          
\hline \hline
\end{tabular}
}
\caption{The impact probabilities on different target bodies in the REF and CASE1B simulations and 
$t_{\rm inst}=0$. Here the probabilities were normalized to one asteroid in the original source region ($1.6<a<3.3$ au).
These results were obtained for a case where the Gaussian mask in $e$ and $i$ was used to set up the initial distribution 
of asteroid orbits (Section 2.5). The impact probabilities reported in rows labeled `direct' were obtained from the 
impacts recorded by the {\it Swift} integrator. The impact probabilities labeled `\"Opik' were obtained from the 
\"Opik code.} 
\end{table}

\clearpage
\begin{table}
\centering
{
\begin{tabular}{lrr}
\hline \hline
                     & REF                & CASE1B           \\      
\hline
$\tau_1$ [Gyr]        & 0.037    & 0.027  \\     
$\tau_2$ [Gyr]        & 0.16     & 0.16   \\  
$\tau_3$ [Gyr]        & 1.5      & 100   \\                
$F_1$ [Myr$^{-1}$]    & 0.50     & 3.9   \\     
$F_2$ [Myr$^{-1}$]    & 0.31     & 0.60   \\  
$F_3$ [Myr$^{-1}$]    & 0.0027   & 0.0006   \\                
\hline \hline
\end{tabular}
}
\caption{An exponential parametrization of the calibrated (differential) impact flux from Eq.~(\ref{exp1}). 
With the parameter values given here, Eq. (\ref{exp1}) can be used to compute the number of impacts
of $D>10$-km asteroids on the Earth. The impact fluxes for other target bodies can be obtained by 
rescaling the flux according to the values listed in Table 2.}
\end{table}

\clearpage
\begin{table}
\centering
{
\begin{tabular}{lrr}
\hline \hline
Case                  & REF                & CASE1B           \\  
                      & (km s$^{-1}$)         & (km s$^{-1}$)          \\    
\hline
Venus                 & 25.8   & 28.6   \\     
Earth                 & 21.0   & 23.5   \\      
Mars                  & 12.0   & 13.7   \\   
Moon                  & 20.7   & 22.9   \\                          
\hline \hline
\end{tabular}
}
\caption{The mean impact speeds on different target bodies in the REF and CASE1B models (focusing 
included). }
\end{table}

\clearpage

\clearpage

\clearpage

\clearpage

\clearpage

\clearpage

\clearpage

\clearpage

\clearpage

\clearpage

\clearpage

\clearpage

\clearpage

\clearpage

\clearpage

\clearpage

\clearpage

\clearpage

\clearpage

\clearpage

\end{document}